%% file: ms_August_2016_astro_ph.tex
\documentclass{aastex6}

\usepackage{epsfig}
\usepackage{hyperref}
\usepackage{graphicx}
\usepackage{epstopdf}
\usepackage{pslatex}
\usepackage{color}
\usepackage[caption=false]{subfig}
\usepackage{float}
\usepackage{subfloat}
\usepackage{cleveref}
 \usepackage{natbib}
 \usepackage[english]{babel}

\newcounter{iso}
\newcommand{\rxn}{\refstepcounter{iso}%
                  (\oldstylenums{\theiso})}

\newcommand\osref[1]{(\oldstylenums{\ref{#1}})}

\begin{document}

\title{Weak and Compact Radio Emission in Early High-Mass Star Forming Regions: I. VLA Observations} 
\bigskip\bigskip

\author{V. Rosero\altaffilmark{1,2}, P. Hofner\altaffilmark{1,}\footnote{Adjunct Astronomer at the National Radio Astronomy Observatory, 1003 Lopezville Road, Socorro, NM 87801, USA.}, M. Claussen\altaffilmark{2}, S. Kurtz\altaffilmark{3}, R. Cesaroni\altaffilmark{4}, E. D. Araya\altaffilmark{5}, C. Carrasco-Gonz\'alez\altaffilmark{3}, L. F. Rodr\'iguez\altaffilmark{3}, K. M. Menten\altaffilmark{6}, F. Wyrowski\altaffilmark{6}, L. Loinard\altaffilmark{3,6} \and S. P. Ellingsen\altaffilmark{7}}

\altaffiltext{1}{Physics Department, New Mexico Tech, 801 Leroy Pl., Socorro, NM 87801, USA}

\altaffiltext{2}{National Radio Astronomy Observatory, 1003 Lopezville Rd., Socorro, NM 87801, USA}
\altaffiltext{3}{Instituto de Radioastronom{\'\i}a y Astrof{\'\i}sica, 
Universidad Nacional Aut\'onoma de M\'exico, Morelia 58090, M\'exico}
\altaffiltext{4}{INAF, Osservatorio Astrofisico di Arcetri, Largo E. Fermi 5, 50125 Firenze, Italy}
\altaffiltext{5}{Physics Department, Western Illinois University, 1 University Circle, Macomb, IL 61455, USA}
\altaffiltext{6}{Max-Planck-Institute f\"{u}r Radioastronomie, Auf dem H\"{u}gel 69, 53121 Bonn, Germany}
\altaffiltext{7}{School of  Physical Sciences, University of Tasmania, Private Bag 37, Hobart, Tasmania 7001, Australia}

%

\begin{abstract}

{We present a high sensitivity radio continuum survey at 6 and 1.3$\,$cm using the  Karl G. Jansky Very Large Array towards a sample of 58 high-mass star forming regions. Our sample was chosen from dust clumps within infrared dark clouds with and without IR sources (CMC--IRs, CMCs, respectively), and hot molecular cores (HMCs), with no previous, or relatively weak radio continuum detection at the $1\,$mJy level. Due to the improvement in the continuum sensitivity of the VLA, this survey achieved map rms levels
of $\sim$ 3--10 $\mu$Jy beam$^{-1}$ at sub-arcsecond angular resolution. We extracted  70 centimeter continuum sources associated with $1.2\,$mm dust clumps. Most sources are weak, compact, and are prime candidates for high-mass protostars. 
Detection rates of radio sources associated with the mm dust clumps for CMCs, CMC--IRs and HMCs are 6$\%$, 53$\%$ and 100$\%$, respectively. This result is consistent with increasing high-mass star formation activity from CMCs to HMCs.
The radio sources located within HMCs and CMC--IRs occur close to the dust clump centers with a median offset from it of 12,000\,AU and $4,000\,$AU, respectively.
We calculated $5 - 25\,$GHz spectral indices using power law fits and obtain a median value of 0.5 (i.e., flux increasing with frequency), suggestive of thermal emission from ionized jets. In this paper we describe the sample, observations, and detections. The analysis and discussion will be presented in Paper II.}

\end{abstract}


\section{Introduction}

High-mass stars  (M ${\scriptstyle\gtrsim}\,$ 8 M$_\odot$) are rare, they evolve on short timescales, and they are usually born in clusters  located in highly
obscured, and distant (typically ${\scriptstyle>}$\,1 kpc) regions.
These observational challenges have made the study of their earliest evolutionary phases difficult. The population of high-mass stars in the Galaxy has been probed by cm-wavelength radio continuum emission at least since the early single-dish 
compact H{\small II} regions surveys  \citep[e.g.,][]{1967ApJ...147..471M}. Subsequent interferometric surveys \citep[e.g.,][]{1989ApJ...340..265W, 1993ApJ...418..368G, 
1994ApJS...91..659K} discovered even more compact emission (the 
so-called ultracompact,  or UCH{\small II} regions), or yet denser regions of ionized gas called hypercompact H{\small II} regions (HCH{\small II}, e.g., \citealp{2003ApJ...597..434W, 2011ApJ...739L...9S}).
All these regions detected in the past -- whether compact or ultra/hypercompact -- share two important characteristics. First, they are fairly bright at radio wavelengths, with cm flux densities ranging from a few mJy to a few Jy. Second, they arise from a 
relatively late stage in the star formation process when nuclear burning likely has already begun, thus producing copious amounts of UV radiation that photoionize the gas surrounding the young star.

Subsequently, with the goal of finding candidates of earlier evolutionary phases of high-mass star formation, researchers turned to the study of dense condensations in molecular clouds. In this paper we will refer to molecular (or dust) {\it clumps}
as structures of $\sim 1\,$pc, which are typically probed by mm single dish studies \citep[e.g.,][]{2002ApJ...566..945B, 2006ApJ...641..389R}. In addition, we will refer to {\it cores} as substructures of size $\sim$0.1 pc that are found within clumps and are
typically probed by radio/mm interferometers \citep[e.g.,][]{2000prpl.conf..299K, 2010A&A...509A..50C}.

Several studies of such condensations in the 1990s  led  to the identification of hot molecular cores (HMCs).
HMCs have a large amount of hot molecular gas ($\sim$ 10$^{2}$\,M$_\odot$, T $>$ 100$\,$K; see Section \ref{Sample_desc}).
HMCs are presumably heated by one or more embedded high-mass protostars, and in general have weak or previously undetectable radio continuum emission at sensitivities of tens of $\mu$Jy
\citep[e.g.,][]{1986ApJ...303L..11G, 1991A&A...252..278C, 1992A&A...256..618C, 1993A&A...276..489O, 2000prpl.conf..299K, 2005IAUS..227...59C, 2010A&A...509A..50C}. They are thus thought to represent an evolutionary phase prior to UC/HC H{\small II} 
regions. 
 
Candidates for an even earlier, possibly pre-stellar, phase were discovered by mm and submm continuum studies of the so-called infrared dark clouds (IRDCs).  IRDCs harbor molecular structures of similar masses and densities as HMCs {(see Table 1 of 
\citealp{2006ApJ...641..389R}). However, the  temperatures are much lower  (T$\sim$10--20 K; \citealp{2006A&A...450..569P, 2010A&A...518L..98P}), which suggests an earlier evolutionary phase than HMCs.  While clearly not all IRDCs are presently
forming stars, the most massive and opaque condensations within IRDCs might form OB stars  \citep{2010ApJ...723L...7K}. 

Several studies have attempted to understand the molecular  condensations where high-mass stars  form and to establish an evolutionary sequence
for them \citep[e.g.,][]{1996A&A...308..573M, 1998A&A...336..339M, 2000A&A...355..617M, 2008A&A...487.1119M}.
Based on recent VLA NH$_3$ \citep{2013MNRAS.432.3288S}, and ATCA H$_2$O maser and cm continuum \citep{2013A&A...550A..21S} observations of a large number of high-mass star forming candidates,
these authors have suggested an evolutionary sequence from quiescent starless cores, with relatively narrow NH$_3$ lines and low temperatures to protostellar cores which already contain IR point sources, and show larger linewidths
and temperatures, as well as the presence of ionized gas. While this classification is reasonable, we still lack an evolutionary scheme for the earliest phases of high-mass star formation.

The present work is an attempt to characterize the earliest phases (prior to UC/HC H{\small II} regions) of high-mass star formation making use of the high continuum sensitivity of the Karl G. Jansky Very Large Array (VLA)\footnote{The National Radio Astronomy 
Observatory is a facility of the National Science Foundation operated under cooperative agreement by Associated Universities, Inc.}. A number of physical processes which cause cm continuum emission have
been suggested to be present during early high-mass star formation. Most of these relate to the disk/flow systems which are expected around high-mass protostars. For instance, \citet{2007ApJ...664..950R} proposed an ionized disk
around the Orion I protostar, \citet{1996ApJ...471L..45N} predict weak cm continuum emission from disk accretion shocks, and \citet{1995ApJ...443..238R}  detected a synchrotron jet in the W3(H$_{2}$O) protostar. Most  low-mass protostars
show molecular flows which are driven by ionized -- mostly thermal -- jets (see \citealp{2010ApJ...725..734G} for a summary), and since molecular outflows are also prevalent  in regions where high-mass protostellar candidates are found  \citep[e.g.,][]{1996ApJ...457..267S, 2001ApJ...552L.167Z, 2002A&A...383..892B}, ionized jets associated with high-mass protostars are also expected. Furthermore, \citet{2007MNRAS.380..246G} interpreted weak and compact continuum sources in S106 and S140 as equatorial ionized 
winds from high-mass protostars, and \citet{2016ApJ...818...52T} have predicted the existence of weak outflow-confined H{\small II} regions in
the earliest phases of high-mass star formation.

At the typical distances of several kpc for high-mass star forming regions, these processes predict flux densities in the $\mu$Jy range and are now accessible to observations with the upgraded VLA. 
We have thus carried out a high sensitivity (rms$\,\sim$ 3--10 $\mu$Jy beam$^{-1}$) VLA survey to search for radio continuum emission at 6 and 1.3 cm with sub-arcsecond angular resolution towards candidates of high-mass star forming  sites
at evolutionary phases earlier than UC and HCH{\small II} regions. In this paper we present our  VLA radio continuum survey of 58 regions selected from the literature which had no previous,  or relatively weak radio continuum detection at the 1 mJy level. 
Below we describe the sample and the observations, present the detections and describe their physical properties. The analysis, interpretation and conclusions of this survey will be presented in Rosero et al. (in preparation; hereafter Paper II). \\

\section{Observations and Data Reduction}\label{observations_sect}

\subsection{Sample Selection} \label{Sample_desc}

The main goal of this survey is to study the centimeter continuum emission from high-mass protostellar candidates.  To ensure an evolutionary phase of our targets earlier
than UC and HCH{\small II} regions,
the most important selection criterion was the non detection, or a very low level ($<$ 1 mJy), of emission at cm wavelengths. In particular, most of our targets are 
non-detections in the CORNISH survey \citep{2008ASPC..387..389P} which has a typical image rms of $0.3\,$mJy beam$^{-1}$ at $6\,$cm. We have further attempted to define a sample which
would represent a progression in high-mass star formation, based on FIR luminosity, mid-IR emission and temperature of the associated
molecular and dust clumps.  This resulted in a total sample of 58 targets  grouped into three categories: hot molecular cores (HMCs; 25 sources),
cold molecular clumps with mid-IR association (CMC$-$IRs; 15 sources), and cold molecular clumps (CMCs; 18 sources) devoid of IR associations.
We note that our sample is not strictly defined in a statistical sense as it contains a number of biases. Rather, a variety of prominent sources were included
to search for radio counterparts.
However, while  our classification cannot be considered completely accurate, it does follow the trend of increasing star forming activity from CMCs to HMCs.
Our approximate classification scheme is similar to the one used by \citet{2013MNRAS.432.3288S}. Below we briefly comment  on the properties of the three groups. 

\begin{enumerate}
\item HMCs: These are high luminosity ($> 10^3\,$L$_\odot$) IRAS sources associated with dense gas, outflows, masers, and have dust and gas
temperatures  $\sim 20\,$K on the $1\,$pc clump scale \citep{1996A&A...308..573M, 2002ApJ...566..931S}. Many HMCs have recently been studied interferometrically in high excitation molecular lines
(e.g., \citealt{2010A&A...509A..50C, 2014ApJ...786...38H}). These studies showed that on the core scale ($0.1\,$pc), HMCs have temperatures $ >$ 100\,K,
masses of $\sim$ 10$^{2}$ M$_\odot$ and densities of $\approx 10^7\,$cm$^{-3}$.  For further information on HMC properties see \citet{2006ApJ...641..389R}, and references therein.
The current interpretation is that HMCs contain highly embedded -- possibly still accreting -- high-mass stars, just prior to the
development of an UC/HCH{\small II} region. 23 HMCs in our sample were selected from  \citet{2002ApJ...566..931S}, one is an IRDC HMC from \citet{2011ApJ...741..120R} and G23.01$-$0.41
is a high-mass protostar with a rotating toroid, 4.5 $\mu$m excess emission and rich maser activity (e.g., \citealt{2008ApJ...673..363F, 2008ApJS..178..330A, 2014A&A...565A..34S}).  

\item CMC--IRs: Representing an earlier evolutionary phase of high-mass star formation, they are cold ($T < 25\,$K), massive ($M > 100\,M_\odot$)  clumps found by MSX and Spitzer IRDC surveys. \citet{2006ApJ...641..389R} presented extensive $1.2\,$mm mapping, and \citet{2009ApJS..181..360C} found that $\approx 50\%$ of these clumps
showed signs of early stellar activity as revealed by Spitzer. CMC$-$IRs are  clumps associated with a 24$\,\mu$m point source and in most cases also
associated with 4.5$\,\mu$m excess 
emission,  likely tracing shocked gas \citep{2006ApJ...645.1264S, 2008AJ....136.2391C}. These clumps are expected to contain high-mass protostellar objects still in the process of accretion, which have not substantially heated the molecular environment (e.g., G11.11$-$0.12P1,  \citealt{2014ApJ...796..130R}). We selected  ten CMC$-$IRs from  \citet{2006ApJ...641..389R} and five additional  CMC$-$IRs were selected from other studies  \citep{2010A&A...518L.123L, 2006A&A...447..929P, 2010A&A...518L..78B, 2010ApJ...715.1132O}. 

\item CMCs: The sources from this target group were mainly selected from the original sample of \citet{2006ApJ...641..389R}, the distinction being that no mid-IR source was
reported by \citet{2009ApJS..181..360C}.  The  flux upper limits at 24 $\mu$m are typically $<$90 mJy (except for  G53.11$+$00.05 mm2; \citealt{2010ApJ...715..310R}).  
The absence of mid-IR sources in these massive cold clumps is consistent  with a pre-stellar phase, possibly at  the onset of collapse. We note that often CMCs are accompanied by
CMC-IRs within the same IRDC, hence star formation is apparently occurring in their neighborhood, which could influence the (future) star formation activity
within the CMCs. In total, we selected 15 CMCs from \citet{2006ApJ...641..389R} and  3 more  were selected from the starless clump candidates in Vulpecula  \citep{2010ApJ...715.1132O}.

\end{enumerate}

\noindent
In Figure \ref{sample_hist} we show the distribution of  kinematic distances, bolometric luminosities (estimated from IRAS, Spitzer and Herschel),  masses,
 and column densities for our sample. The data are taken from the literature; references are given in Table \ref{obs}.
The majority of the  sources are at distances between 1 and 6 kpc, but some HMCs 
 are as distant as $\sim$ 12 kpc (see Figure \ref{sample_hist}). The bolometric luminosity spans  4 orders of magnitude, with a clear separation between
 the low luminosity CMCs and the high luminosity HMCs. The CMC--IRs straddle this range. The HMCs
 tend to be more massive, whereas the CMCs and CMC--IRs have a similar mass distribution.  The range of column densities in the three types of sources is 1.8$\times 10^{22}$ -- 8.5$\times 10^{23}$ cm$^{-2}$. The majority of sources have sufficiently high column density to warrant high-mass star formation \citep{2008Natur.451.1082K, 2010A&A...517A..66L}.

\subsection{VLA Observations}
VLA continuum observations (project codes 10B$-$124 and 13B$-$210) at 6 and 1.3 cm were made for all sources in the sample.  To achieve similar resolution and ({\it u,v}) coverage, we observed with scaled arrays, using the A-configuration for  6 cm and the B-configuration for 1.3 cm. A summary of the VLA observational parameters is given in Table \ref{VLA_parameters}. 
A list of phase calibrators used to make the observations at 6 and 1.3 cm  is given in Table \ref{cal}. 3C286 was used as flux density calibrator for all sources except for
LDN1657A$-$3 and UYSO1, for which 3C147 and 3C48 were used at 6 cm and 1.3 cm, respectively.

The 58 sources that we observed are summarized in Table \ref{obs}. Column 1 gives the region name, column 2 gives the band frequency, while columns 3, 4, 5 and 6 give the R.A, Decl., Galactic longitude and latitude of the pointing center. In some cases, several sources were sufficiently close to each other that they could be observed in a single pointing within the $25.5\,$GHz primary beam; in Table \ref{obs} this is indicated with a footnote.  Columns 9, 10, 11 and 12 list whether we detected cm continuum  sources  coincident with the dust clumps mapped by \citet{2006ApJ...641..389R} and  \citet{2002ApJ...566..945B} (see Section \ref{core_association}),  the core/clump type, distance to the  region and bolometric luminosity, respectively. Column 13 gives references for the distance and luminosity of each region. \\

\newpage
\subsubsection{6 cm Observations}
The 6 cm observations were made in the A configuration between 2011 June and August, providing a typical angular resolution of about  0\rlap.$^{\prime \prime}$4. Two $1\,$GHz wide basebands (8--bit samplers) were employed, 
centered at 4.9 and 7.4 GHz.
Each baseband was divided into 8 spectral windows (SPWs), each  with a bandwidth of 128 MHz.
Therefore, the data were recorded in 16 unique SPWs, each comprised of 64  $2\,$MHz wide channels, resulting in  a total bandwidth of 2048 MHz.
The SPWs were configured to avoid the strong methanol maser emission at 6.7 GHz.
The observations were taken in scheduling  blocks of about four regions per block, performing
alternating observations on a target source for 720 s and a phase calibrator for 90 s.
The typical total on-source time  was  $\sim\,$40 minutes.

The data were processed  using NRAO's Common Astronomy Software 
Applications (CASA)\footnote{http://casa.nrao.edu} package.  
Eight channels at the edges of each baseband  
were flagged due to  substantial band roll-off (and therefore loss of sensitivity). In addition, we inspected  the data for radio frequency interference (RFI) or other problems, performing ``flagging'' when needed. The flux density scale was set via standard NRAO models for the flux  calibrators and using the Perley--Butler (2010)  flux scale. We used the {\tt gencal}  task to  check for antenna position corrections and also to apply a gain curve and antenna efficiency factors.
Delay and  bandpass solutions were formed based on observations of the flux density calibrator. These solutions were applied when solving for the final amplitude and phase calibration using the task {\tt gaincal} over the full bandwidth. 
 We measured the flux density of the phase calibrators using the task {\tt fluxscale}. The amplitude, phase, delay and bandpass solutions were applied to the target sources using the task {\tt applycal}. 
The images were made using the {\tt clean} task and  Briggs ${\tt ROBUST}=0.5$ weighting. Because most of the detections have low S/N 
($<$ 20), no self-calibration was attempted. 

As a consistency check, and to ensure the absence of line contamination or RFI, we imaged and 
inspected each SPW separately. Moreover, each 1~GHz baseband was imaged separately to provide a better estimate 
of spectral index. Finally, a combined image was made, including the data from both basebands. All maps were primary beam corrected.
The synthesized beam size and position angle (P. A.)  and rms noise of the combined image for each region are shown in columns 7 and 8 of  Table \ref{obs}. \\

\subsubsection{1.3 cm Observations}
The 1.3 cm observations were made in the B configuration, acquiring the first half of the data  between 2010 November and  2011 May,  and the second half between 2013 November  and 2014 January. The correlator set-up was the same as that used at 6 cm, with the two basebands centered at 21 and 25.5 GHz.
The observations were mostly taken in scheduling blocks of  two regions each, performing
alternating observations on a target source for 240 s and a phase calibrator for 60 s.
The typical total on-source time  was  $\sim\,$40 minutes. 
Pointing corrections using the referenced pointing procedure were obtained  every hour and were applied during the observations.\\

The data reduction was done in the same fashion as for the 6 cm observations.  
In addition, we corrected for atmospheric opacity using the weather station information from the {\tt plotWeather} task.
The images were made using the {\tt clean} task using natural  weighting to provide the best sensitivity in the maps. As at 6 cm, we imaged each baseband individually (for spectral index) and together 
(for morphology and improved S/N). All  maps were primary beam corrected.
The synthesized beam size and position angle (P. A.)  and rms noise of the combined image for each region are shown in columns 7 and 8 of  Table \ref{obs}. \\

\section{Results}\label{results_section}

Radio sources were identified in each of the baseband-combined images described in Section~\ref{observations_sect}. The criterion adopted for defining a radio detection is that the peak intensity $I_{\nu}$ is $\geq 5$ times the image rms ($\sigma$) in either C or K-band. 
Subsequently, in each $1\,$GHz wide band we determined the peak position, flux density $S_{\nu}$ and peak brightness $I_{\nu}$ by  enclosing each radio source in a  box using the task {\tt viewer} of CASA.
Table \ref{Parameters} reports the parameters for all detected radio sources within the 1\rlap.$^{\prime}$8  FWHM primary beam at 25.5 GHz.  This table, for each observed region, lists each detected radio source, its
frequency, peak position, flux density, peak intensity, 
morphology, millimeter clump association, and spectral index (see below).  We describe the morphology either as compact (C) if the detection is apparently unresolved or shows no structure on the scale of a few synthesized beams,
or resolved (R) if the detection is more extended. Several sources in our survey are very weak; the occasional presence of image artifacts (e.g., the emission lying in a negative bowl) inhibited in some cases an accurate measurement of the flux density $S_{\nu}$. In 
these cases we only report the
peak intensity $I_{\nu}$ (see section \ref{alpha}). Several sources were only detected in one of the baseband-combined maps; in this case we report a 3$\sigma$ limit value for the peak intensity in each $1\,$GHz baseband.

Since the goal of this survey was to detect radio emission from high-mass protostars we restrict the discussion to the radio sources that are associated with the dust clumps mapped by \citet{2006ApJ...641..389R} and  \citet{2002ApJ...566..945B} (see below). 
In Table \ref{obs} column 9 we indicate
whether there are radio sources  coincident with these mm clumps. To distinguish these sources further, in Table \ref{Parameters} for each region we label radio detections which are located within the mm--clump by capital letters, i.e.,  ``A'', ``B'', ``C'', etc., from east to west.
Sources that are within the 25.5 GHz FWHM primary beam but outside of the mm--clump are labeled by their Galactic coordinates. Contour images for all radio sources associated with mm--clumps are shown in Figure \ref{fig:maps}.\\

At the very low rms values of this survey many  radio sources were detected; this is particularly the case in the $6\,$cm maps due to their large primary beam (9\rlap.$^{\prime}$2 at 4.9 GHz). Hence it is necessary to consider background contamination 
by 
extragalactic radio sources.  At $6\,$cm, above a 5$\,\sigma$ flux density of 25 $\mu$Jy, we expect 0.52 extragalactic radio sources per arcmin$^{-2}$ \citep{1991AJ....102.1258F}. Indeed, several of these radio sources, mainly outside of the primary beam, show double lobed 
morphologies typical of radio galaxies. At the typical dust clump size of $\sim$30$^{\prime \prime}$, we thus expect to detect about 0.13 sources on average in the 6 cm maps.  Since we observed a total of 58 clumps, we expect to have about 8 extragalactic sources in the entire sample. 

To quantify this number for the $1.3\,$cm bands, we use the 2 cm source counts model from \citet{2005A&A...431..893D} scaled to the 10C survey source counts by \citet{2011MNRAS.415.2708A}. The average rms noise from our data at 1.3 cm is $\sim$9 $\mu
$Jy beam$^{-1}$  and from this model we predict that we should see 0.12 sources arcmin$^{-2}$, or $\sim$0.4 radio sources within the 1\rlap.$^{\prime}$8 FWHM primary beam of the 25.5 GHz maps above a 5$\sigma$ flux density of 45 $\mu$Jy. Therefore,  
within the mm-clump regions at an average size of $\sim$30$^{\prime \prime}$ the likelihood of detecting extragalactic sources is low ($\sim 0.03$ radio sources per clump) at 1.3 cm. Since we observed a total of 58 clumps, we expect to have about 2 extragalactic sources in the entire sample. 

Of the 58 dust clumps we observed, we did not detect  radio sources toward 24 of  them.
For the remaining 34 dust clumps, we report a total of  70 radio sources associated with the dust clumps (see Section~\ref{core_association}), 
often finding two or more sources within a single clump (see Section~\ref{multiplicity}).  
We also report a total of  62 sources\footnote{Due to a large offset of the dust clump from the pointing center, the 
19282$+$1814 region has been excluded from the statistics. However, we include contour maps at $6\,$cm and report the radio sources detected in Table \ref{Parameters}.} that are outside the radius of the mm-clump but within the 25.5 GHz primary beam. 
The detection statistics for the sample are summarized in Table~\ref{Summary_det}.

\subsection{Association with Millimeter Clumps}\label{core_association}

Among the detections we have sources associated with dense gas traced by dust mm emission, and that are therefore likely part of the star forming region.  The 1.2 mm clump data were taken from \citet{2006ApJ...641..389R} and \citet{2002ApJ...566..945B} for 
the IRDC clumps and HMCs, respectively, except for some individual sources. We classified a radio source as associated with the $1.2\,$mm clumps if it was located within the FWHM of the dust emission.
Toward the dust mm clumps we detected radio continuum emission in 1/18 (6$\%$) of CMCs, 8/15 (53$\%$) of CMC$-$IR and 25/25 (100$\%$) of HMCs, with a  total of  70 detections.
This detection statistics follows the expected trend of increasing star formation activity from 
CMCs to HMCs, i.e., in more evolved regions the probability of detecting weak and compact radio sources is higher than in less evolved regions.  

Our data, in conjunction with the $1.2\,$mm maps, allow us to check where star formation occurs within the clumps. Figure \ref{offset_mm_radio} shows the distribution of position offsets between radio sources and the peak of the $1.2\,$mm emission.  
The FWHM median angular sizes for each type of clump in this sample are 31$^{\prime \prime}$, 25$^{\prime \prime}$ and 18$^{\prime \prime}$ for CMCs, CMC--IRs and HMCs, respectively.
When no mm size for the clump is available in the literature, we use the FWHM median size for the specific type of clump. 
Multiple radio sources located closer than 3$^{\prime \prime}$ are counted as a single radio source in this histogram. 
From Figure \ref{offset_mm_radio} we see that for HMCs the radio source distribution is strongly peaked toward the center of the mm clumps. The median distance of the radio sources from the center of the mm dust clumps in HMC sources is  $0.06\,$pc,
or about $10^4\,$AU, and the distribution peaks at $0.04\,$pc, or about $8000\,$AU. 
For CMC--IR sources we see a flatter distribution, which due to the lower
number of radio detections could still be consistent with the HMC distribution. In fact, using  a Kolmogorov-Smirnov test to compare the spatial offset distribution of the HMCs and CMC--IRs we estimated a p-value$=$ 0.34, i.e., there is no
strong evidence that for these two core/clump types the offset distribution is different. Finally, for CMC sources only 1 radio source is detected coincident with the dust clumps, thus no statement can be made about their general offset distribution.

 \subsection{Spectral Indices}\label{alpha}
 
Because the observations were made in scaled arrays, the $6\,$cm and the $1.3\,$cm images are sensitive to the same angular scales. Hence, these images  allow
a determination of the source spectral indices $\alpha$, defined via $S_{\nu}$ 
$\propto$ $\nu^{\alpha}$, independent of source extent. The spectral index for each component 
was calculated using the flux density at the four central observed frequencies: 4.9 and 7.4 GHz  at 6 cm and 20.9 and 25.5 GHz at 1.3 cm.  Therefore, for each source $\alpha$ is 
calculated over a wide frequency range ($\sim$ 20 GHz),  providing a characterization of the frequency behavior of the radio sources,
even for low signal-to-noise  detections. For very weak sources in the presence of image artifacts, the spectral index  was determined using the
peak intensity $I_{\nu}$ instead of the total flux density. Figure \ref{fig:spix} shows the cm SEDs and power law fits to the data. The uncertainty of the spectral index 
given by the power law fit was computed  adopting a uniform flux error for all sources. We assumed that the measured flux density (or the 
peak intensity) is accurate to within 10$\%$, added in quadrature with an assumed 10$\%$ error in calibration. These errors are likely underestimated for weak sources, and 
overestimated for  strong sources.  Table \ref{Parameters}  reports in column 10 the calculated spectral indices including the uncertainties for each source. For components not 
detected at 6 (1.3) cm, a  lower (upper) limit in the spectral index is calculated by using a value of $S_{\nu}$ of 3$\sigma$. 
Our calculation assumes that there is no significant source 
variability in the 3 year interval between the $6\,$cm observations and half of the observations at $1.3\,$cm (e.g., \citealt{2003A&A...398..901C}).
 As can be seen in Figure \ref{fig:spix}, in some cases the spectral behavior of the two data points within an observing band suggests an opposite trend than the
derived spectral index. This can be accounted for by a variety of effects such as calibration errors, image artifacts, etc. Especially for the weakest sources detected just above $5\,\sigma$
we found that in-band spectral indices are not always reliable. Therefore, the spectral index calculated for all data points over a broader frequency span should provide a  
better characterization of the spectral behavior of the continuum emission. A further explanation for  inconsistencies in the spectral index estimated within an  intra-band with respect to the one estimated from the broader band of frequencies could be due to unresolved sources of opposite spectral behavior within the beam. 

Figure \ref{alpha_hist} shows the distribution of the spectral indices for  components within the mm--clump. The lower and upper limits refer to the estimated $\alpha$ if there is no 
detection at 6 or 1.3 cm, respectively. The distribution of the detections (i.e., 
excluding  the limits) has a median value of 0.5 and an 
average value of  0.4, suggesting thermal emission for most sources. These values are consistent with the typical spectral indices found in low-mass thermal radio jets \citep{1998AJ....116.2953A}. We also note that we detect a significant number of negative spectral indices: the fraction of the detected sources (excluding limits) with a spectral index less than $-$0.25 is 
 10$\%$. These latter sources are strong candidates for non-thermal emission. \\

Thermal free-free emission at centimeter wavelengths in high-mass star forming regions can have distinct origins, such as H\textrm{II} regions, photoionized internally or externally, 
ionized accreting flows \citep{2002ApJ...568..754K, 2003ApJ...599.1196K, 
2007ApJ...666..976K}, stellar winds \citep{2015Sci...348..114C}, or shocks produced by the collision 
of  thermal  jets  with  surrounding material (see \citealp{2012ApJ...755..152R} and \citealp{2013ApJ...766..114S} for an extensive  list of different ionization processes in high-mass 
YSOs). The non-thermal components could be protostars with an active 
magnetosphere \citep{2013A&A...552A..51D} or high-mass binary stars producing synchrotron 
radiation in the region where their winds collide \citep{2012ApJ...755..152R}. This emission could also arise from fast shocks in disks and jets, although only few of these non-thermal jets 
have been detected (e.g., HH 80$-$81: \citealp{2010Sci...330.1209C}, Serpens: \citealp{2016ApJ...818...27R}).
Further discussion  of the physical nature of the emission will be presented in Paper II.

\subsection{Source Multiplicity}\label{multiplicity}

High-mass stars are expected to form mainly in clusters and on average to have more physical companions than low-mass stars. For instance, in Orion, the Trapezium cluster has four OB type stars 
in close proximity, and the companion star fraction of high-mass stars is 3 times larger than in low-mass stars  \citep{2007ARA&A..45..481Z}. Hence, a key  issue in  
high-mass star formation is the density distribution and the multiplicity in massive clumps. Furthermore, the presence of multiple sources  in different evolutionary phases also describes the time 
sequence of high-mass star formation. In this regard our data show an interesting dichotomy. We detected multiple radio sources  primarily for HMCs: about half of HMCs showed two or more 
radio sources within the FWHM of the mm clump, and the other half show single detections. For CMC--IRs, only 20$\%$ of the regions with detections have two or more continuum sources
within the clump.  Another interesting observational result is that 5 out of the 25 HMCs observed have an extended  radio source reminiscent of a UCH{\small II} region
(e.g., Wood \& Churchwell 1989).  These bright and extended radio continuum components are only seen towards the HMCs of our sample.  These regions will be discussed further in paper II.

\section {Summary}\label{summary_section}
We have carried out  deep VLA observations of radio continuum at 6 and $1.3\,$cm towards a sample of 58 high-mass star forming region candidates selected from the literature, and with no previous  or relatively weak radio continuum detection at the $1\,$mJy level. Our 
sample has been classified according to the infrared properties of the targets and other signposts of star formation activity into three types: CMCs, CMC--IRs and HMCs. In total, we detected  70 radio sources associated with dusty millimeter clumps in these regions.
The detected weak and compact radio sources are prime candidates for high-mass protostars. In this paper we have presented the sample and our observations. Some of the important aspects to highlight from our observations are the following: 

\begin{itemize}
\item[--] The detection rates of radio sources associated with the mm dust clumps for CMCs, CMC--IRs and HMCs are 6$\%$, 53$\%$ and 100$\%$, respectively. This result is consistent with increasing high-mass star formation activity from CMCs to HMCs.

\item[--] Radio sources located within HMCs and CMC--IR occur close to the dust clump center with a median offset from the clump center of $10,000\,$AU and 4,000 AU, respectively.

\item[--] The median 5$-$25 GHz spectral index for radio sources associated with the mm clumps is $0.5$, which suggests thermal emission from ionized jets, and is very similar to what is observed in low-mass protostars. Additionally, we found that at least  10$\%$ of sources show negative spectral indices, indicative of non-thermal emission. 

\item[--] HMCs  often have multiple cm continuum sources within the surrounding dust clump, whereas CMC--IRs generally have only one or not detectable radio source.
 
\end{itemize}

\noindent
The analysis and discussion of our detections  will be presented in Paper II. 

\acknowledgements

We thank H. Beuther for providing $1.2\,$mm maps for the HMCs in our sample. We thank J. Marvil for stimulating technical discussions about the VLA capabilities. V.R. is supported by the NRAO Grote Reber Doctoral Fellowship.  P.H. acknowledges support from NSF grant AST$-$0908901 for this project. L.L. acknowledges the financial support of CONACyT, Mexico, DGAPA, UNAM, and the von Humboldt Foundation.
This research made use of APLpy, an open-source plotting package for Python hosted at http://aplpy.github.com. We thank the anonymous referee whose comments
improved this manuscript.

Software: CASA, APLpy



\clearpage
\tabletypesize{\scriptsize}
\begin{deluxetable}{l c c c c c c  c c c c c c}	
\tablecaption{Observed Clumps \label{obs}}  
\tablewidth{0pt}				  						  
\tablehead{					  
\colhead{Region} &			
\colhead{Band}    &	  
\colhead{R.A.} &				  
\colhead{Dec.} &
\colhead{$\ell$} &				  
\colhead{\it{b}} &				  
\colhead{Beam Size} &
\colhead{rms} &
\colhead{Detection} &
\colhead{Type} &
\colhead{Distance}&
\colhead{L} &
\colhead{Ref.}
\\				  
\colhead{} &	
\colhead{} &				  
\colhead{(J2000)} &					  
\colhead{(J2000)} &	
\colhead{(deg)} &					  
\colhead{(deg)} &
\colhead{($^{\prime \prime}$$\times$$^{\prime \prime}$, deg)} &	
\colhead{($\mu$Jy beam$^{-1}$)} &	
\colhead{} &
\colhead{} &
\colhead{(kpc)} &		  
\colhead{($10^{3}$ L$_{\odot}$)} &			  
}						  			  
\startdata  
\setcounter{iso}{0}	
\input{latex2_v3.txt}
\enddata
\end{deluxetable}

\clearpage
\tabletypesize{\scriptsize}
\begin{deluxetable}{l c c }	
\tablecaption{VLA Parameters \label{VLA_parameters}}  
\tablewidth{0pt}				  
						  
\tablehead{					  
\colhead{Parameter} &			
\colhead{6 cm}    &	  
\colhead{1.3 cm} 		  			  
}						  					  
\startdata           
Center Frequency (GHz)                                &  4.9, 7.4                                                &   20.9, 25.5                                           \\[3pt]  
Configuration                                                  & A                                                           &    B                                                          \\[3pt]  
Spectral Windows (per frequency)                 &8($\times 128$ MHz)                          &   8($\times 128$ MHz)                       \\[3pt]  
Bandwidth  (MHz)                                          & 2048           &  2048                                                 \\[3pt]
Primary Beam  FWHM                                  & 9\rlap.$^{\prime}$2, 6\rlap.$^{\prime}$1         &  2\rlap.$^{\prime}$2, 1\rlap.$^{\prime}$8        \\[3pt]
Typical synthesis beam FWHM                      & $\sim$0\rlap.$^{\prime \prime}$4$\times$0\rlap.$^{\prime \prime}$3        &    $\sim$0\rlap.$^{\prime \prime}$4$\times$0\rlap.$^{\prime \prime}$3                   \\[3pt]  
Typical rms ($\mu$Jy beam$^{-1}$)               &     5                                                          &    9                                                           
\enddata
\tablecomments{Reported typical values of the synthesized beam and rms apply to the combined image at each band, i.e., including all data from both $1\,$GHz wide basebands.}

\end{deluxetable}

\tabletypesize{\scriptsize}
\begin{deluxetable}{l c l }	
\tablecaption{VLA Calibrators \label{cal}}  
\tablewidth{0pt}				  					  
\tablehead{					  
\colhead{Calibrator} &			
\colhead{Astrometry Precision\tablenotemark{a}}    &	  
\colhead{Sources calibrated} 		  	  
}						  			  
\startdata           
J1832$-$1035    &     C     &   18089$-$1732, G15.05$+$00.07 mm1, 18151$-$1208, G15.31$-$00.16 mm3, 18182$-$1433, IRDC 18223$-$3, \\
                          &              &   G18.82$-$00.28 mm3, 18345$-$0641, 18264$-$1152, G25.04$-$00.20 mm3, G22.35$+$00.41 mm1, G22.73$+$00.11 mm1,\\  
                           &             &   G23.60$+$00.00 mm6, G23.01$-$0.41, G24.33$+$00.11 mm4, 18337$-$0743 \\[2pt]  
J1851$+$0035  &      C      &  G27.75$+$00.16 mm2, G28.23$-$00.19 mm1, G28.53$-$00.25 mm1, 18437$-$0216, 18440$-$0148, \\
                          &              &  G30.97$-$00.14 mm2, 18470$-$0044, G34.43$+$00.24 mm1, 18517$+$0437, 18521$+$0134, G35.39$-$00.33 mm2,\\
                          &              &  G35.59$-$00.24 mm2, 18553$+$0414, 18566$+$0408, 19012$+$0536, G38.95$-$00.47 mm1, 19035$+$0641\\[2pt]                         
J2007$+$4029  &     B       &   20126$+$4104, 20216$+$4107, 20293$+$3952, 20343$+$4129                                                  \\[2pt]      
J2015$+$3710\tablenotemark{b}   &     C       &   20126$+$4104, 20216$+$4107, 20293$+$3952, 20343$+$4129                   \\
\enddata
\tablenotetext{\text{a}}{Astrometry precision A, B and C correspond to positional accuracy of $<$2 mas, 2$-$10 mas and 0.01$-$0.15 arcsecs, respectively.}	

\tablecomments{The phase calibrators listed were used for observations at 6 and 1.3 cm bands, unless otherwise indicated. 3C 286 was used as flux calibrator for all sources, except for
LDN1657A$-$3 and UYSO1 for which 3C 147 and 3C 48 were used at 6 cm and 1.3 cm, respectively.}
\end{deluxetable}


\newpage

\begin{deluxetable}{l c c c c c c c c c}
\centering
\tabletypesize{\scriptsize}
\tablecaption{Continuum Parameters \label{Parameters}}
\tablewidth{0pt}
\tablehead{
Region   & Radio & $\nu_{c}$ & RA       & Dec       & $S_{\nu}$   & $I_{\nu}$              & Morphology &  mm  & Spectral  \\[2pt]
              &    Source              & (GHz)       &(J2000)& (J2000) & ($\mu$Jy)   & ($\mu$Jy\,beam$^{-1}$)  &                     &   assoc        & Index\\[-20pt]\\}
\startdata
\input{latex1_v3.txt}
\enddata
\tablenotetext{\text{a}}{1.2 mm continuum data is not available for this source. The association was based on a typical size for each type of clump.}
\tablenotetext{\text{b}}{Non-detection. Upper limit is 3$\sigma$ value in map.}
\tablenotetext{\text{c}}{Components have a bridge blending emission.}
\tablenotetext{\text{d}}{Map at this frequency is not available.}
\tablenotetext{\text{e}}{Contains embedded component  B.}

\tablecomments{In the morphology column C stands for compact and R for resolved.  The uncertainty of the spectral index is 1$\sigma$ statistical error from the fit. Several sources in our survey are very weak; the occasional presence of image artifacts (e.g., the emission lying in a negative bowl) inhibited in some cases an accurate measurement of the flux density $S_{\nu}$. In 
these cases the spectral index  was determined using the
peak intensity $I_{\nu}$ instead of the total flux density. }

\end{deluxetable}

\newpage

\begin{deluxetable}{l c c c}
\tabletypesize{\scriptsize}
\tablecaption{Summary of Detections   \label{Summary_det}}
\tablewidth{0pt}

\tablehead{
                                                                                                               & CMC            & CMC$-$IR             & HMC           \\[-20pt]\\}
\startdata
Number of observed regions                                                                        &   18                       &            15           &      25               \\[2pt]  
Number of regions with radio detections\tablenotemark{a}                          &   8                        &            9             &     25     \\[2pt]  
Number of regions with radio detections within the mm-clump\tablenotemark{b}                             &   1                        &            8             &     25     \\[2pt]  
\hline
Number of  radio sources detected at both 1.3 and 6 cm                  &    2(0)                    &          11(10)       &      54(40)$\tablenotemark{c}$                       \\[2pt]  
Number of radio sources only detected at 1.3 cm                           &    7(1)                   &           2(2)           &      9(6)$\tablenotemark{c}$                       \\[2pt]  
Number of radio sources only detected at 6 cm                              &     7(0)                  &           7(1)           &       33(10)$\tablenotemark{c}$                         \\[2pt]    														
\enddata
\tablenotetext{\text{a}}{Any detection within the 25.5 GHz primary beam (1\rlap.$^{\prime}$8). }
\tablenotetext{\text{b}}{Only detections located within the  FWHM size of the mm-clump. }
\tablenotetext{\text{c}}{Numbers do not include radio sources found in 19282$+$1814.}

\tablecomments{Values in parentheses correspond to the number of
sources that are within the millimeter clump, which are a total of   70 sources. }

\end{deluxetable}


\clearpage
\begin{figure}[htbp]
\centering
\begin{tabular}{cc}

\includegraphics[width=0.52\linewidth, trim = 35 150 15 100, clip]{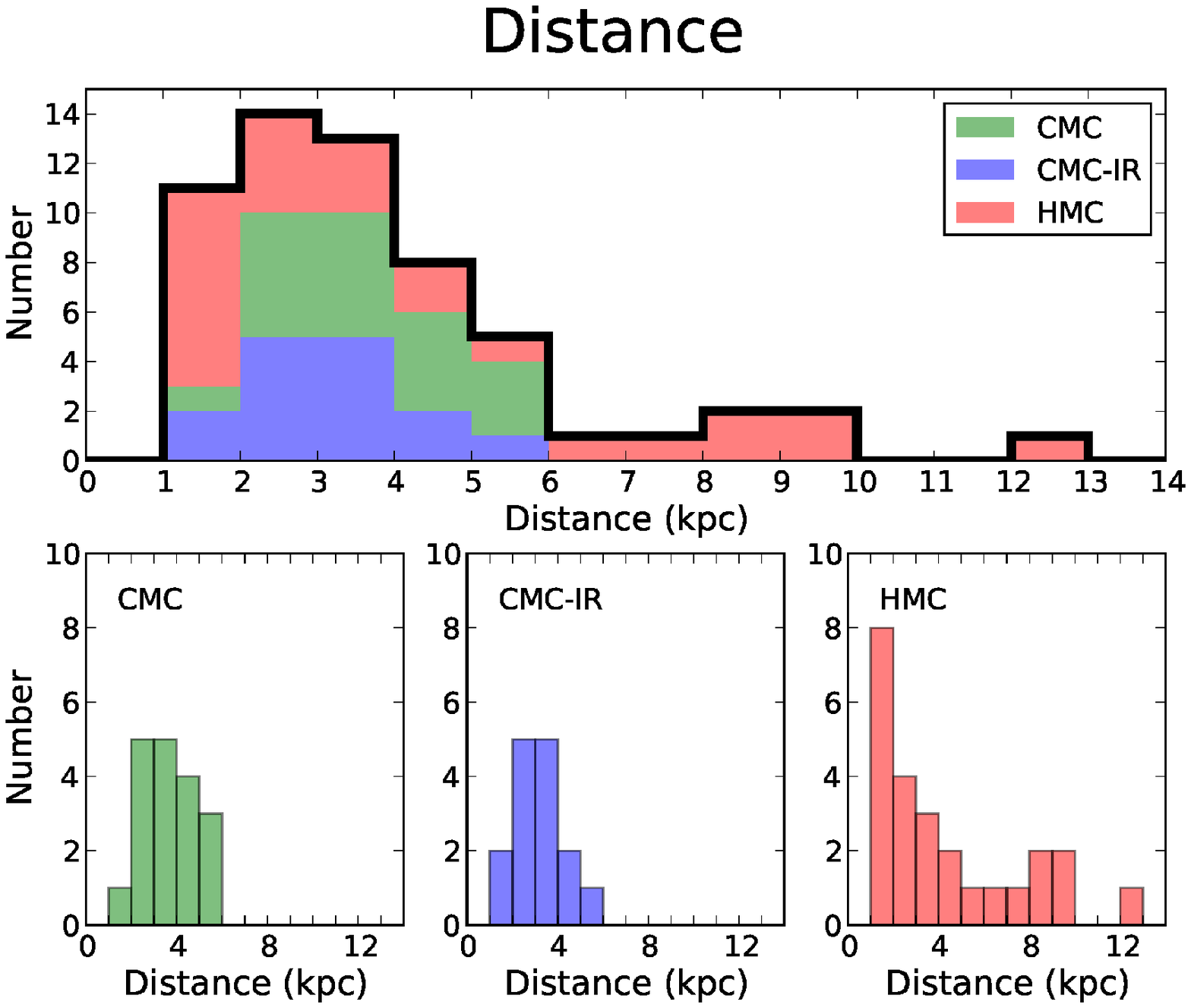} &
\includegraphics[width=0.52\linewidth, trim = 35 150 15 100, clip]{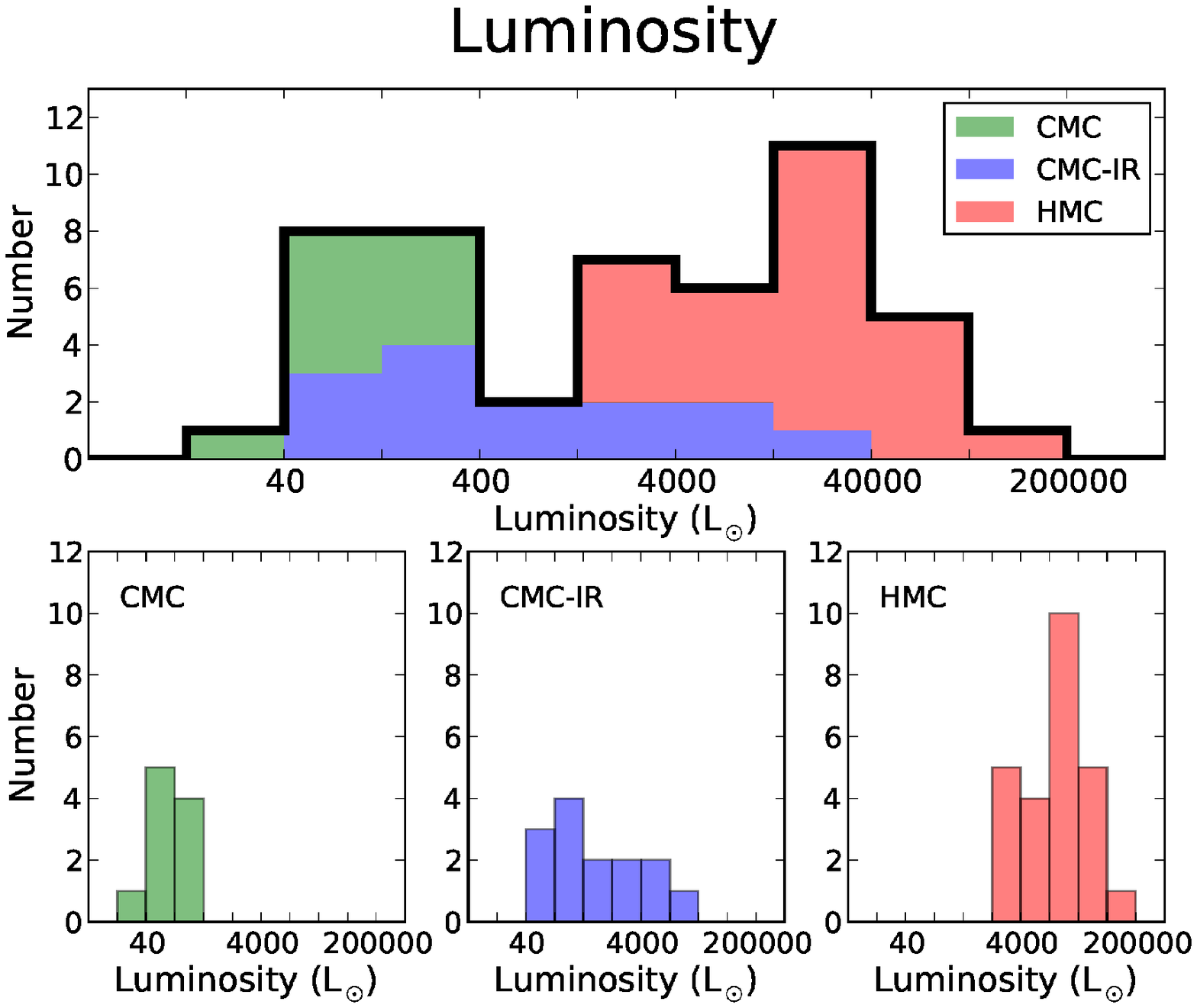} \\
\includegraphics[width=0.52\linewidth, trim = 35 150 15 100, clip]{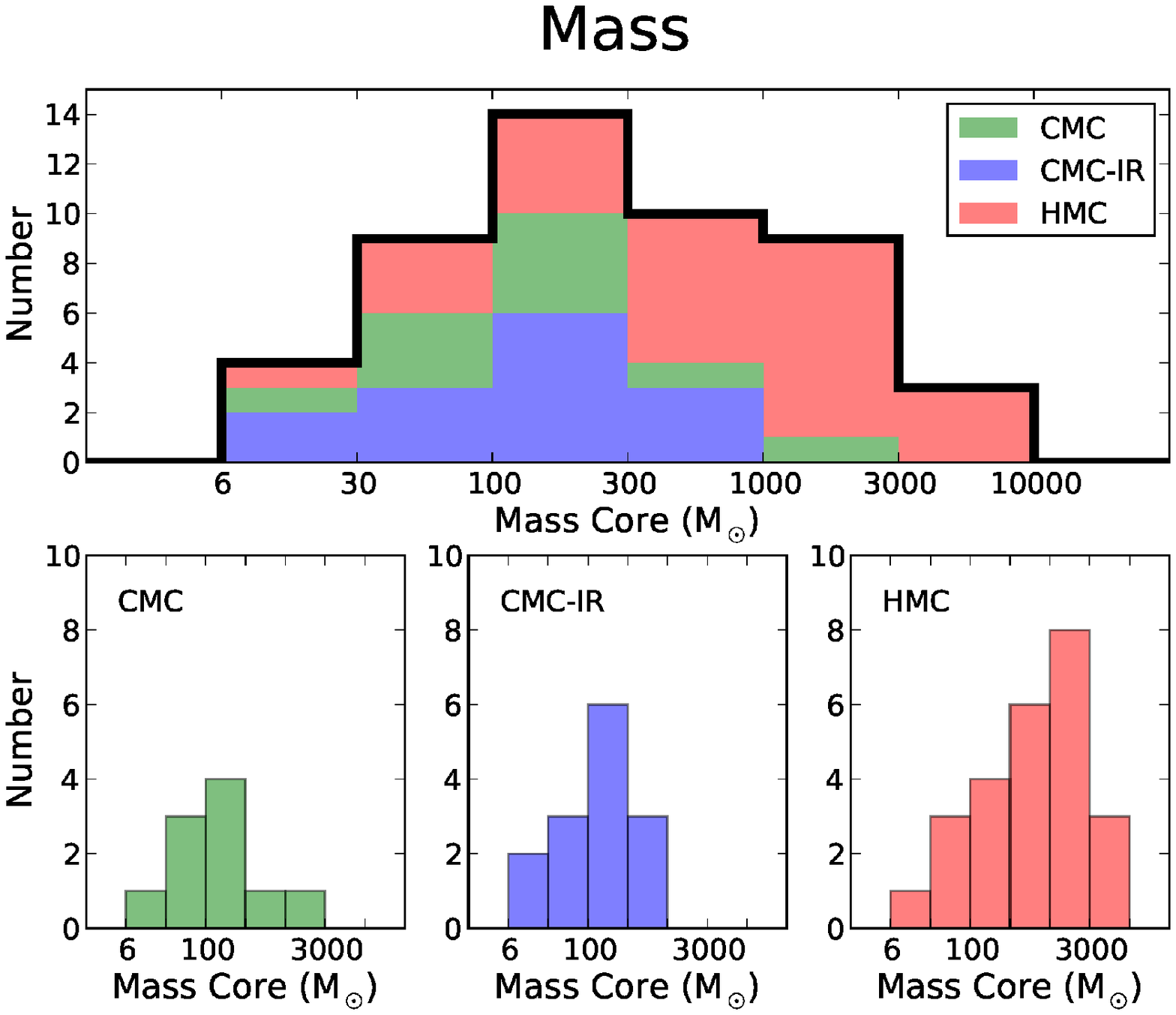}&
\includegraphics[width=0.52\linewidth, trim = 35 150 15 100, clip]{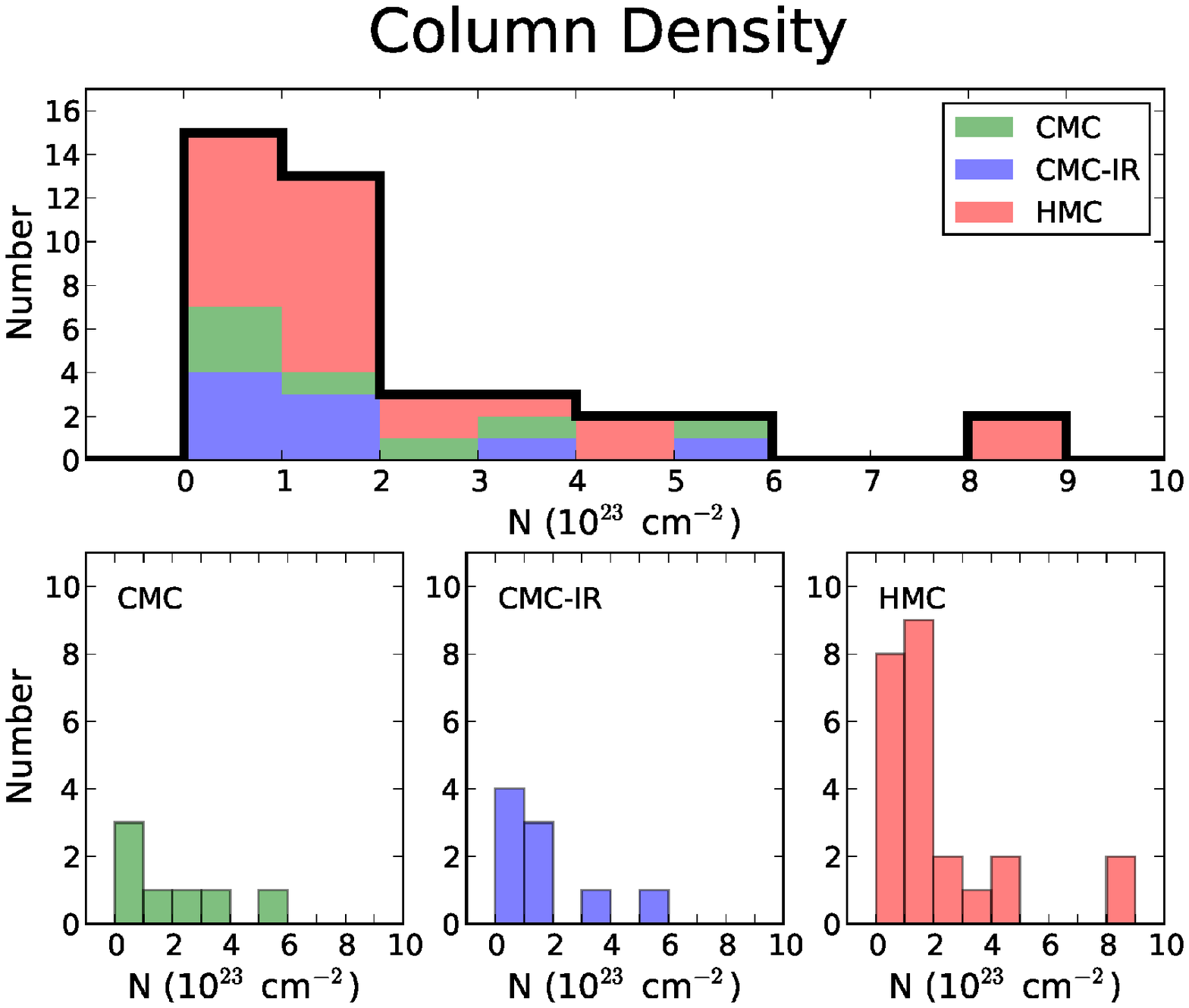}

\end{tabular}

 \hspace*{\fill}%
  \vspace{-0.7cm}  \caption{\small{Distribution of kinematic distance, bolometric luminosity,  mass,  and column density (N) for our targets. All values were taken from the literature (see Table \ref{obs} for references). In each case the top panels correspond to the distribution of the sum of the corresponding parameter, and the bottom panels show the distribution for each type of clump independently. We use the near distance for 5 sources which have a kinematic distance ambiguity. Values of masses from \citet{2002ApJ...566..945B} were scaled following \citet{2005ApJ...633..535B}. For the total sample, nine sources do not have available values for mass and luminosity.}}%
    \label{sample_hist}%
\end{figure}

%
%

\begin{figure}[htbp]
\centering
\begin{tabular}{cc}
 
\hspace*{\fill}%

 \includegraphics[width=0.44\textwidth, trim = 3 60 10 160, clip, angle = 0]{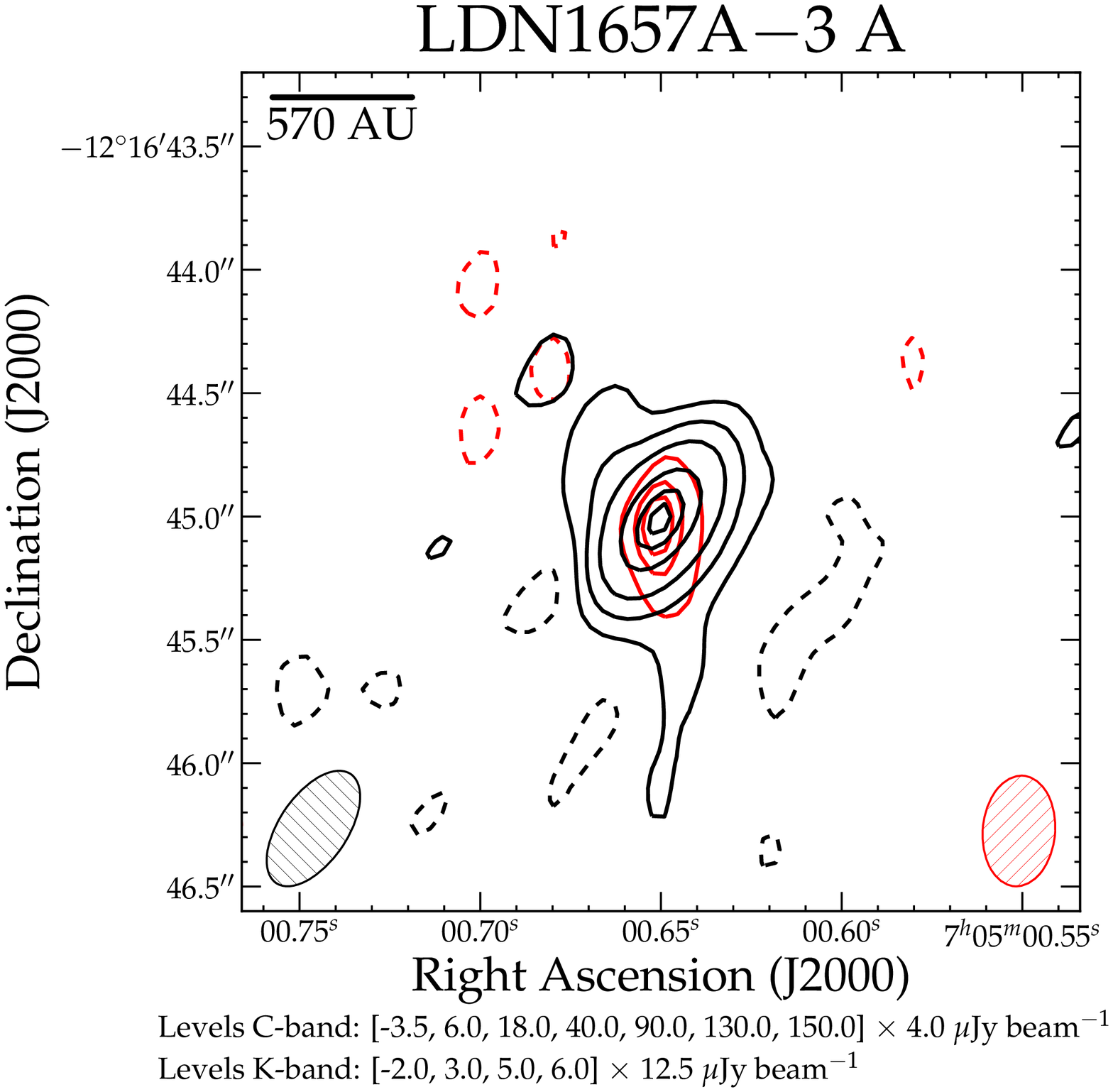} &
 \includegraphics[width=0.44\textwidth, trim = 3 60 10 160, clip, angle = 0]{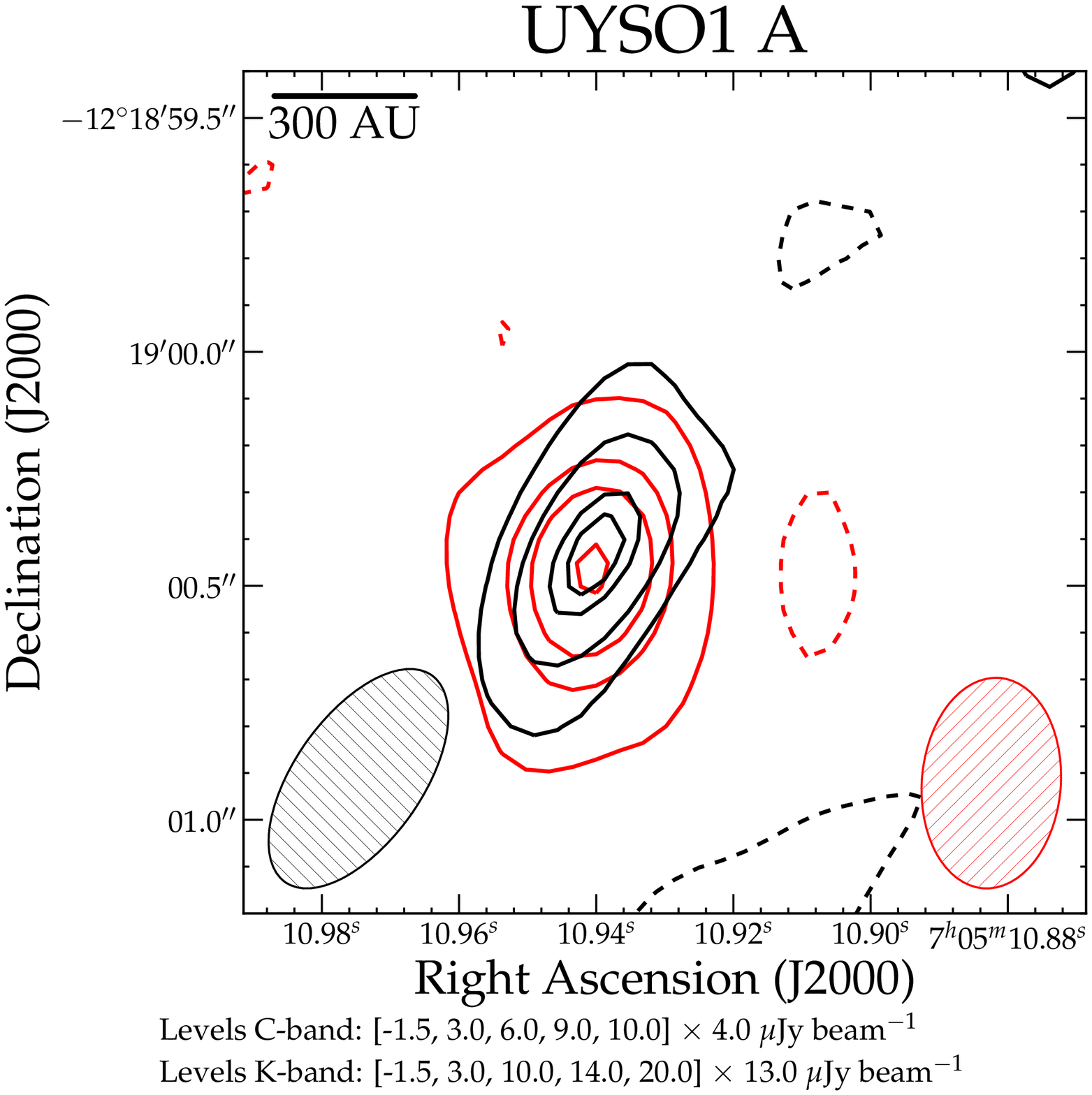} \\
  \includegraphics[width=0.44\textwidth, trim = 3 60 10 160, clip, angle = 0]{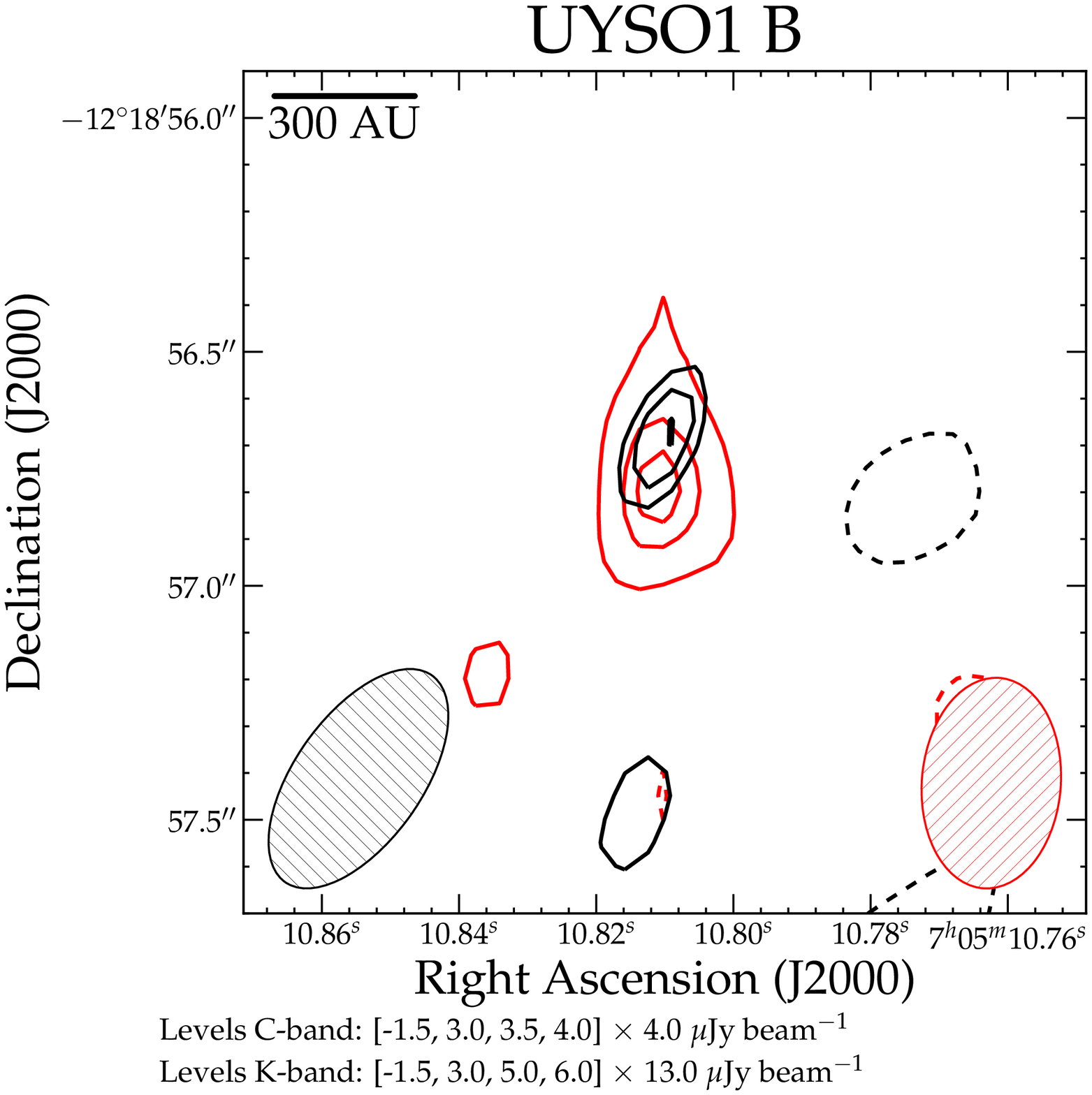}&
 \includegraphics[width=0.44\textwidth, trim = 3 60 10 160, clip, angle = 0]{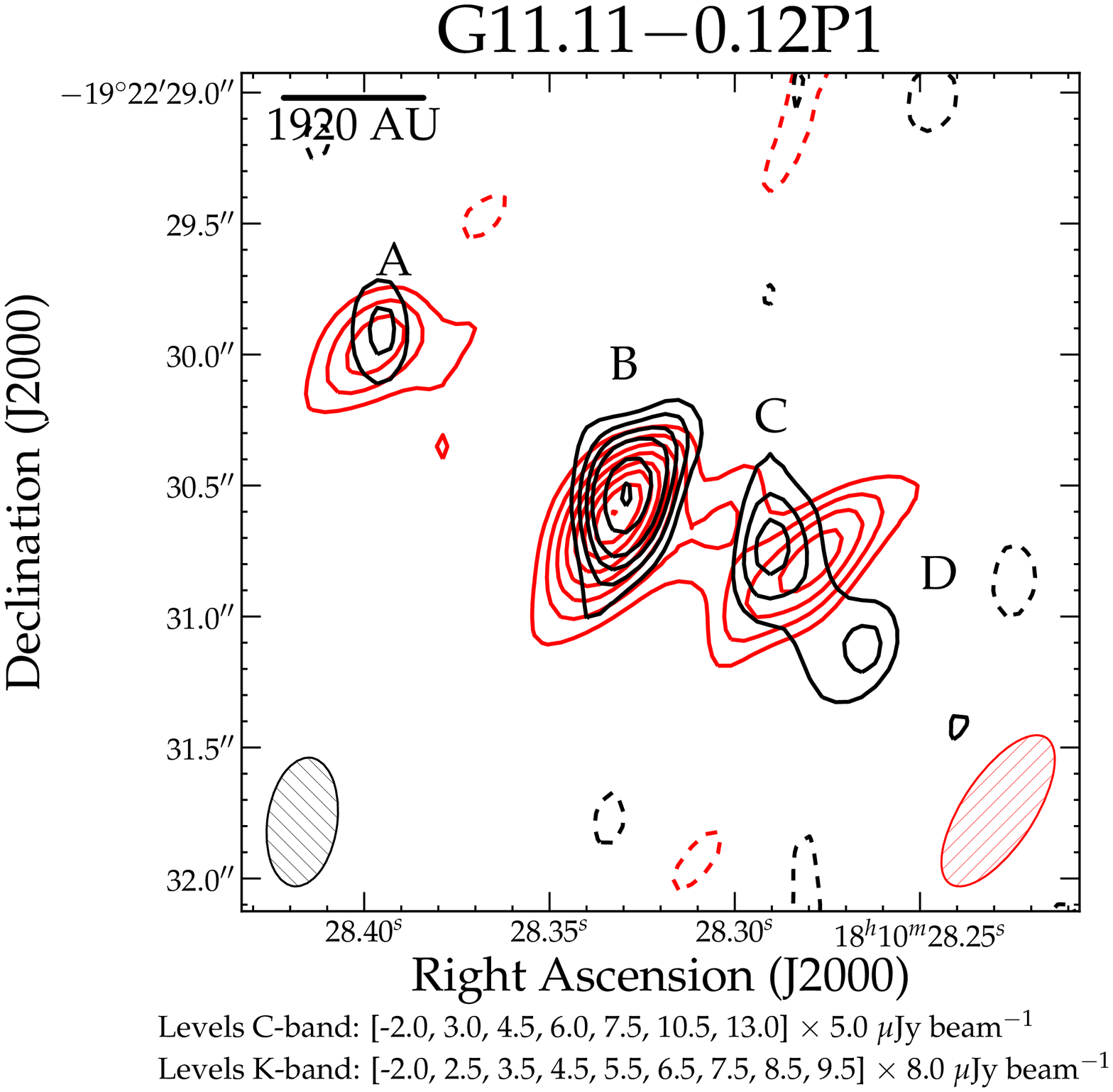} \\
 \includegraphics[width=0.44\textwidth, trim = 3 60 10 160, clip, angle = 0]{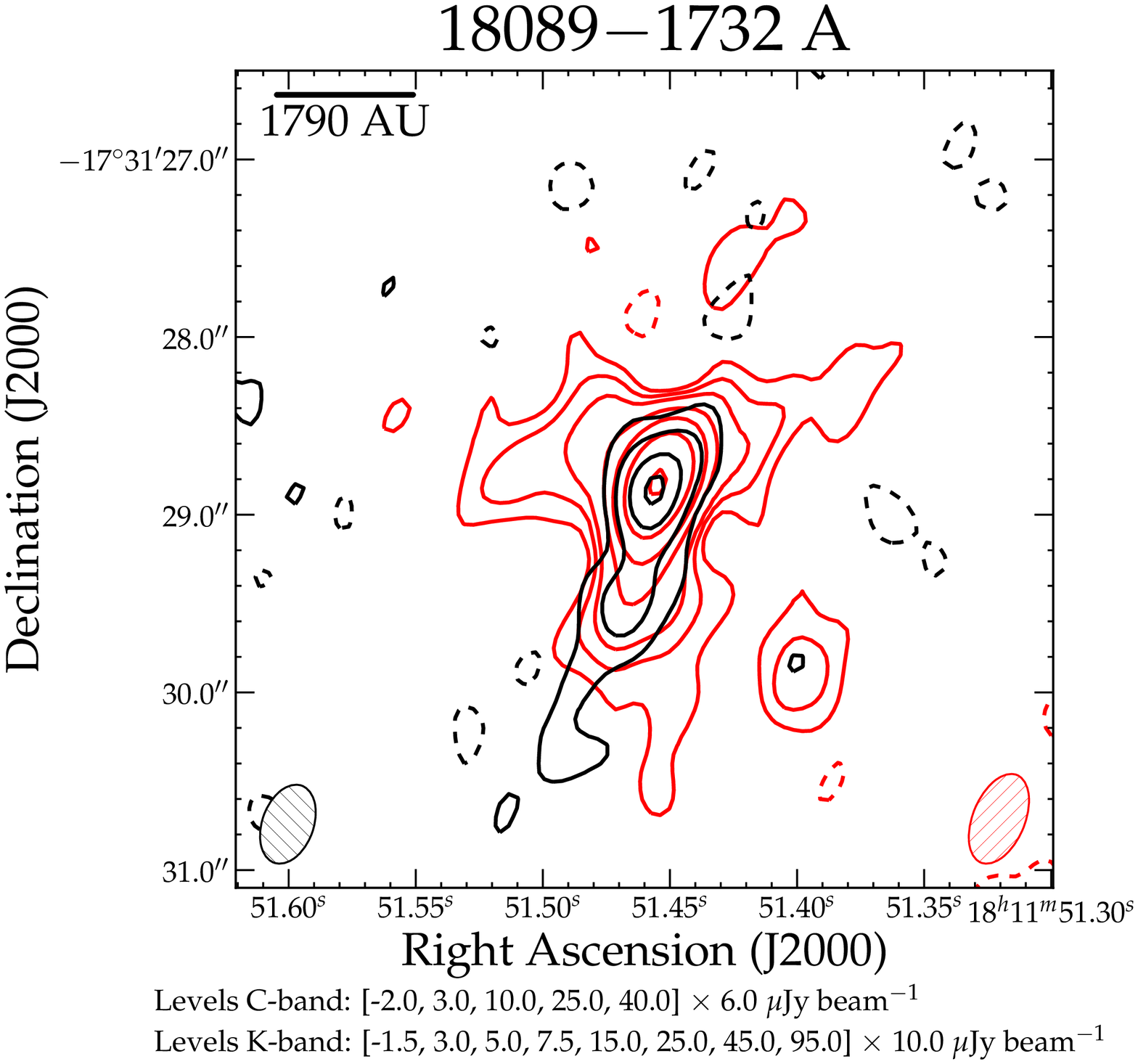} &
  \includegraphics[width=0.44\textwidth, trim = 3 60 10 160, clip, angle = 0]{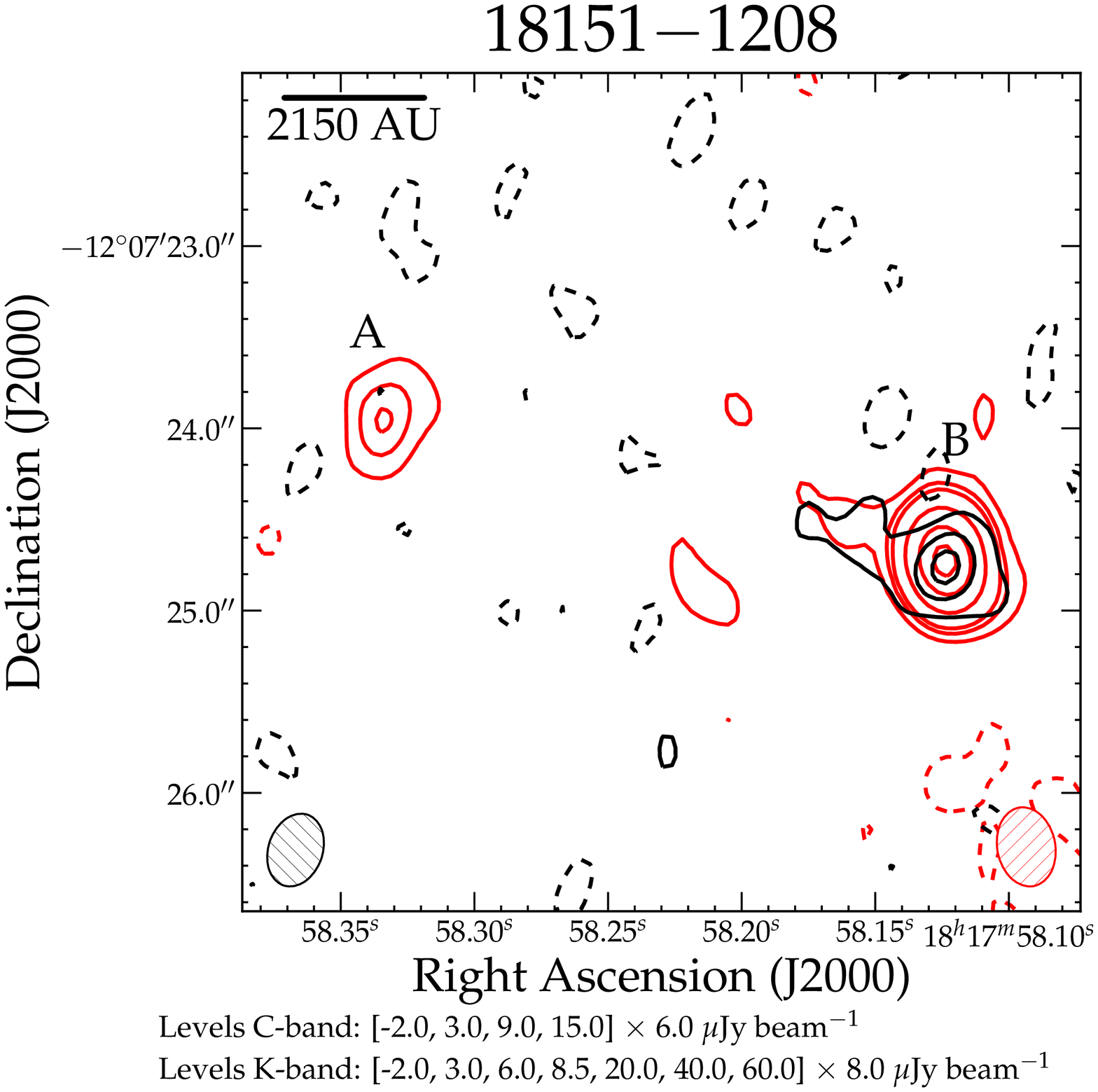}

\end{tabular}

 \hspace*{\fill}%
   \caption{\small{VLA contour plots of the 6 (black) and 1.3  (red) cm combined maps toward all sources associated with mm dust clumps. The synthesized beams are shown in the lower left and right corner for 6 and 1.3 cm, respectively. A scale bar in units of AU is shown in the upper left. We are using the near distance for sources with distance ambiguity. There is not available map at 1.3 cm for 19282$+$1814 A.}}
 \label{fig:maps}
\end{figure}

\begin{figure}[htbp]
\centering
\begin{tabular}{cc}
  \ContinuedFloat
\hspace*{\fill}%
 \includegraphics[width=0.44\textwidth, trim = 3 60 10 160, clip, angle = 0]{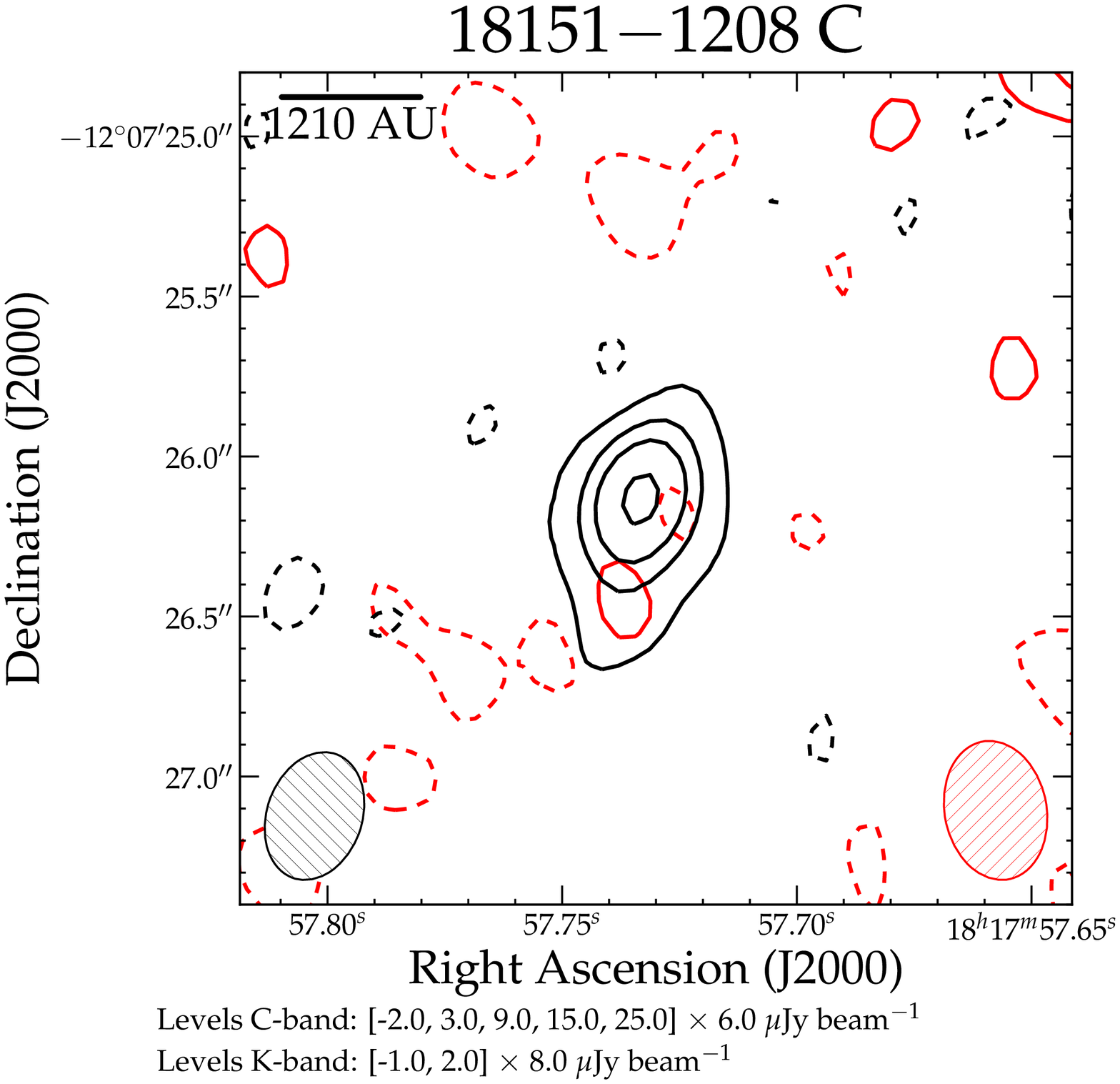} &
 \includegraphics[width=0.44\textwidth, trim = 3 60 10 160, clip, angle = 0]{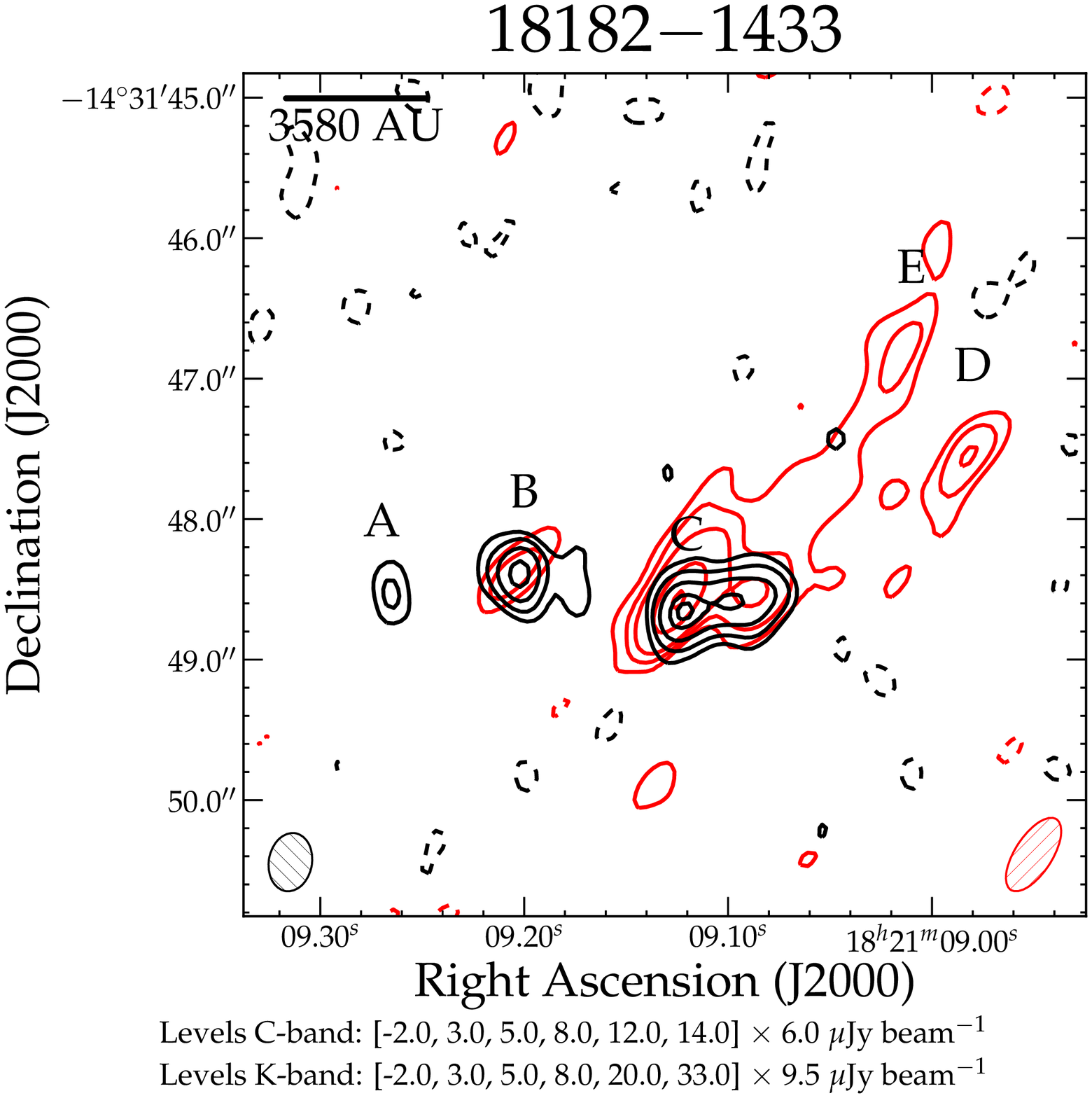} \\
  \includegraphics[width=0.44\textwidth, trim = 3 60 10 160, clip, angle = 0]{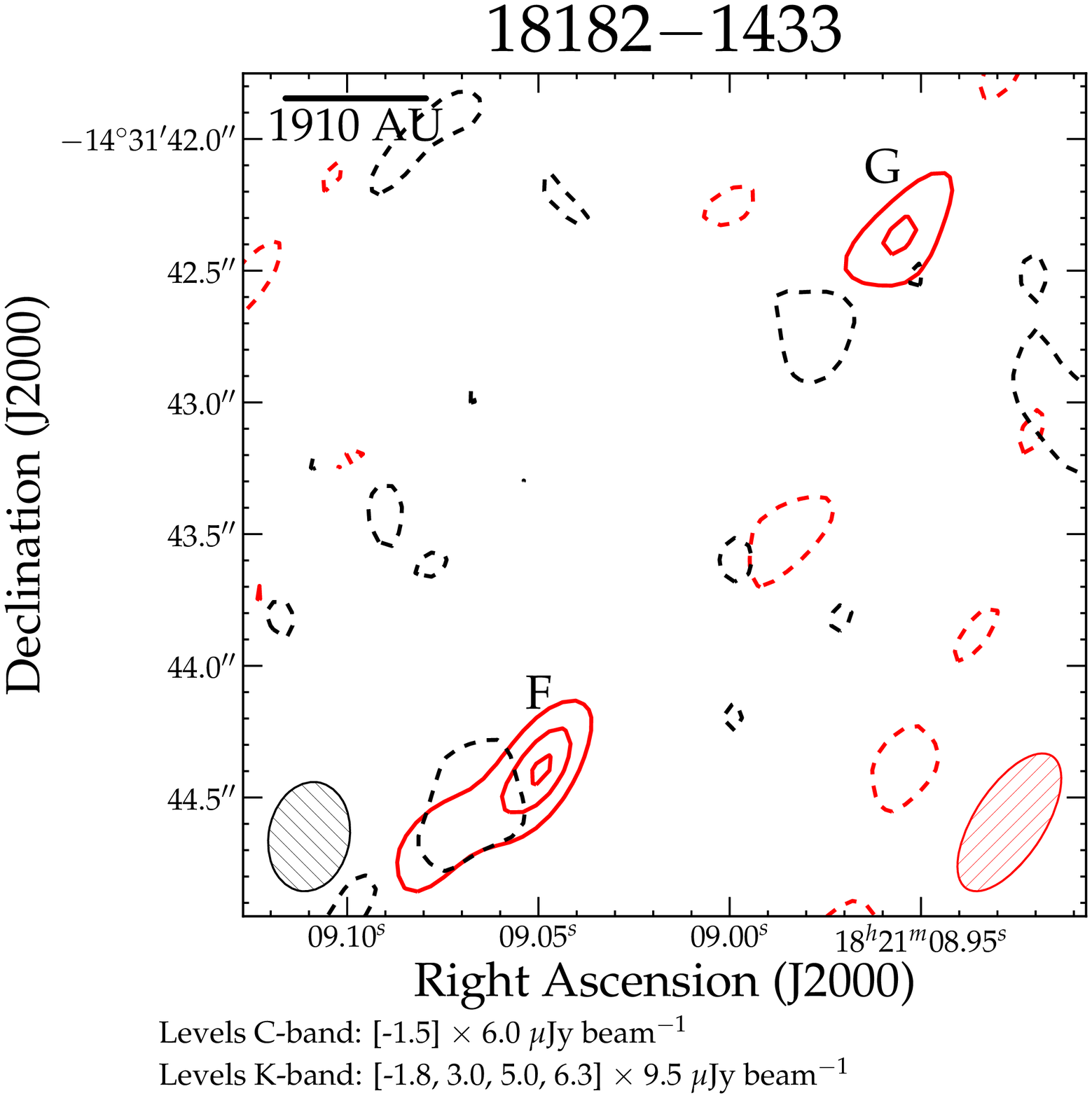}&
 \includegraphics[width=0.44\textwidth, trim = 3 60 10 160, clip, angle = 0]{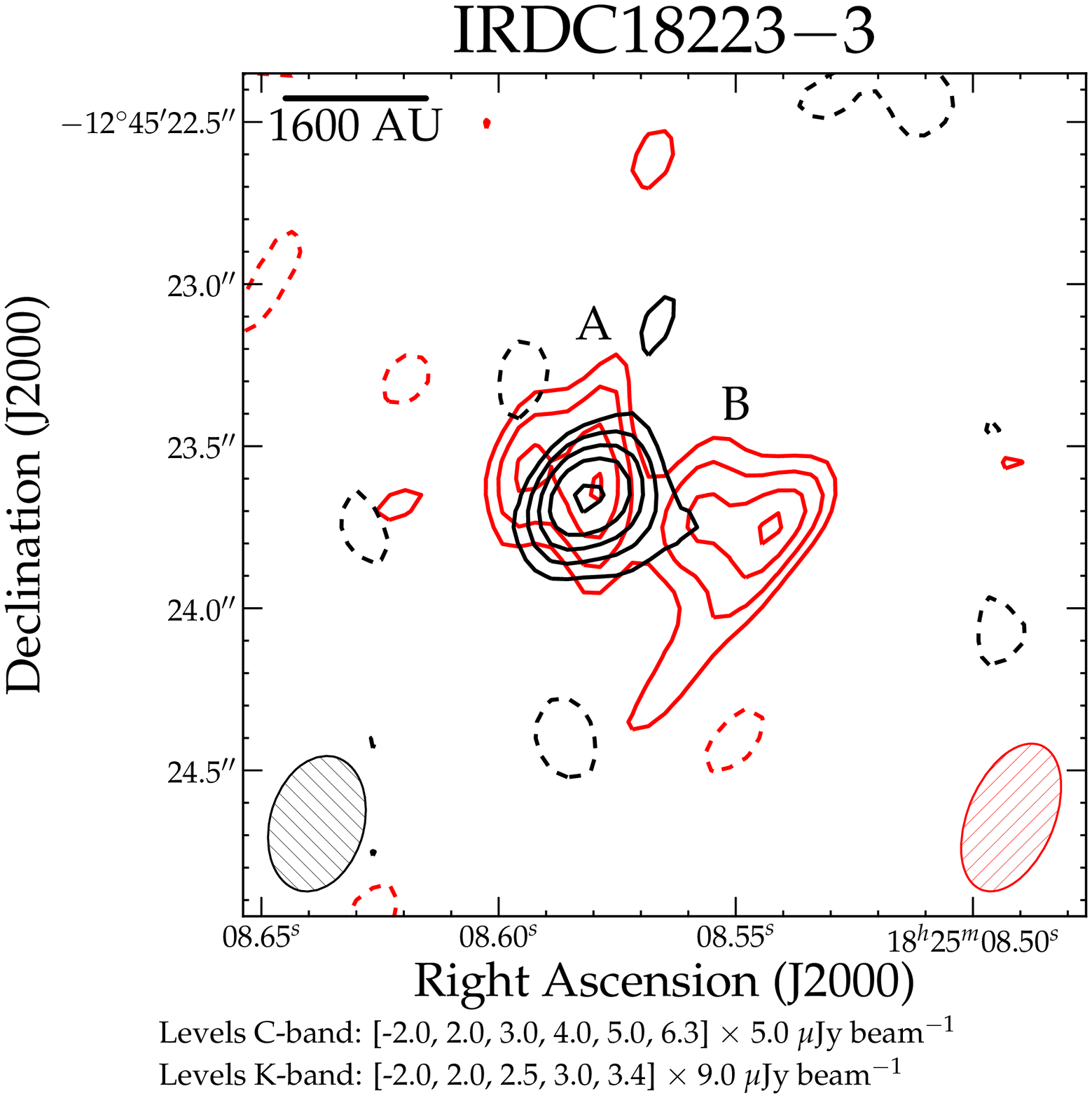} \\
 \includegraphics[width=0.44\textwidth, trim = 3 60 10 160, clip, angle = 0]{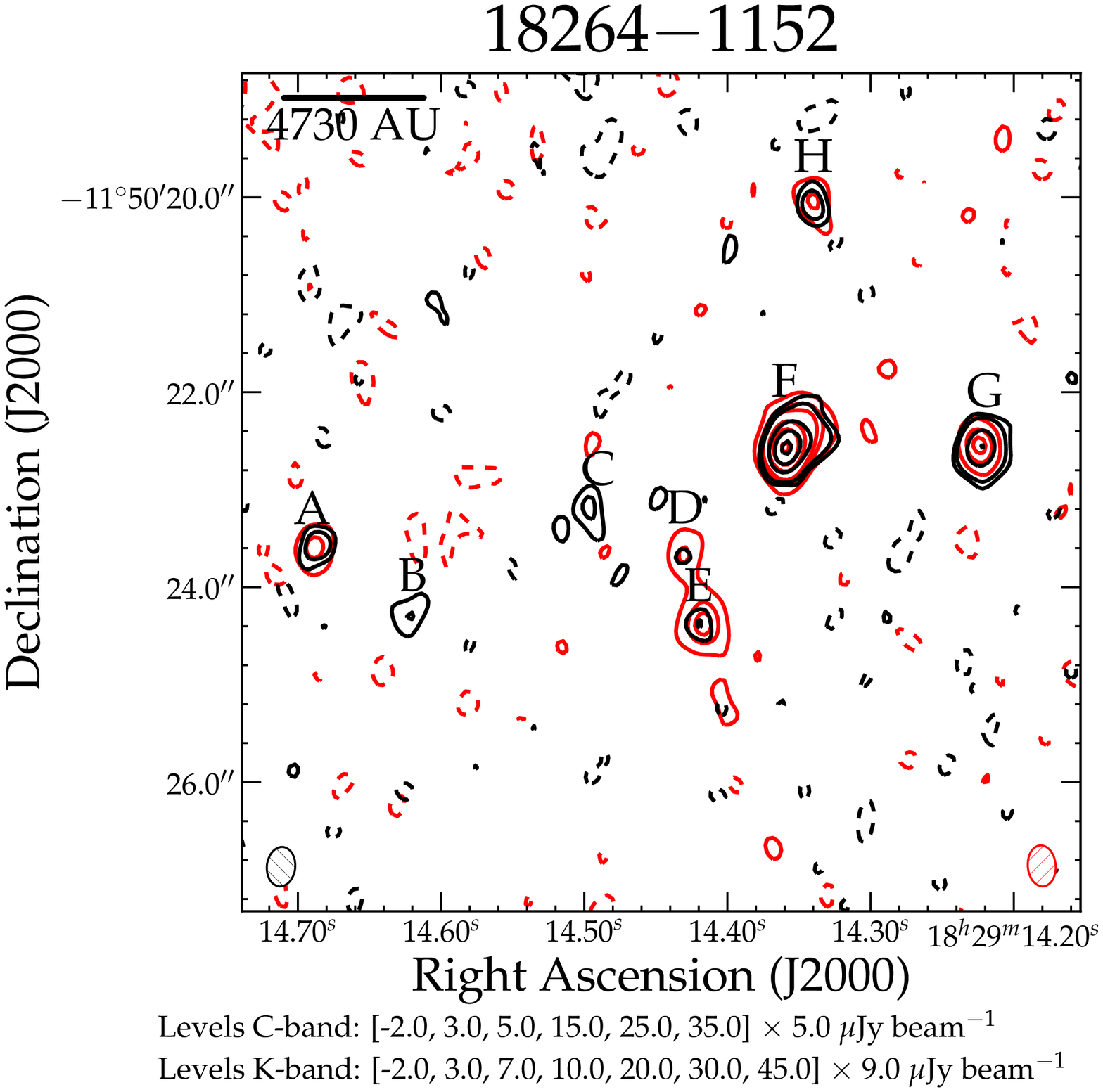} &
  \includegraphics[width=0.44\textwidth, trim = 3 60 10 160, clip, angle = 0]{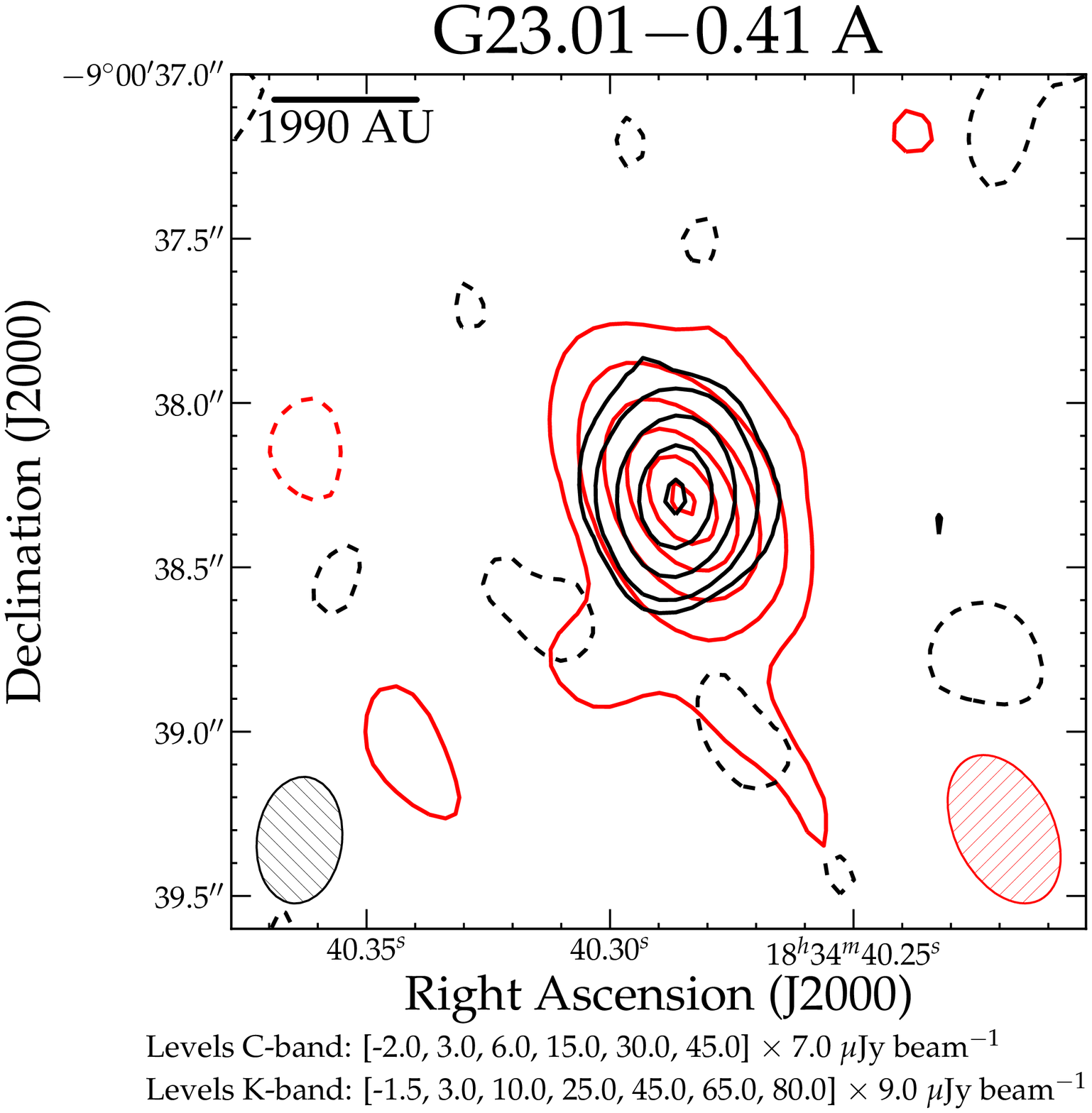}

\end{tabular}

 \hspace*{\fill}%
\caption{\small{Continued.}}
 \label{fig:maps}
\end{figure}

\begin{figure}[htbp]
\centering
\begin{tabular}{cc}
  \ContinuedFloat
\hspace*{\fill}%
 \includegraphics[width=0.44\textwidth, trim = 3 60 10 160, clip, angle = 0]{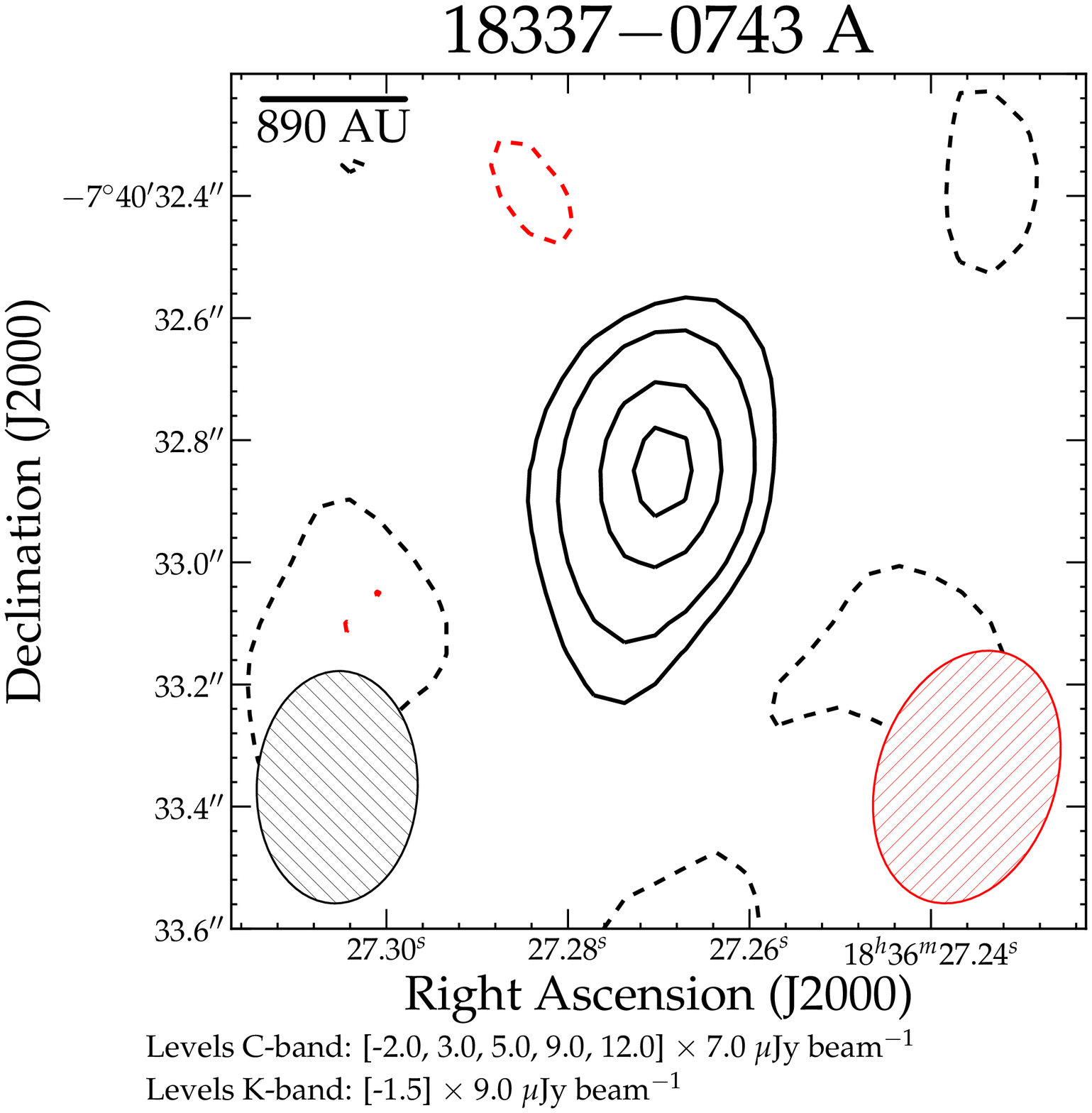} &
 \includegraphics[width=0.44\textwidth, trim = 3 60 10 160, clip, angle = 0]{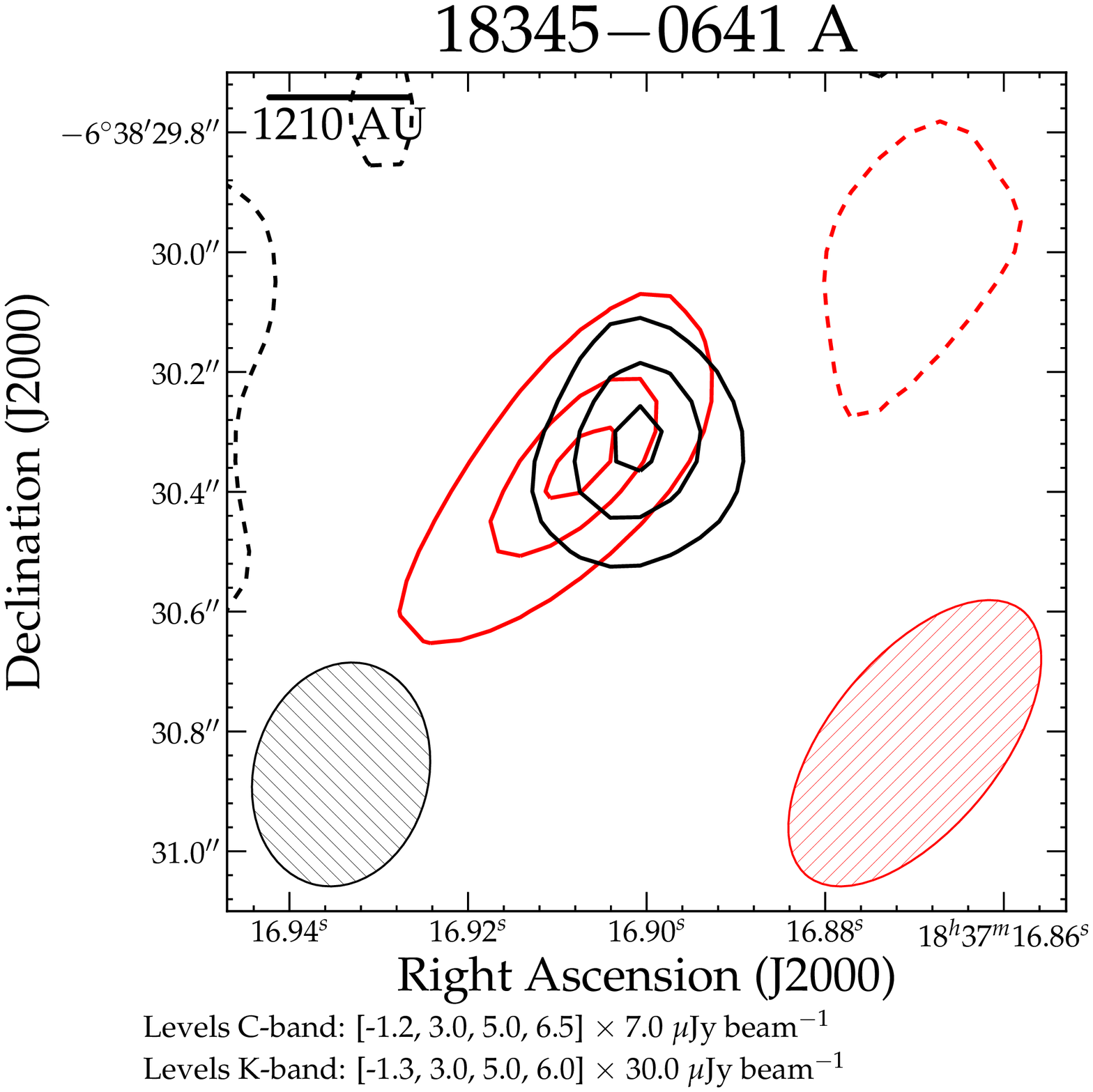} \\
  \includegraphics[width=0.44\textwidth, trim = 3 60 10 160, clip, angle = 0]{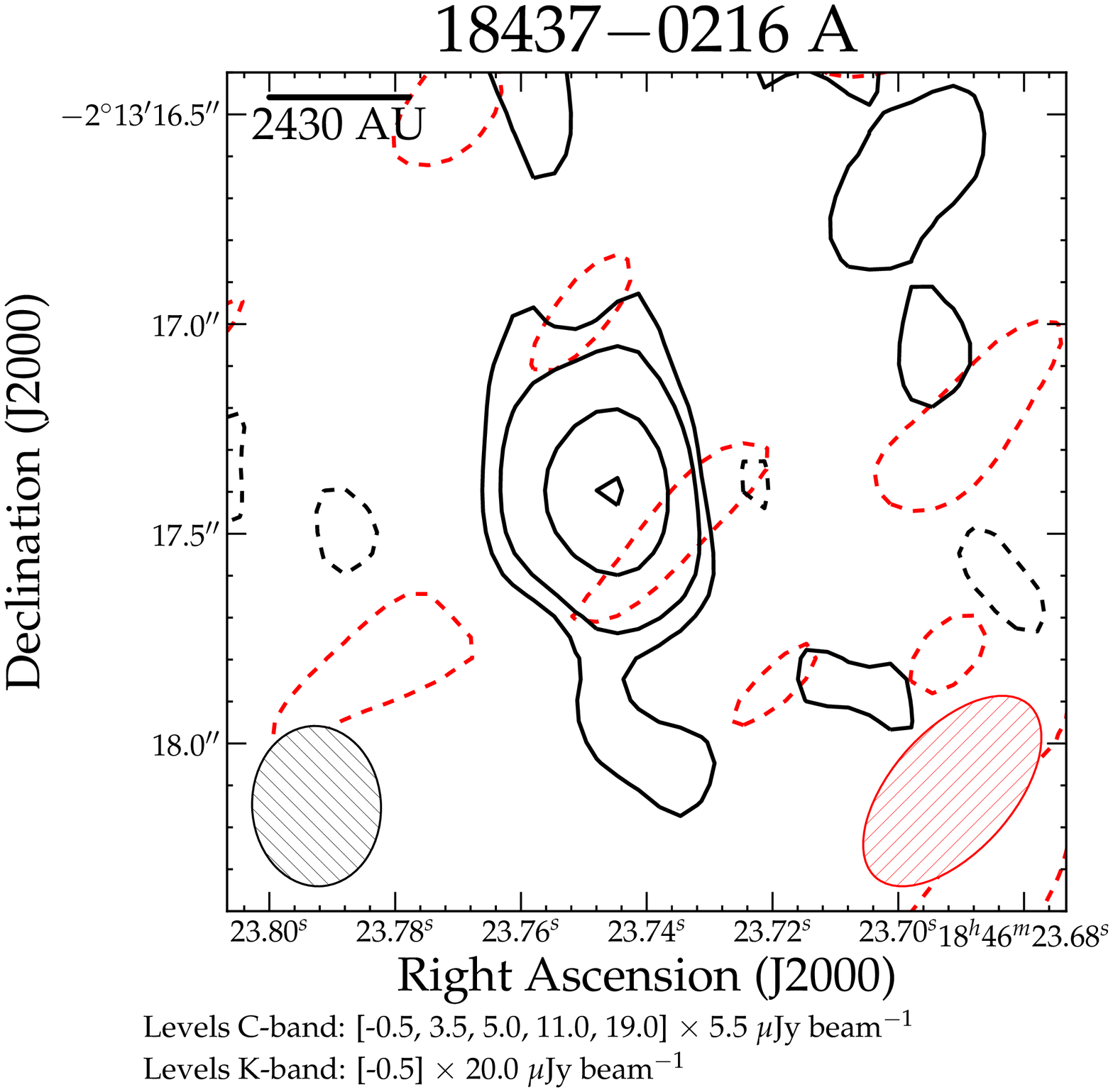}&
 \includegraphics[width=0.44\textwidth, trim = 3 60 10 160, clip, angle = 0]{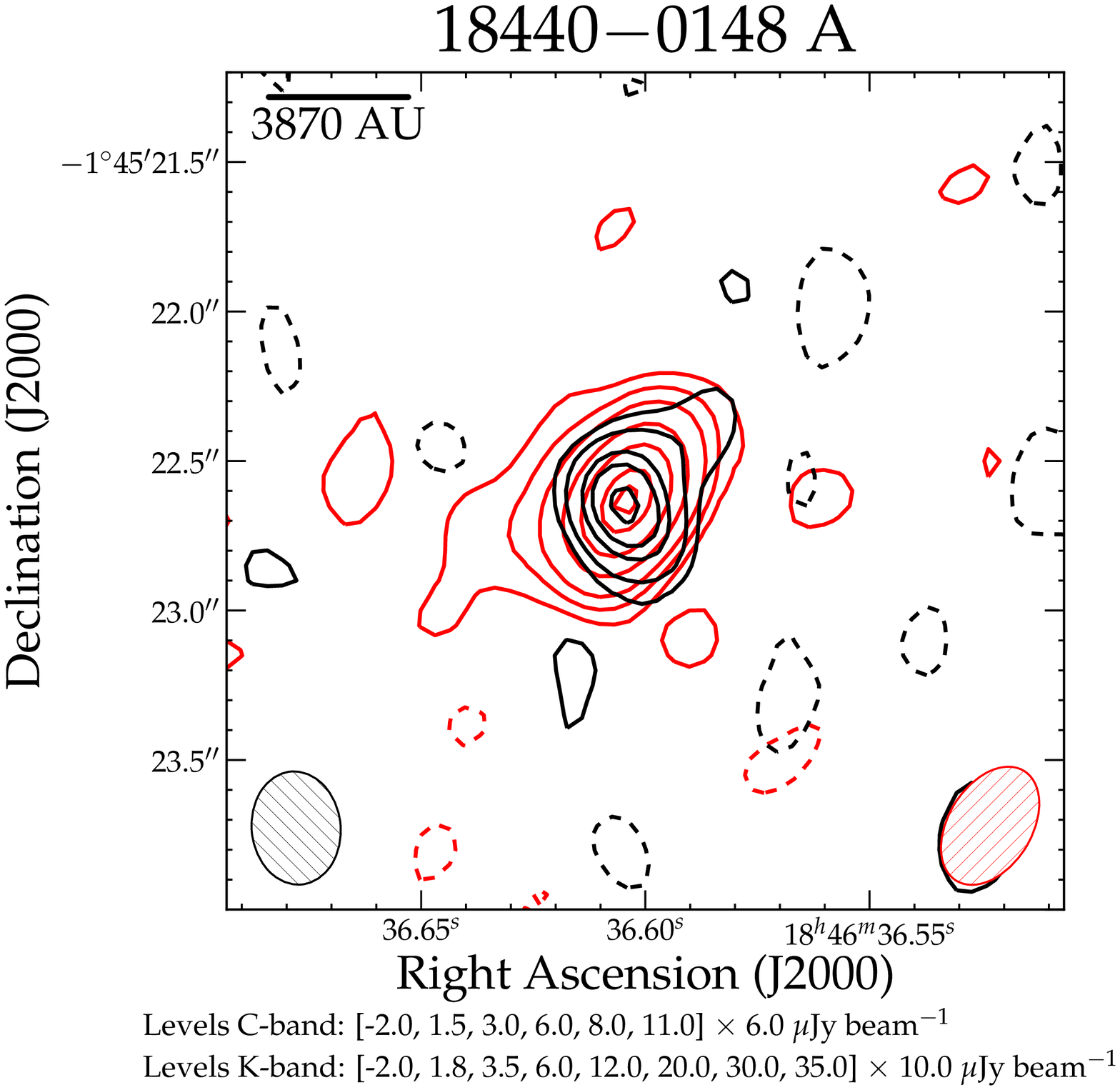} \\
 \includegraphics[width=0.44\textwidth, trim = 3 60 10 160, clip, angle = 0]{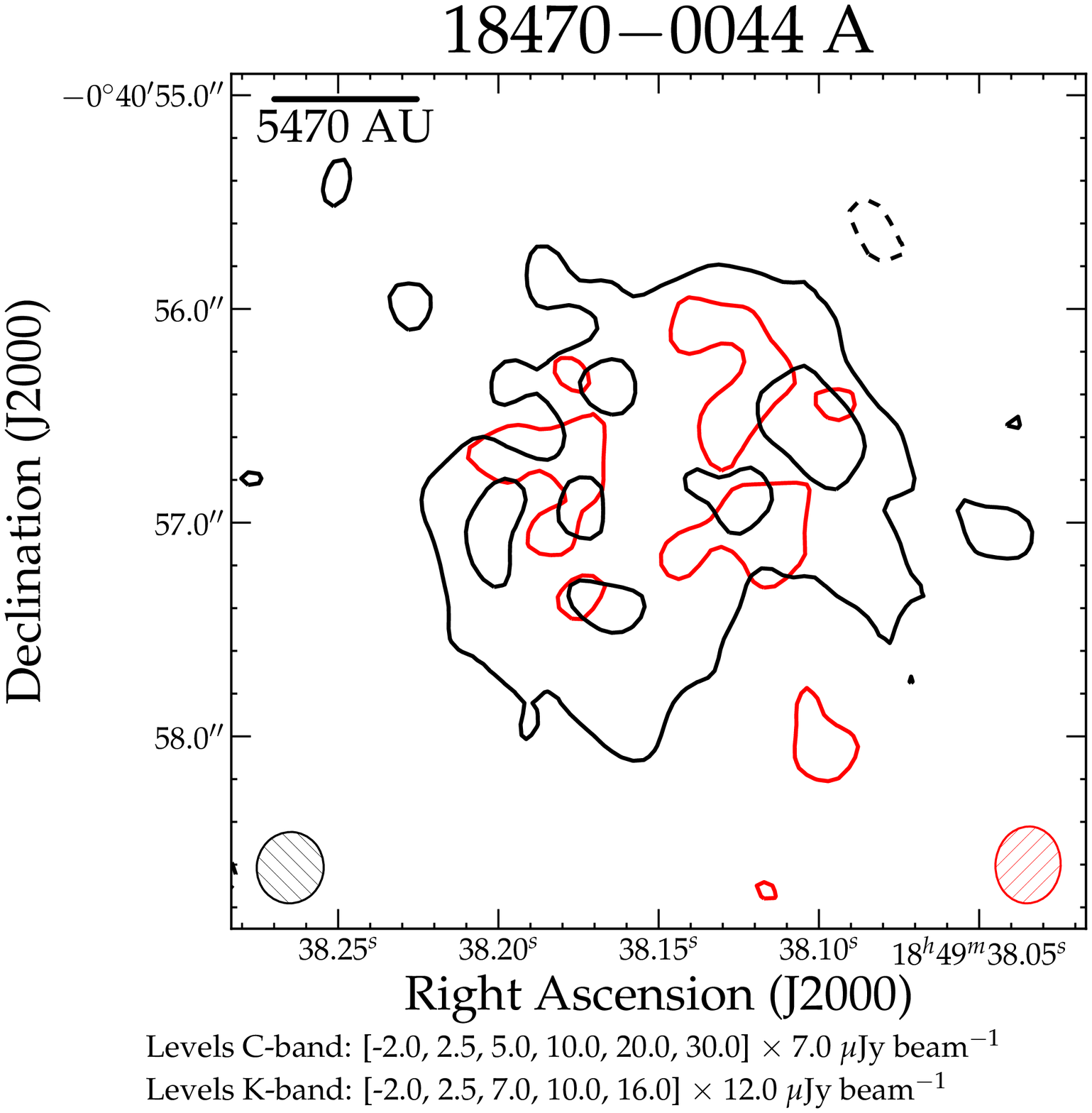} &
  \includegraphics[width=0.44\textwidth, trim = 3 60 10 160, clip, angle = 0]{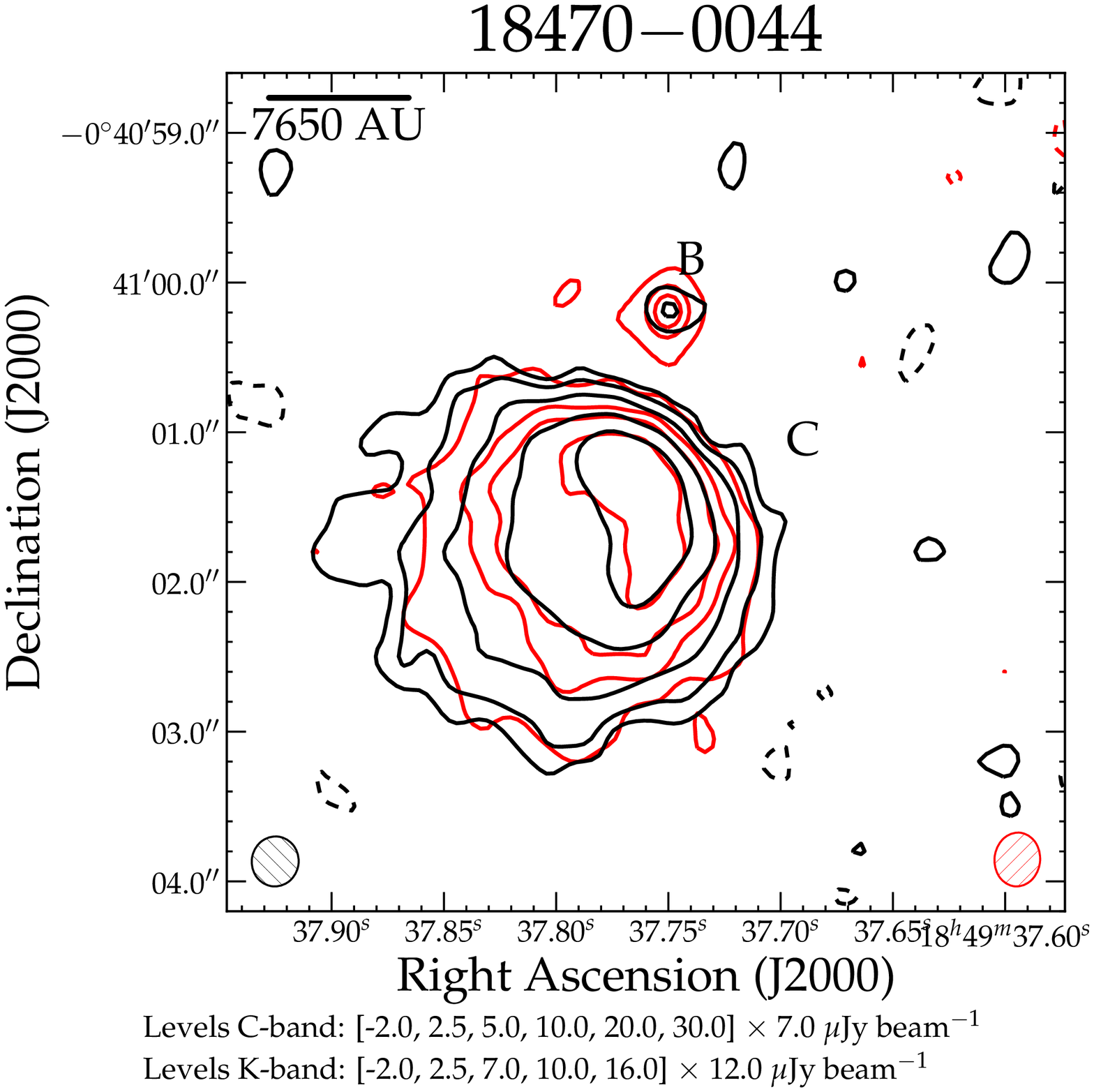}

\end{tabular}

 \hspace*{\fill}%
\caption{\small{Continued.}}
 \label{fig:maps}
\end{figure}

\begin{figure}[htbp]
\centering
\begin{tabular}{cc}
  \ContinuedFloat
\hspace*{\fill}%
 \includegraphics[width=0.44\textwidth, trim = 3 60 10 160, clip, angle = 0]{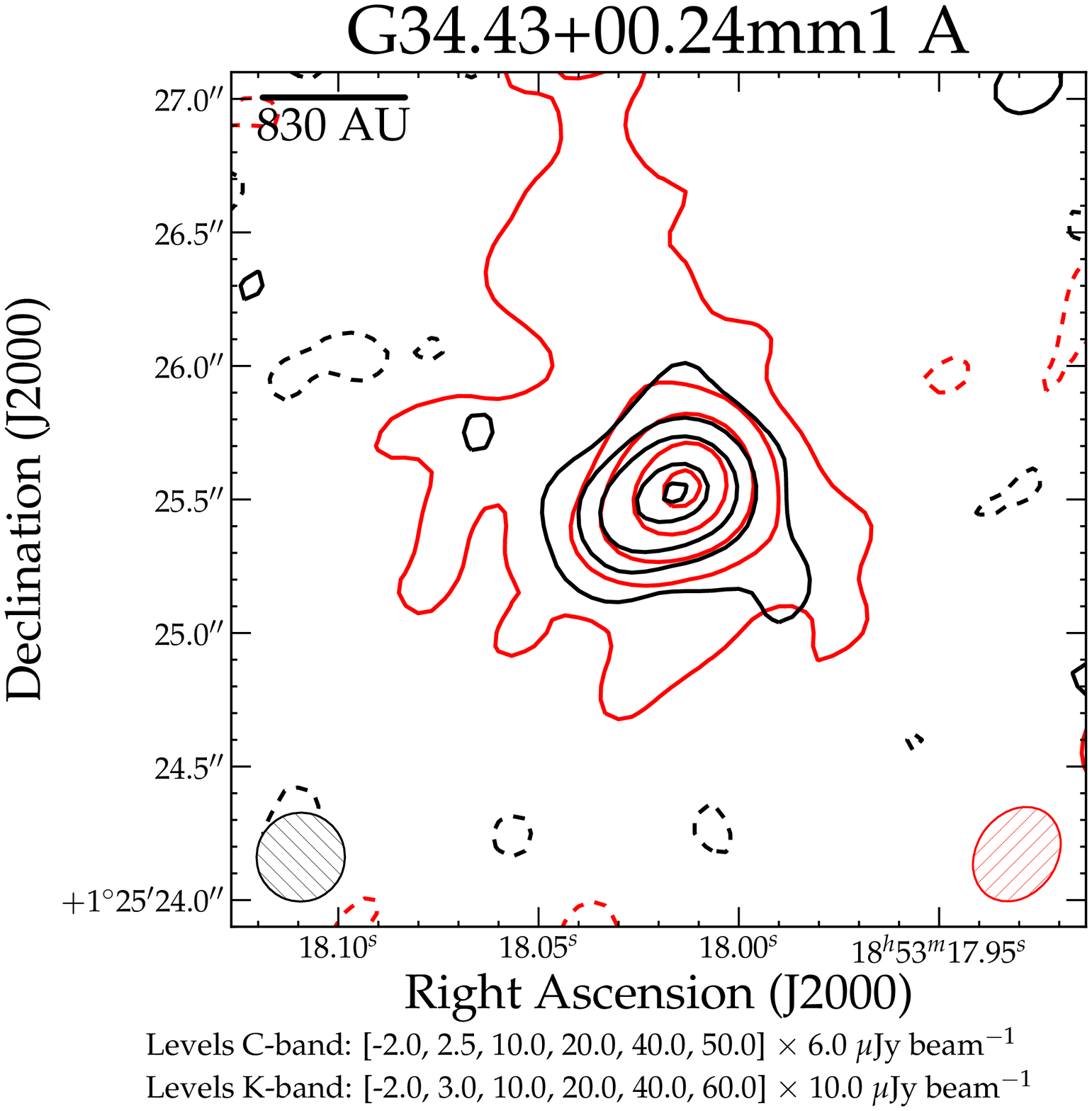} &
 \includegraphics[width=0.44\textwidth, trim = 3 60 10 160, clip, angle = 0]{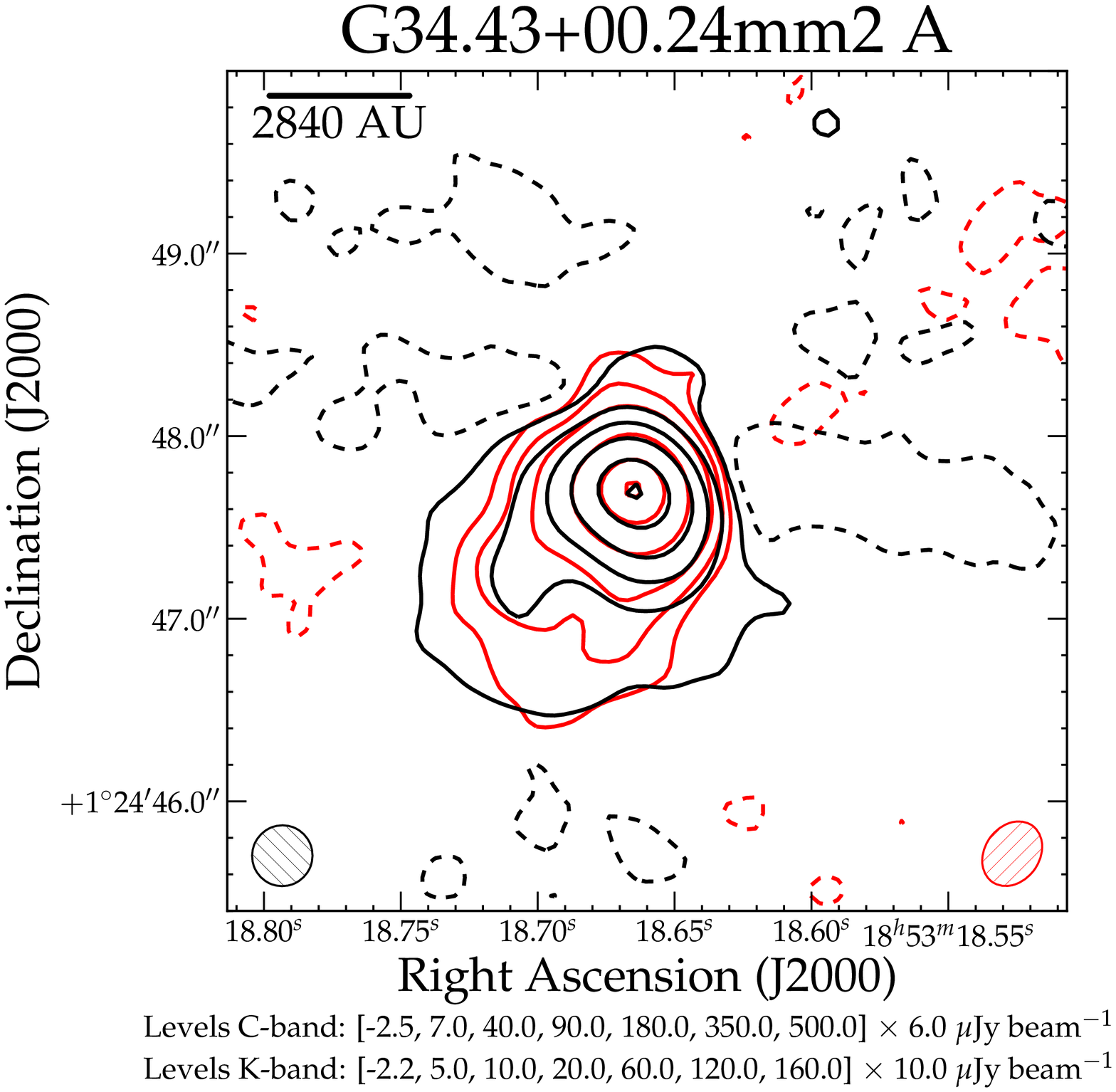} \\
  \includegraphics[width=0.44\textwidth, trim = 3 60 10 160, clip, angle = 0]{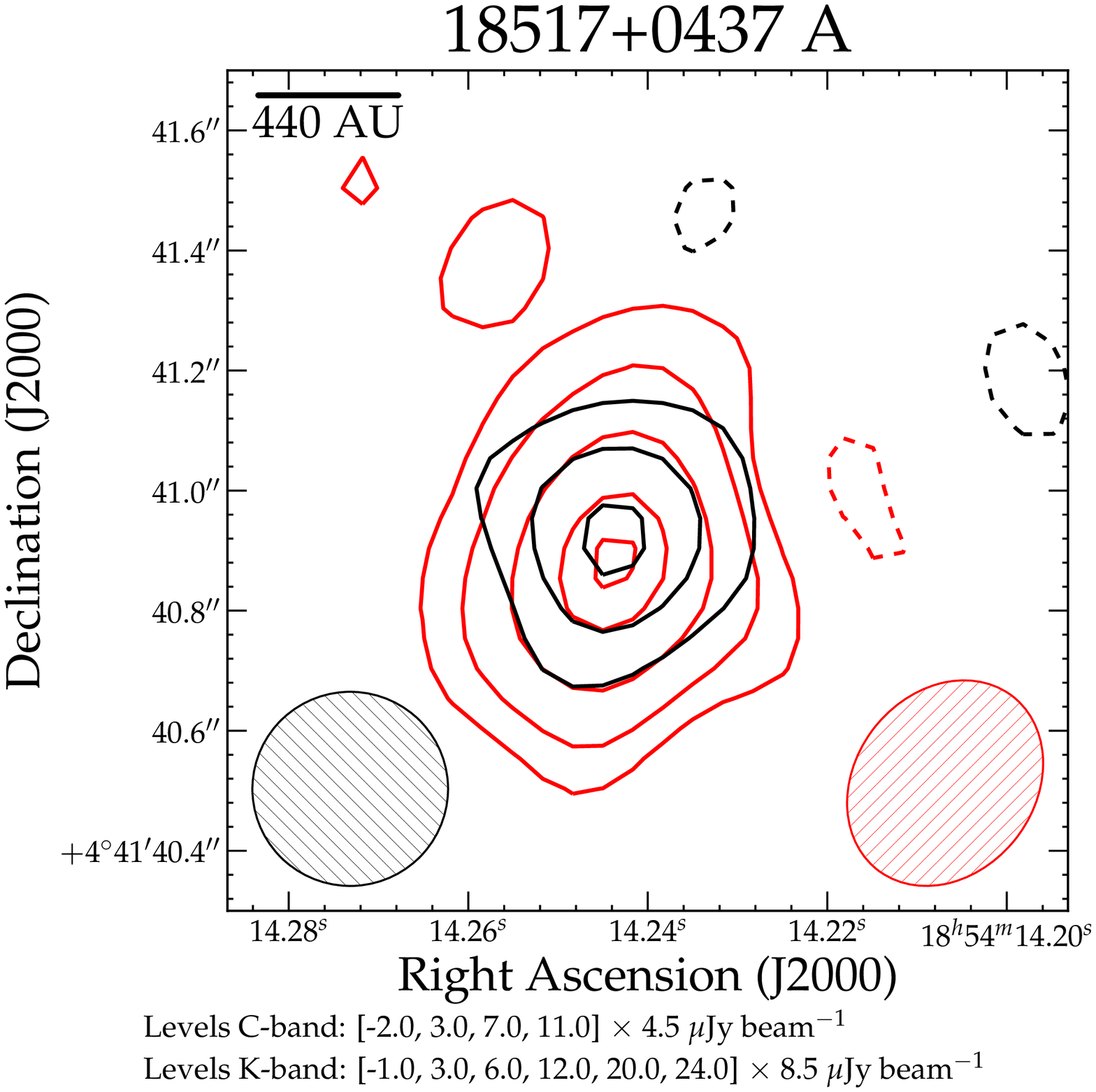}&
 \includegraphics[width=0.44\textwidth, trim = 3 60 10 160, clip, angle = 0]{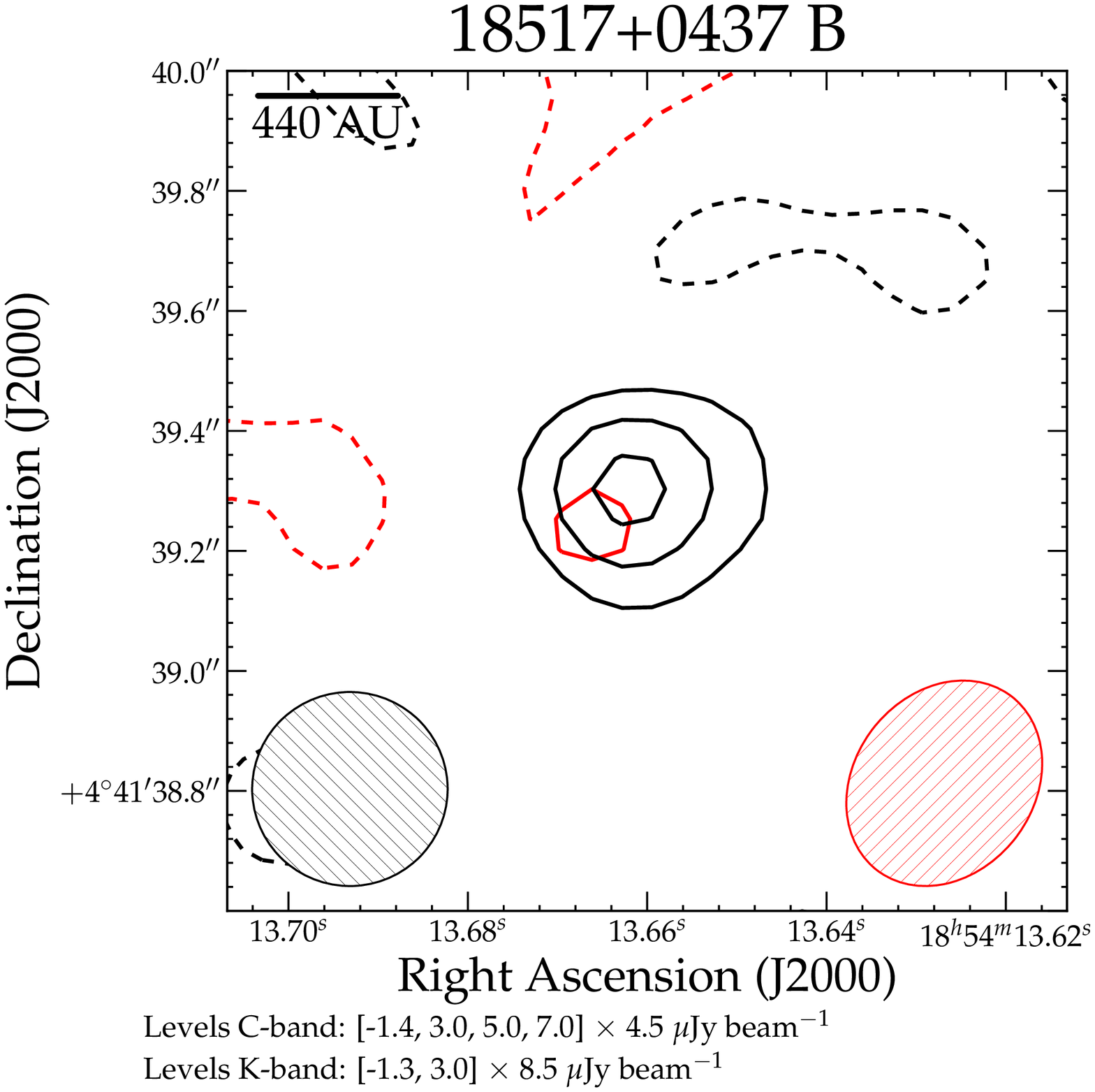} \\
 \includegraphics[width=0.44\textwidth, trim = 3 60 10 160, clip, angle = 0]{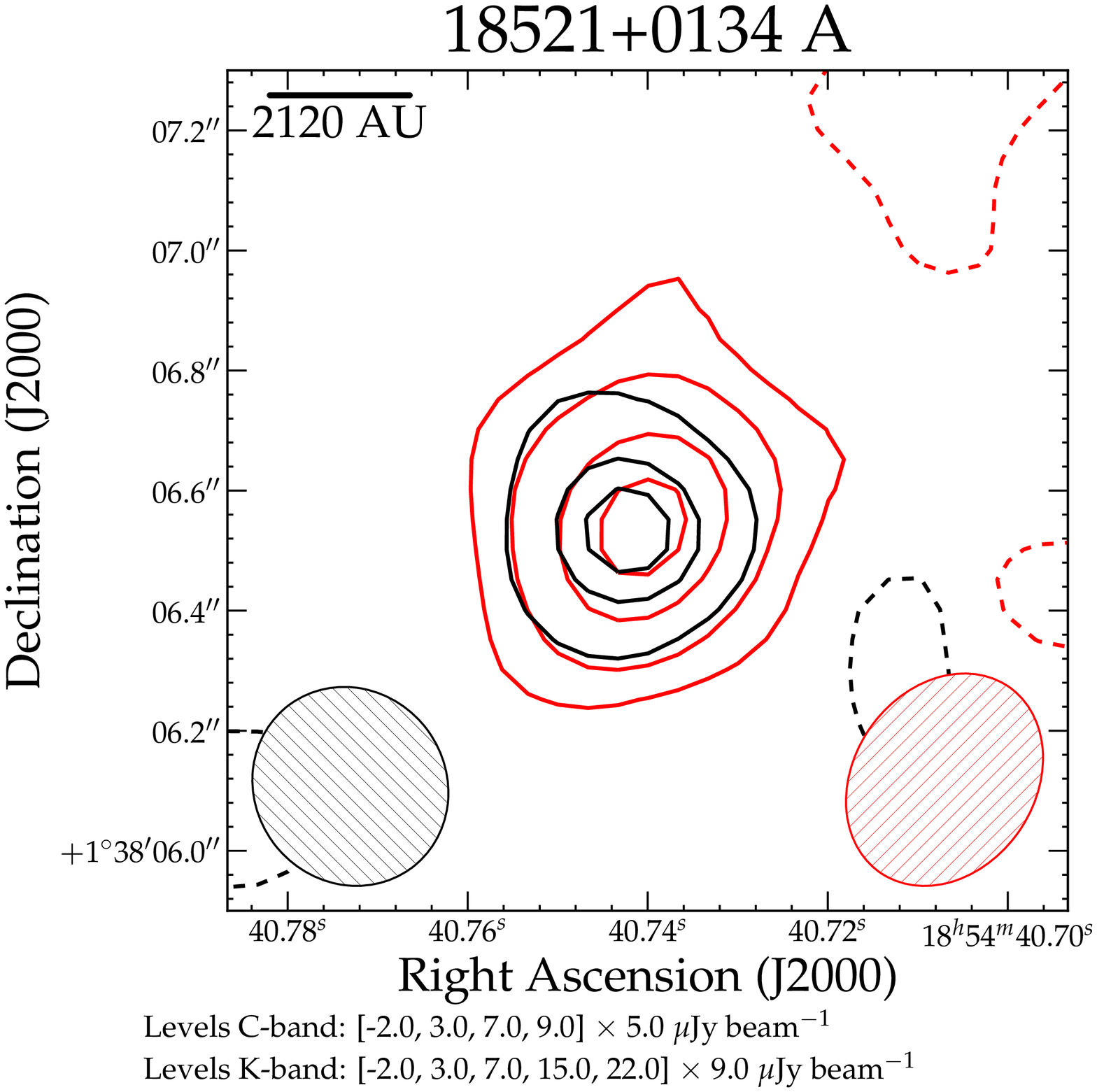} &
  \includegraphics[width=0.44\textwidth, trim = 3 60 10 160, clip, angle = 0]{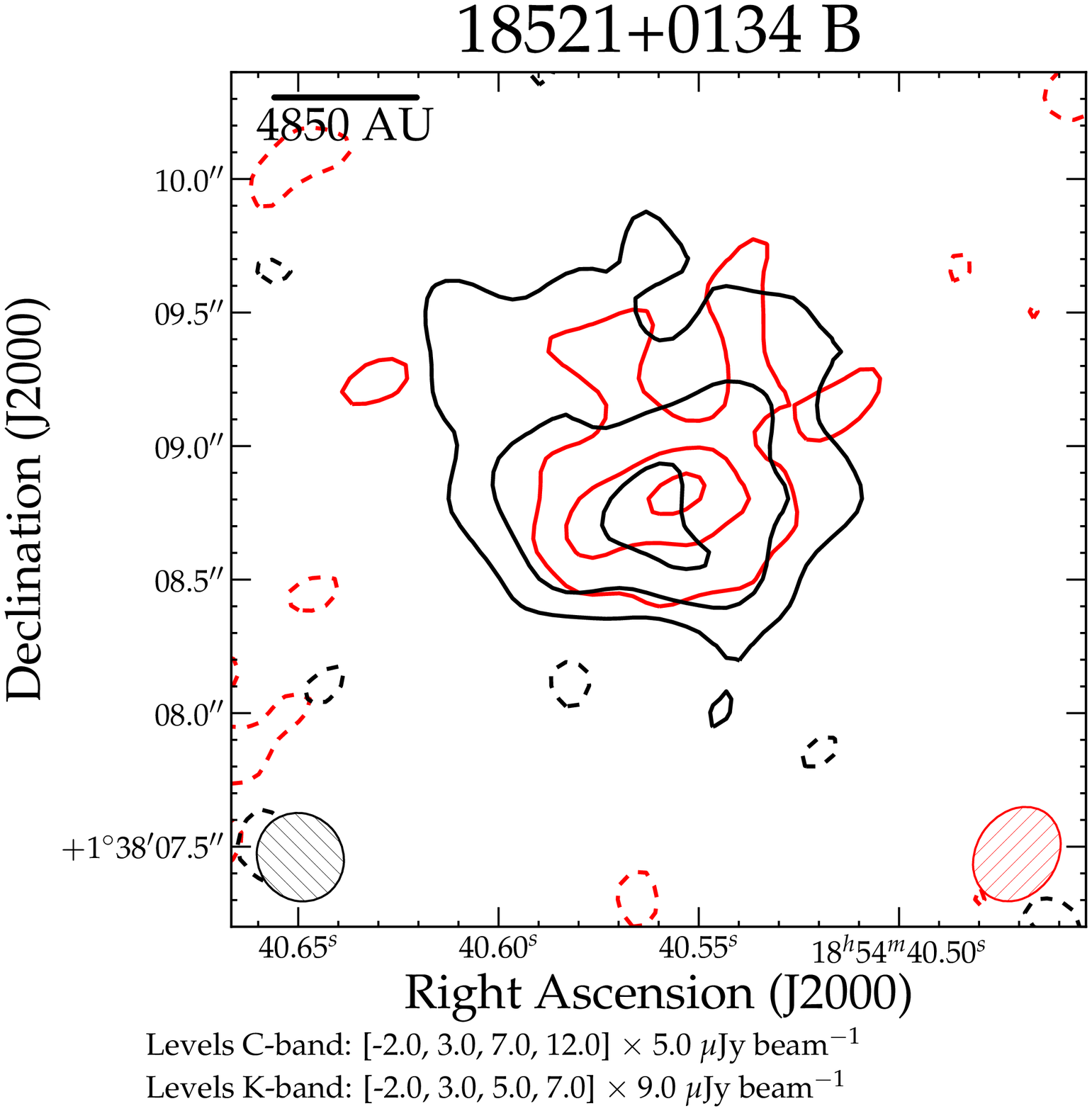}

\end{tabular}

 \hspace*{\fill}%
\caption{\small{Continued.}}
 \label{fig:maps}
\end{figure}

\begin{figure}[htbp]
\centering
\begin{tabular}{cc}
  \ContinuedFloat
\hspace*{\fill}%
 \includegraphics[width=0.44\textwidth, trim = 3 60 10 160, clip, angle = 0]{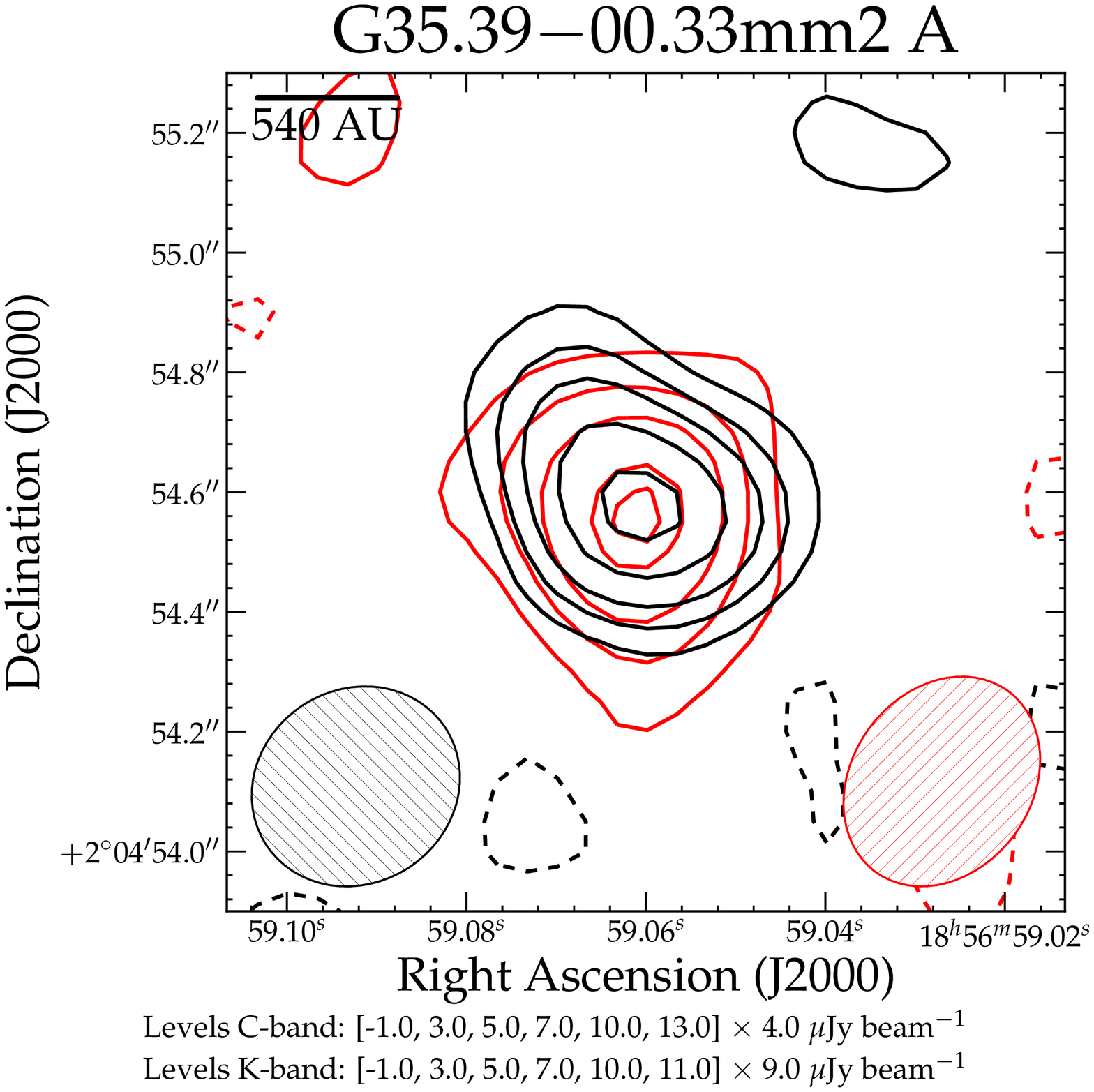} &
 \includegraphics[width=0.44\textwidth, trim = 3 60 10 160, clip, angle = 0]{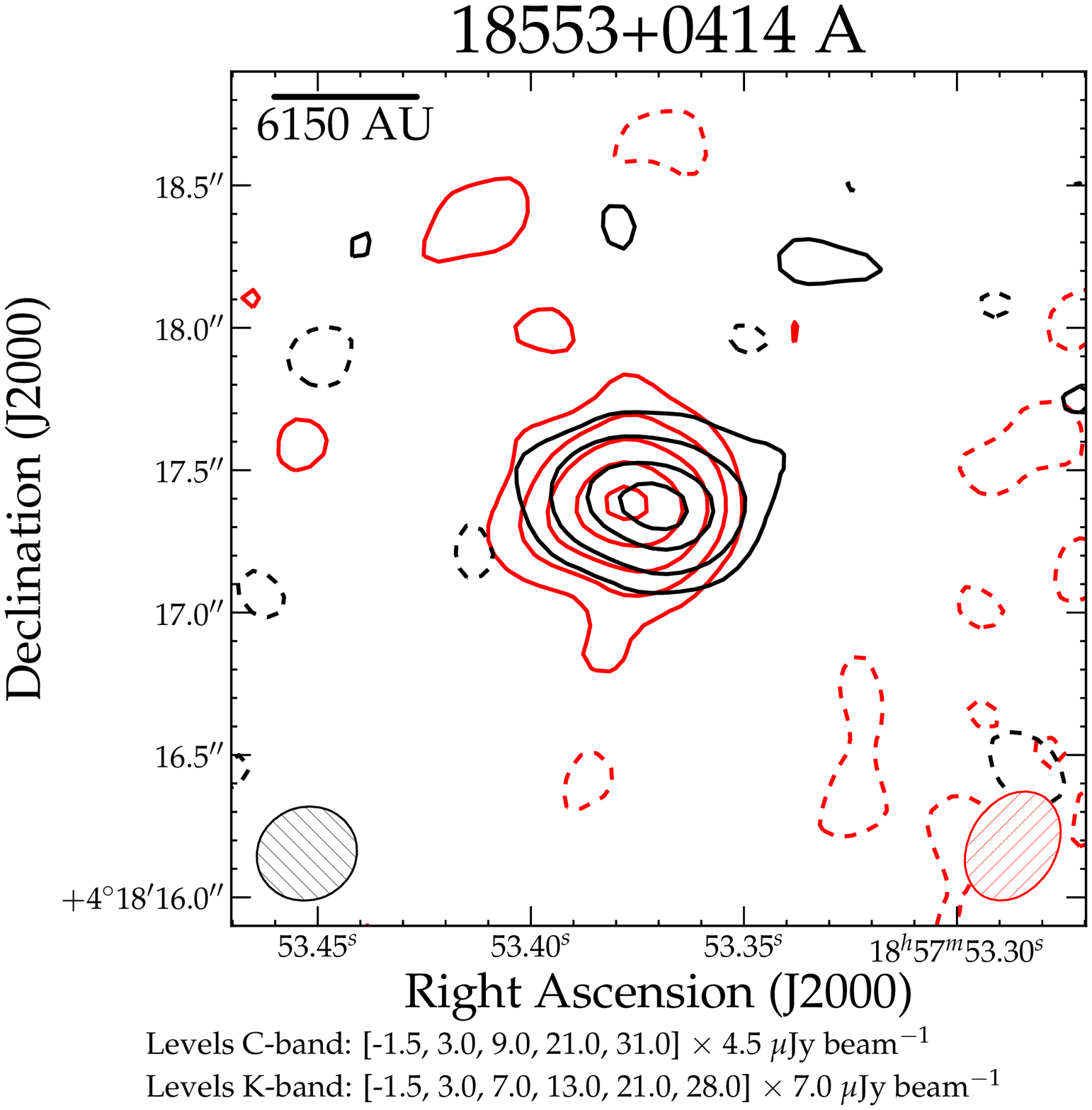} \\
  \includegraphics[width=0.44\textwidth, trim = 3 60 10 160, clip, angle = 0]{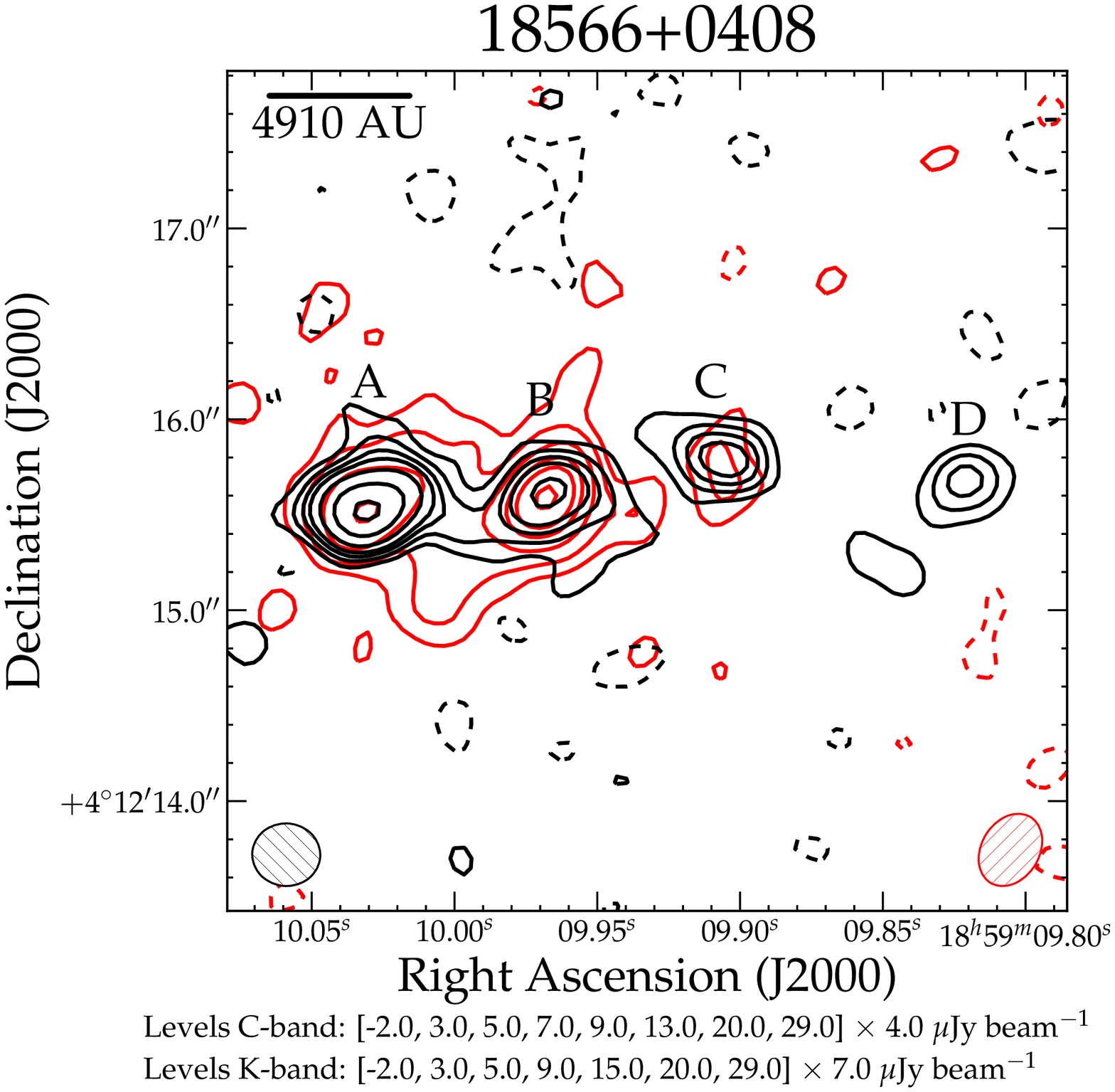}&
 \includegraphics[width=0.44\textwidth, trim = 3 60 10 160, clip, angle = 0]{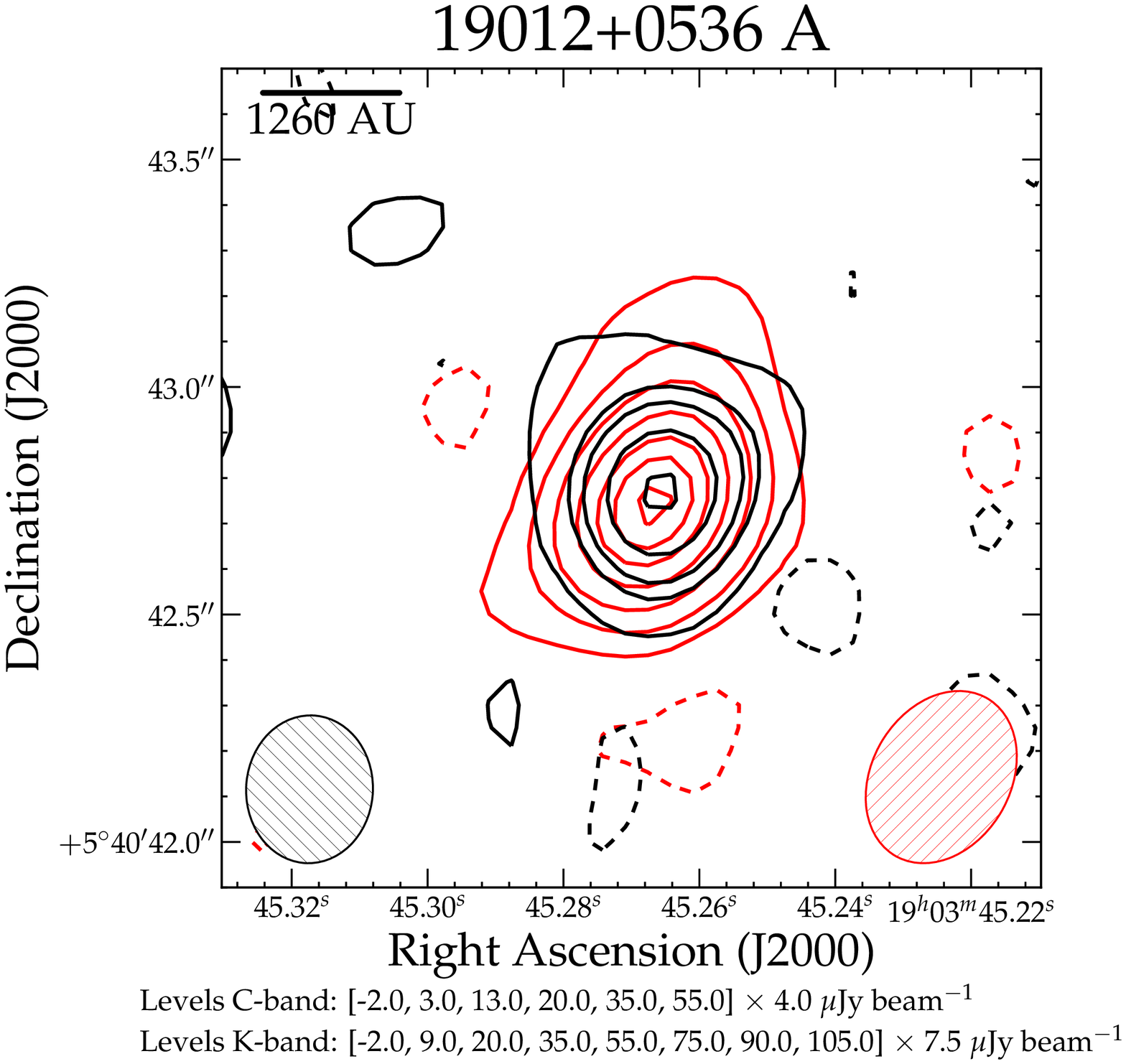} \\
 \includegraphics[width=0.44\textwidth, trim = 3 60 10 160, clip, angle = 0]{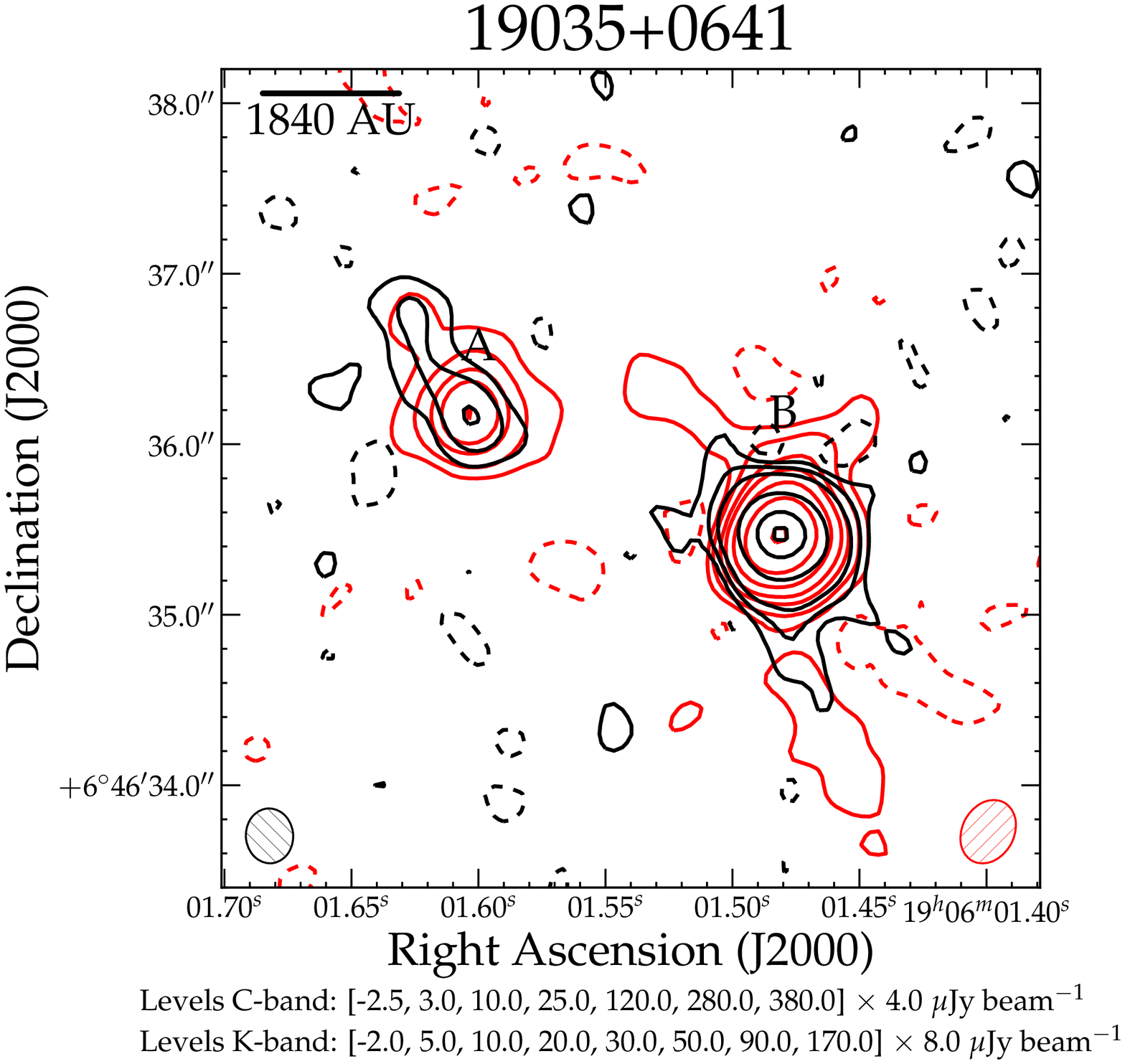} &
  \includegraphics[width=0.44\textwidth, trim = 3 60 10 160, clip, angle = 0]{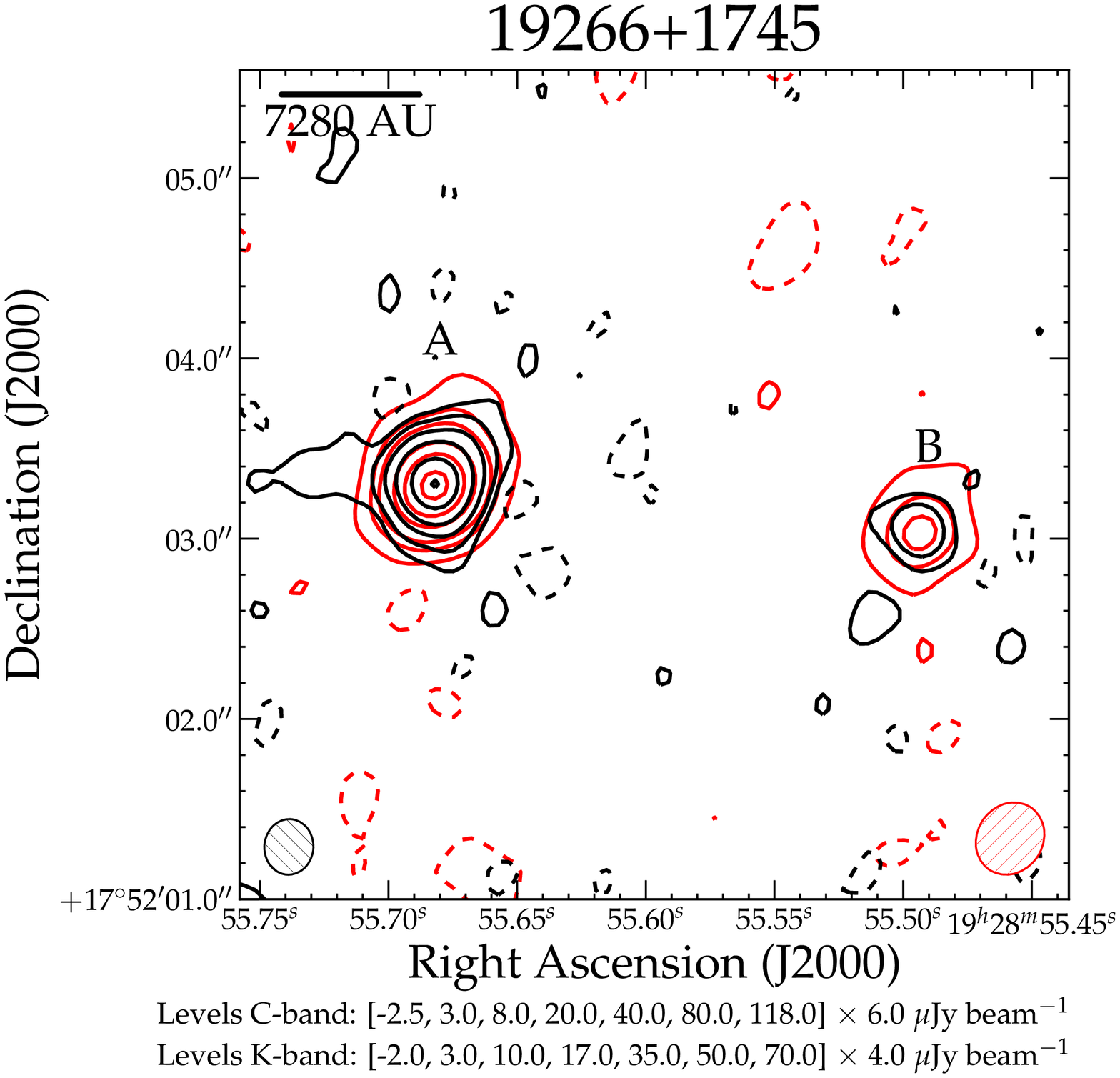}

\end{tabular}

 \hspace*{\fill}%
\caption{\small{Continued.}}
 \label{fig:maps}
\end{figure}

\begin{figure}[htbp]
\centering
\begin{tabular}{cc}
  \ContinuedFloat
\hspace*{\fill}%
 \includegraphics[width=0.44\textwidth, trim = 3 60 10 160, clip, angle = 0]{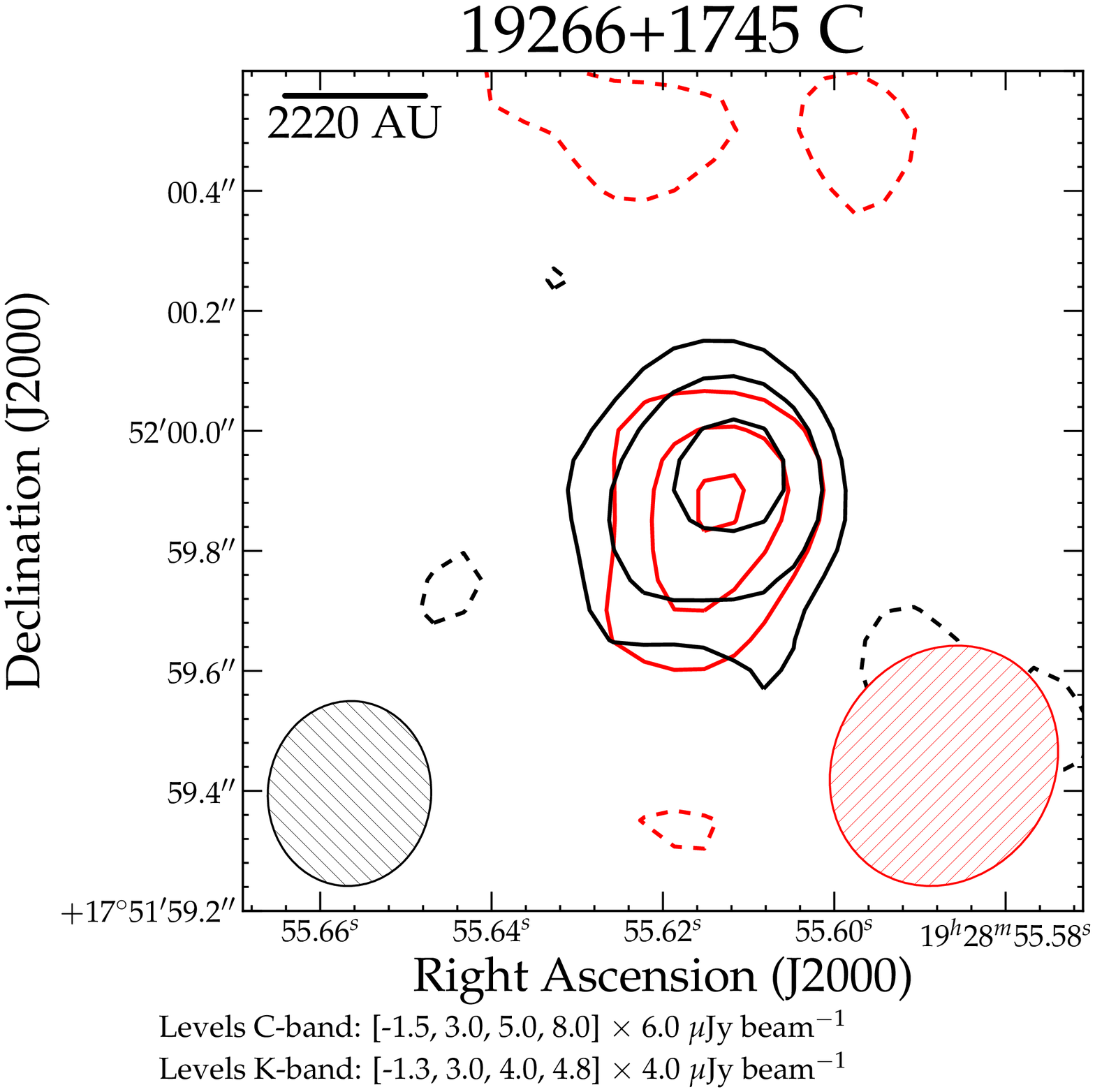} &
 \includegraphics[width=0.44\textwidth, trim = 3 60 10 160, clip, angle = 0]{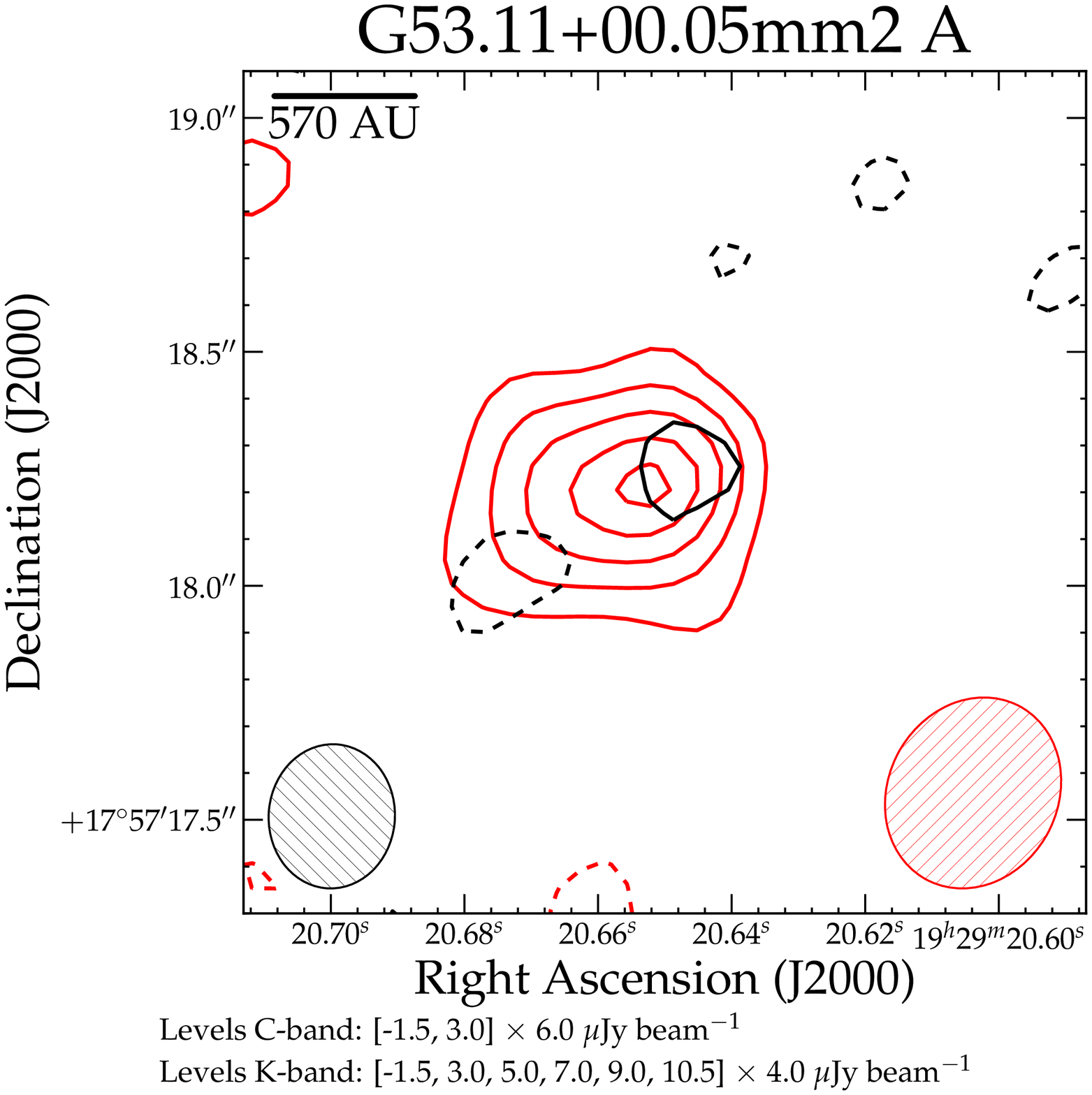} \\
  \includegraphics[width=0.44\textwidth, trim = 3 60 10 160, clip, angle = 0]{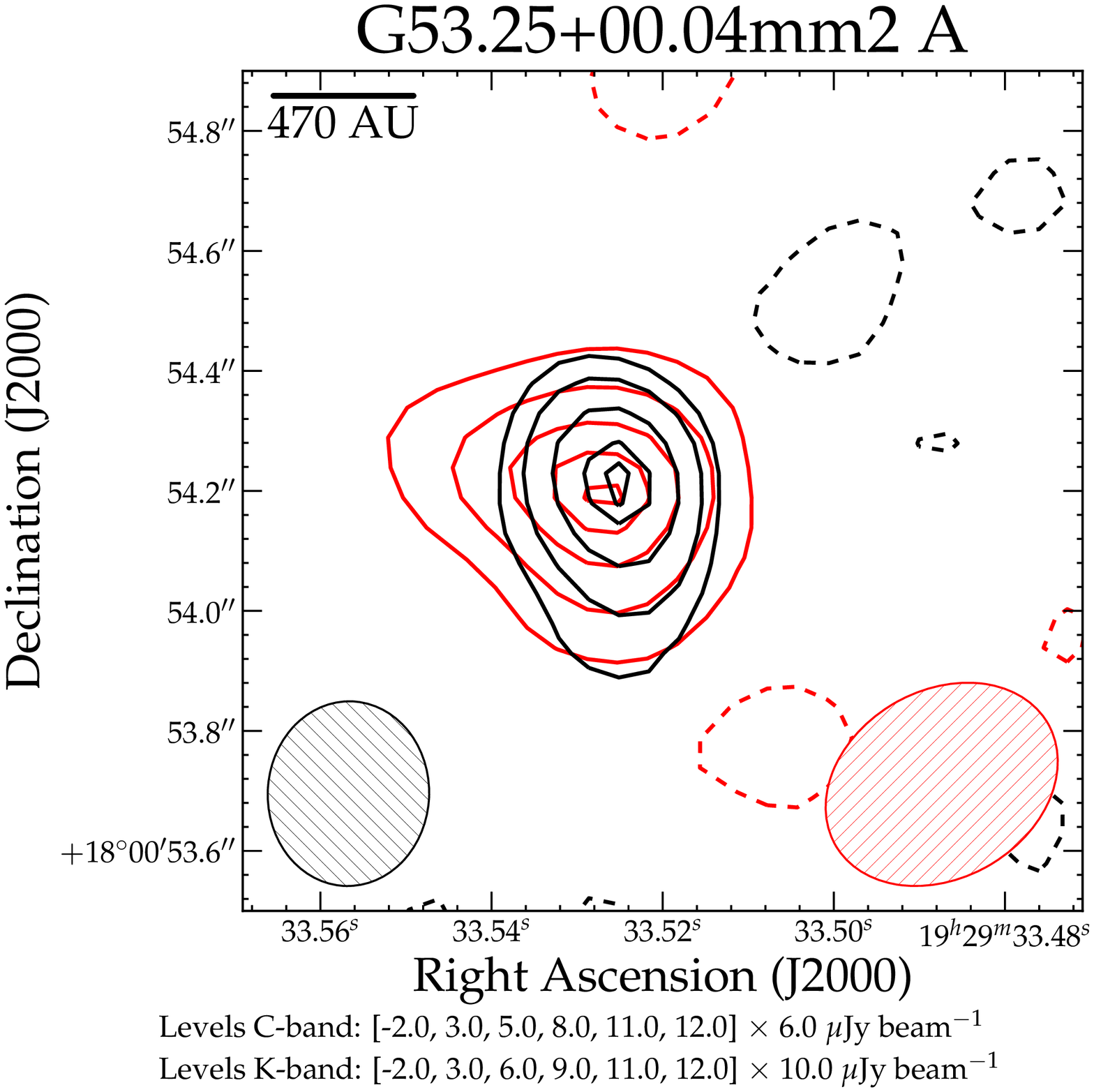}&
 \includegraphics[width=0.44\textwidth, trim = 3 60 10 160, clip, angle = 0]{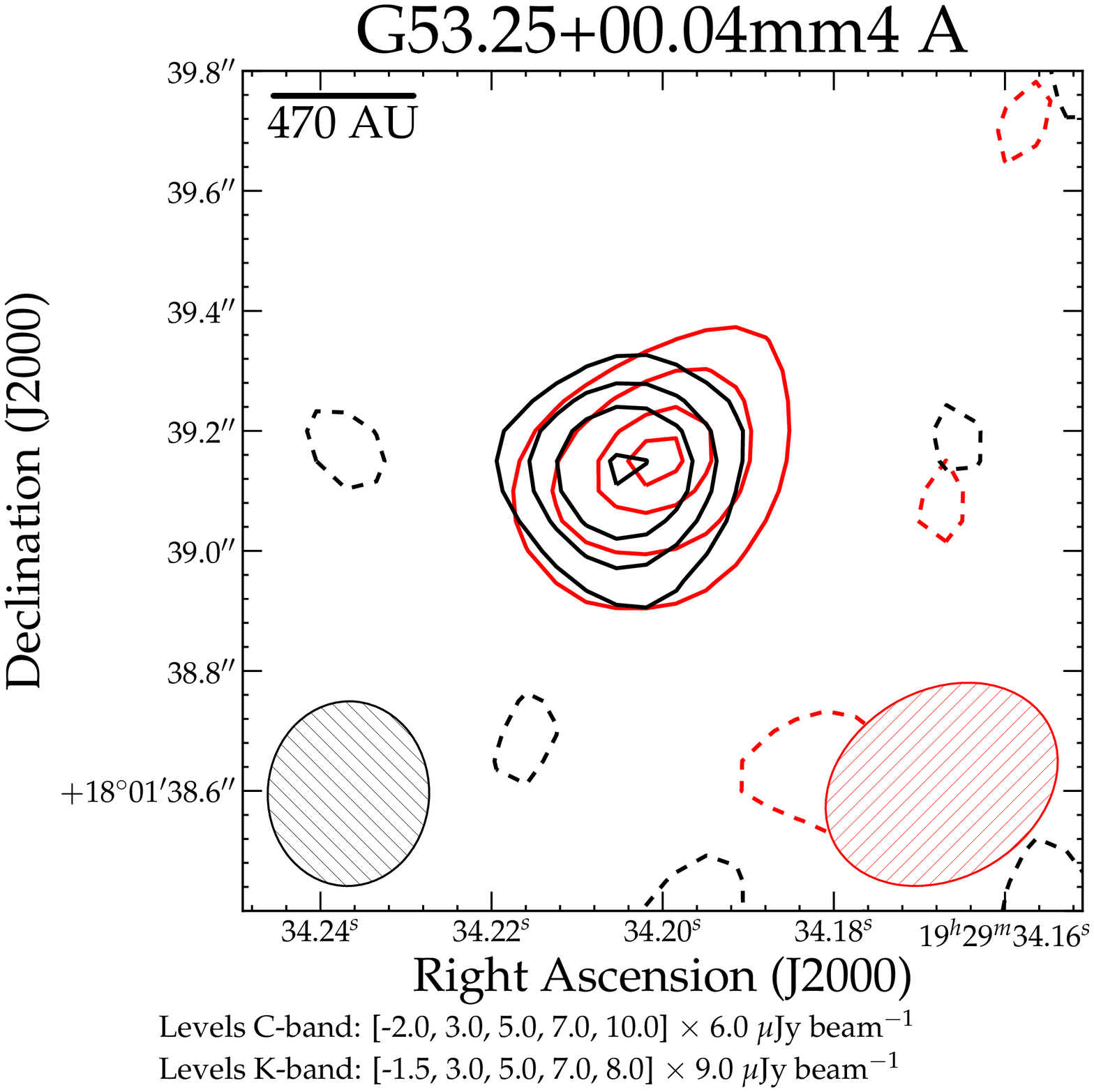} \\
 \includegraphics[width=0.44\textwidth, trim = 3 60 10 160, clip, angle = 0]{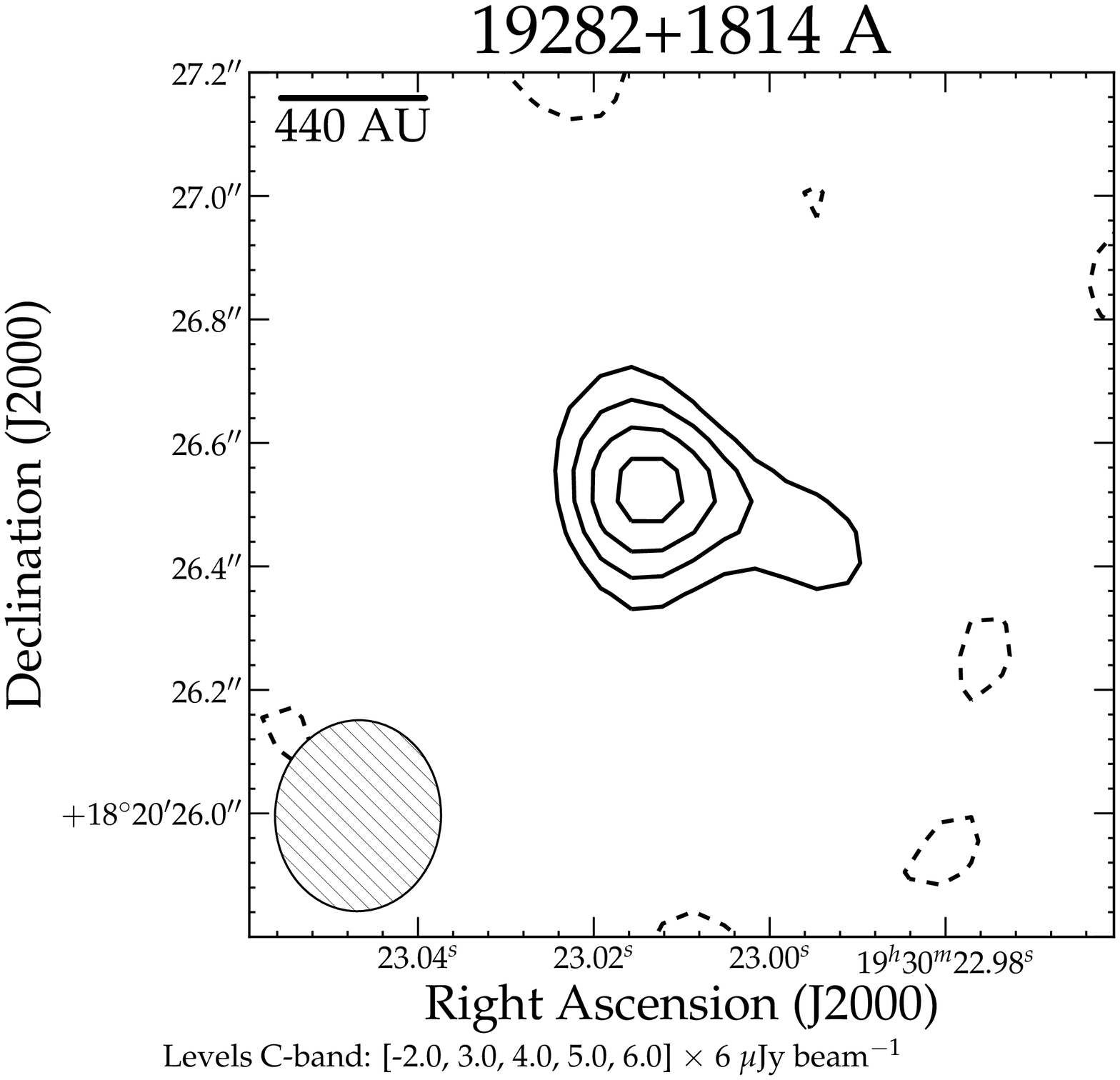} &
  \includegraphics[width=0.44\textwidth, trim = 3 60 10 160, clip, angle = 0]{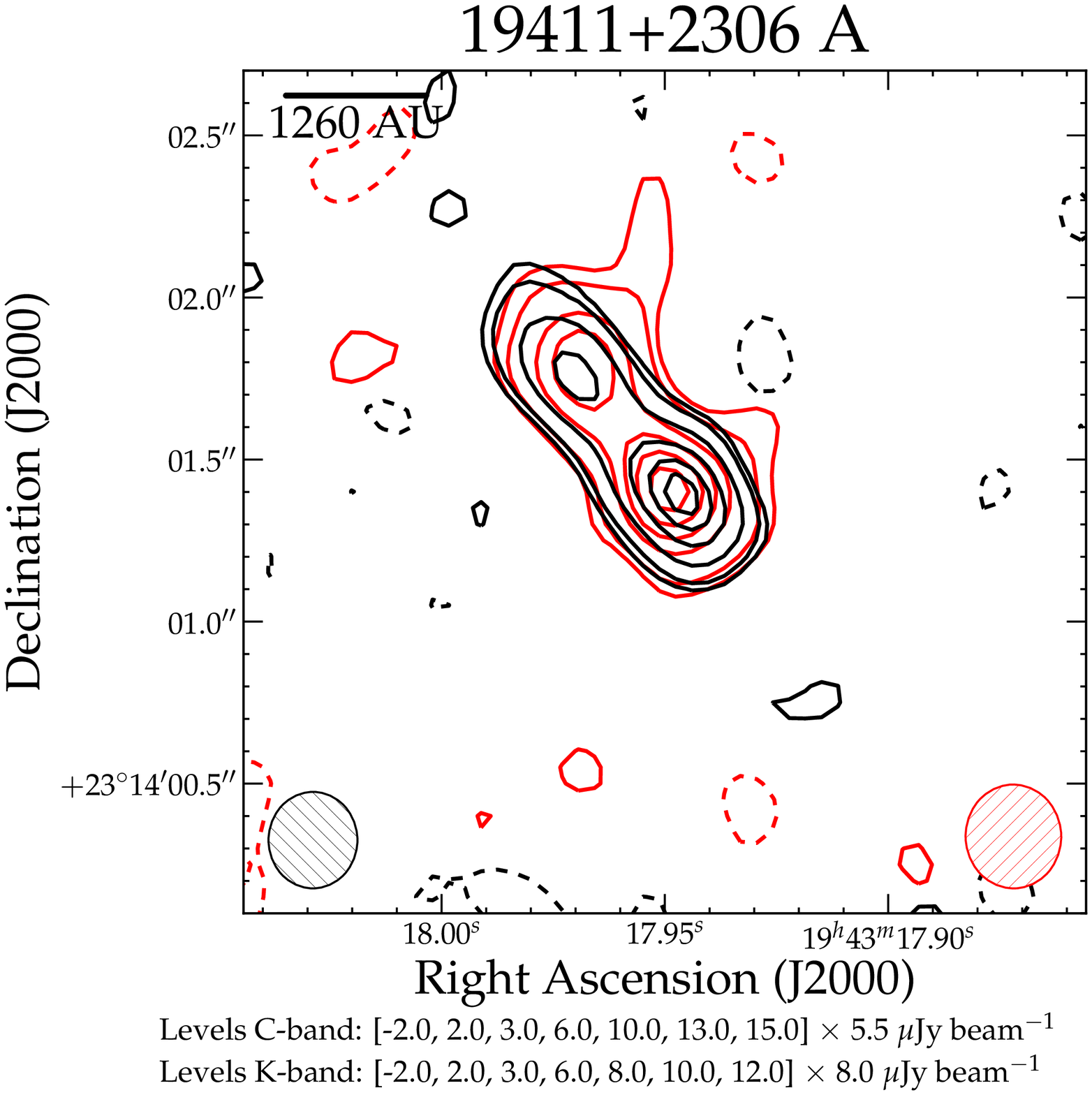} 

\end{tabular}

 \hspace*{\fill}%
\caption{\small{Continued.}}
 \label{fig:maps}
\end{figure}

\begin{figure}[htbp]
\centering
\begin{tabular}{cc}
  \ContinuedFloat
\hspace*{\fill}%
 \includegraphics[width=0.44\textwidth, trim = 3 60 10 160, clip, angle = 0]{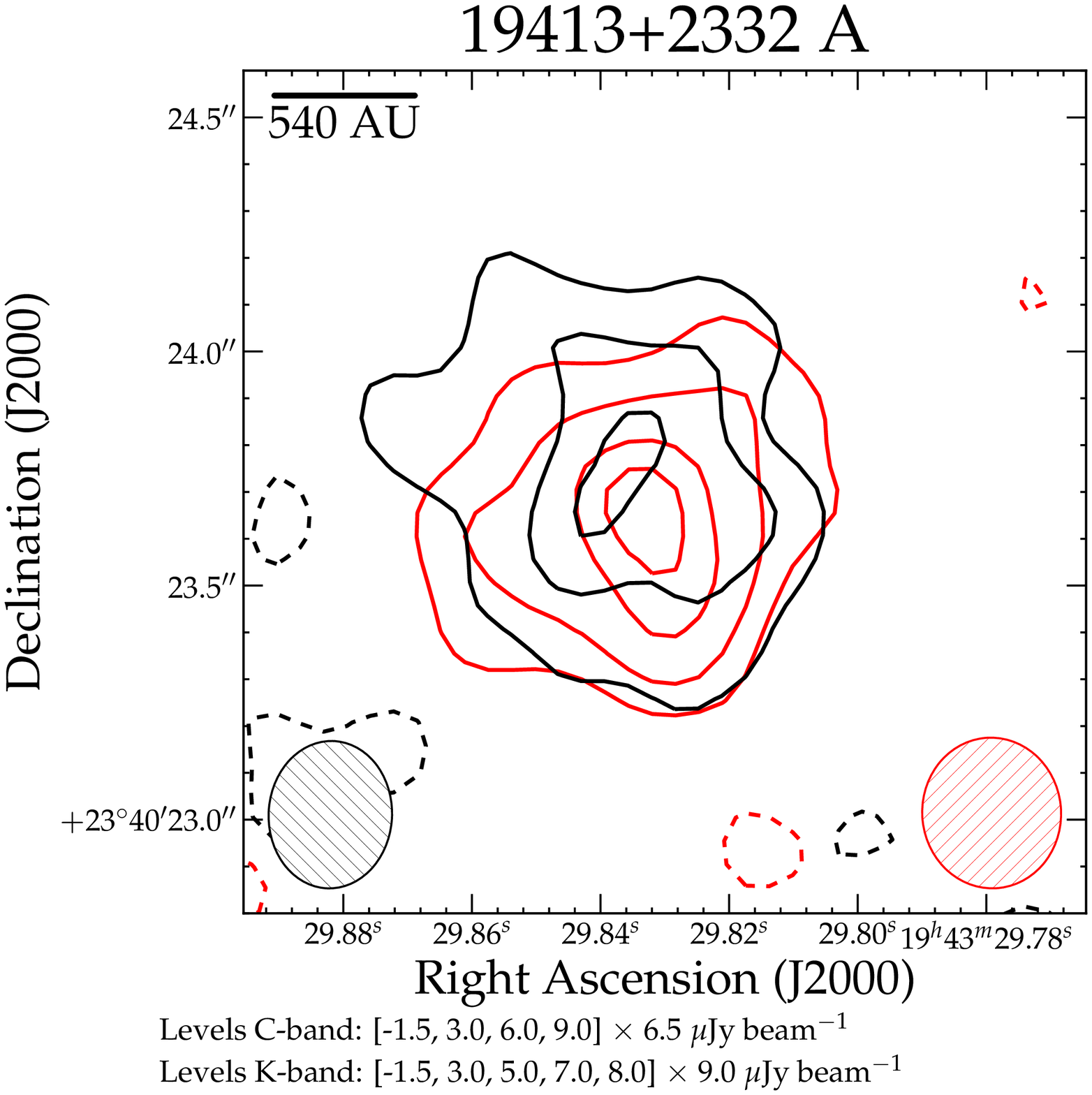} &
 \includegraphics[width=0.44\textwidth, trim = 3 60 10 160, clip, angle = 0]{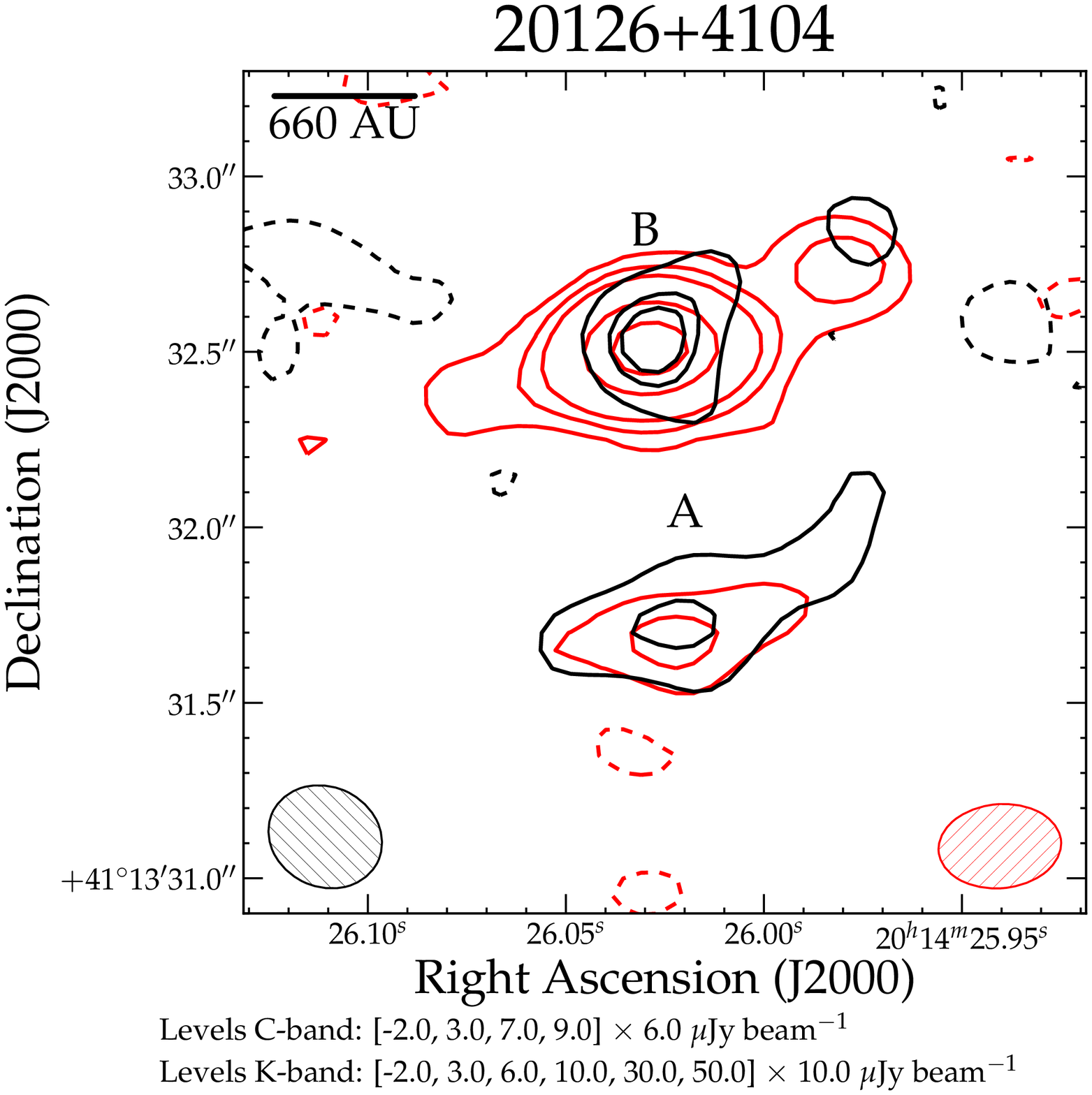} \\
  \includegraphics[width=0.44\textwidth, trim = 3 60 10 160, clip, angle = 0]{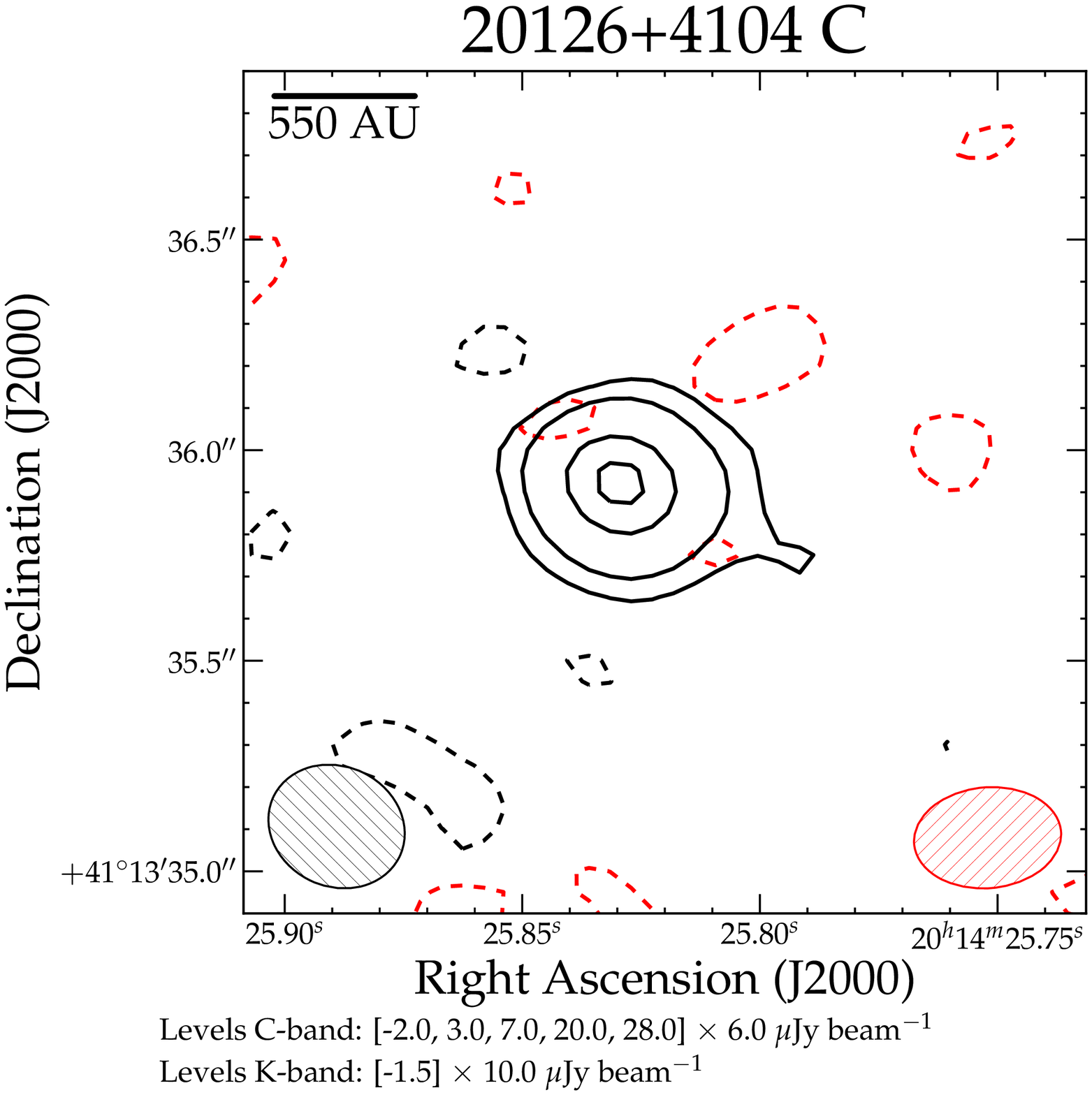}&
 \includegraphics[width=0.44\textwidth, trim = 3 60 10 160, clip, angle = 0]{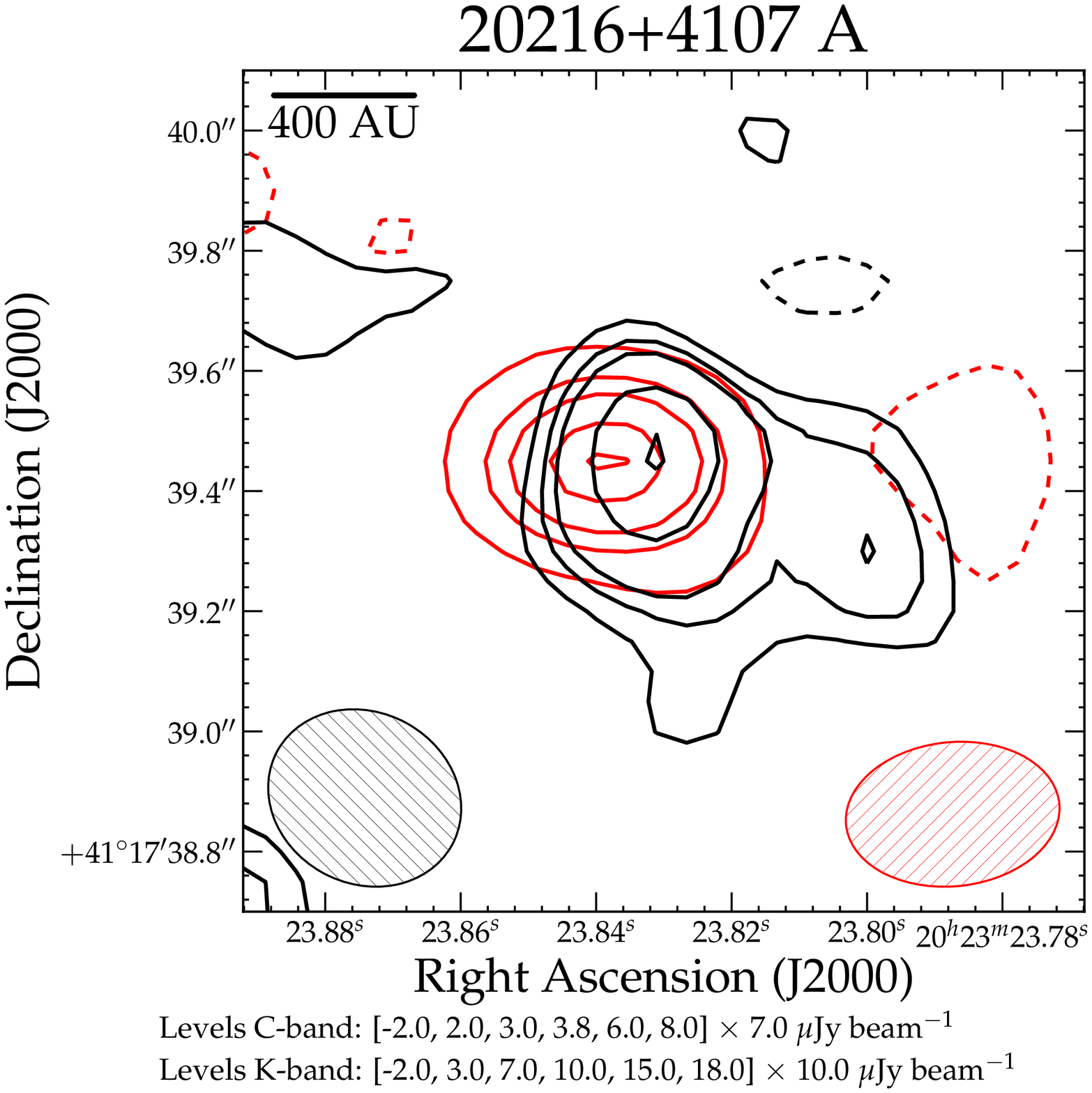} \\
 \includegraphics[width=0.44\textwidth, trim = 3 60 10 160, clip, angle = 0]{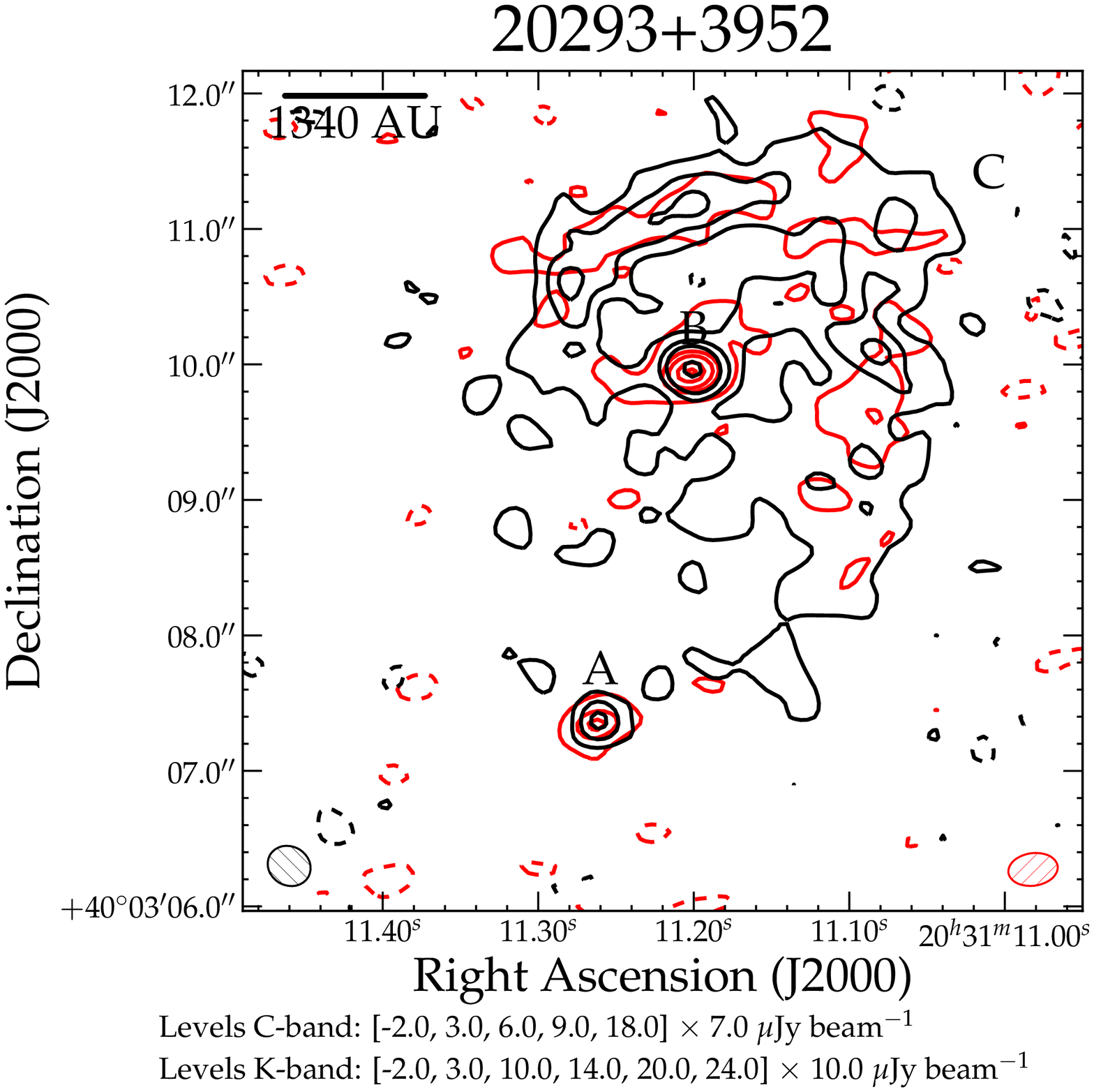} &
  \includegraphics[width=0.44\textwidth, trim = 3 60 10 160, clip, angle = 0]{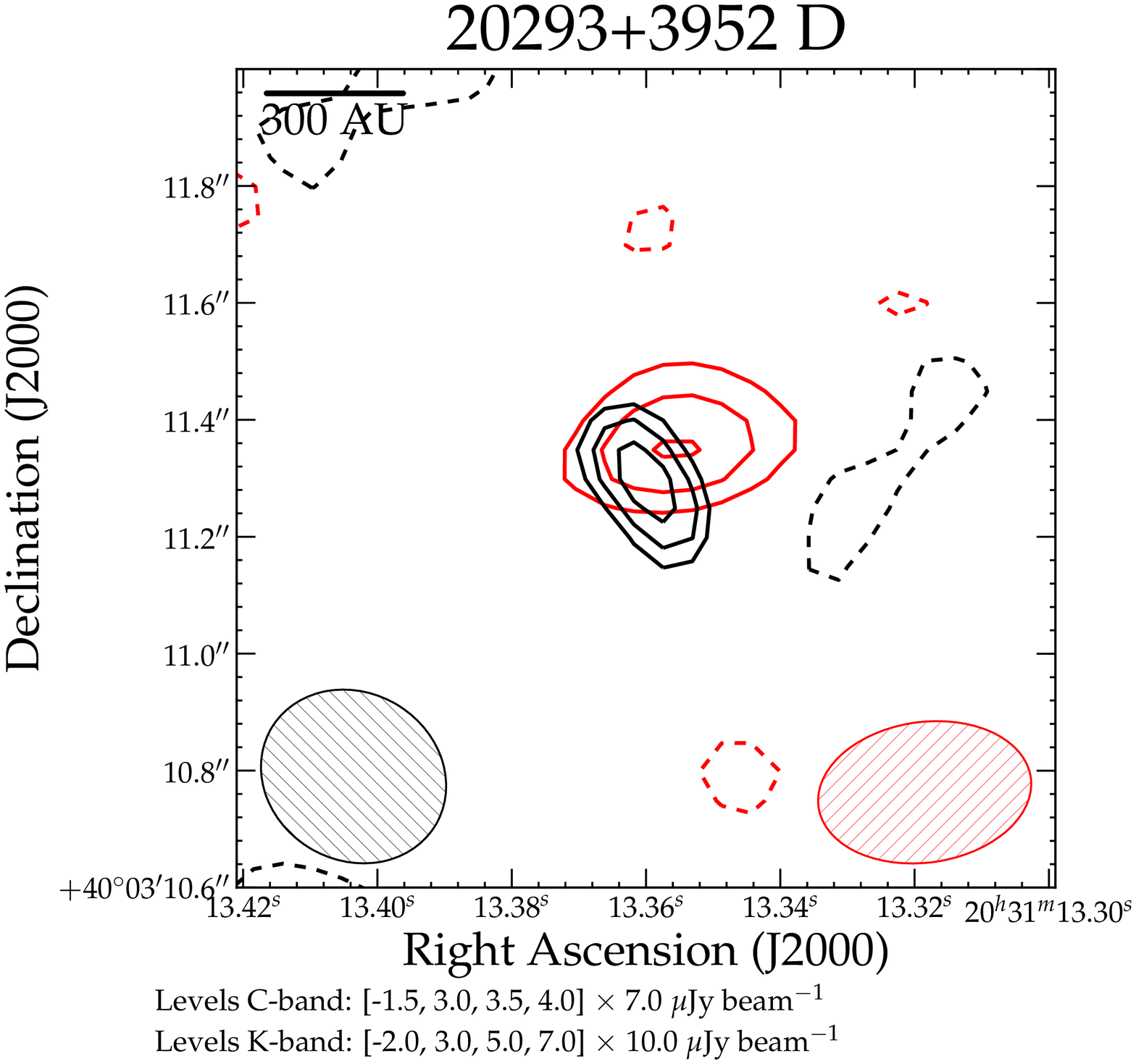} 

\end{tabular}

 \hspace*{\fill}%
\caption{\small{Continued.}}
 \label{fig:maps}
\end{figure}

\begin{figure}[htbp]
\centering
\begin{tabular}{cc}
  \ContinuedFloat
\hspace*{\fill}%
 \includegraphics[width=0.44\textwidth, trim = 3 60 10 160, clip, angle = 0]{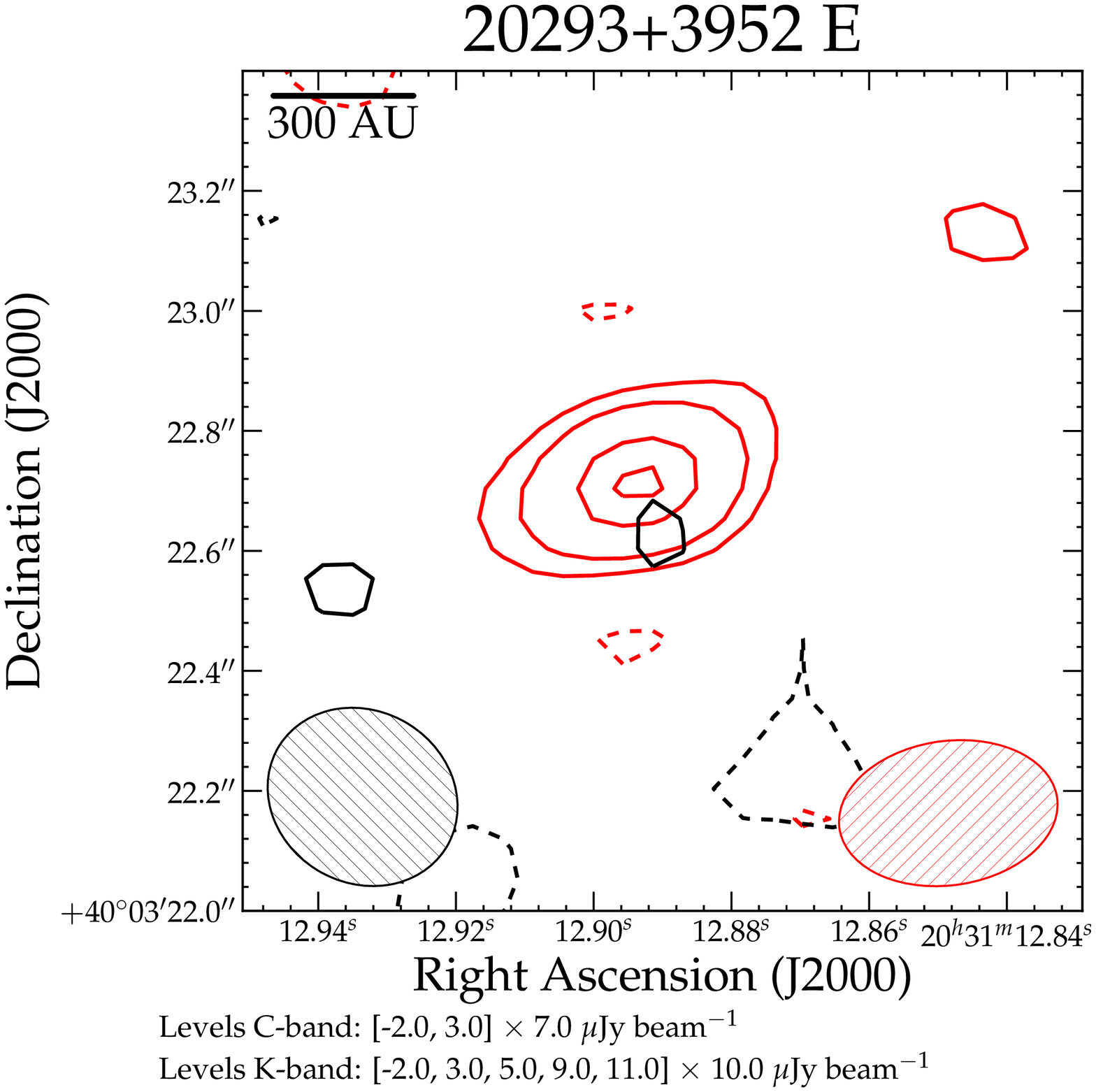} &
 \includegraphics[width=0.44\textwidth, trim = 3 60 10 160, clip, angle = 0]{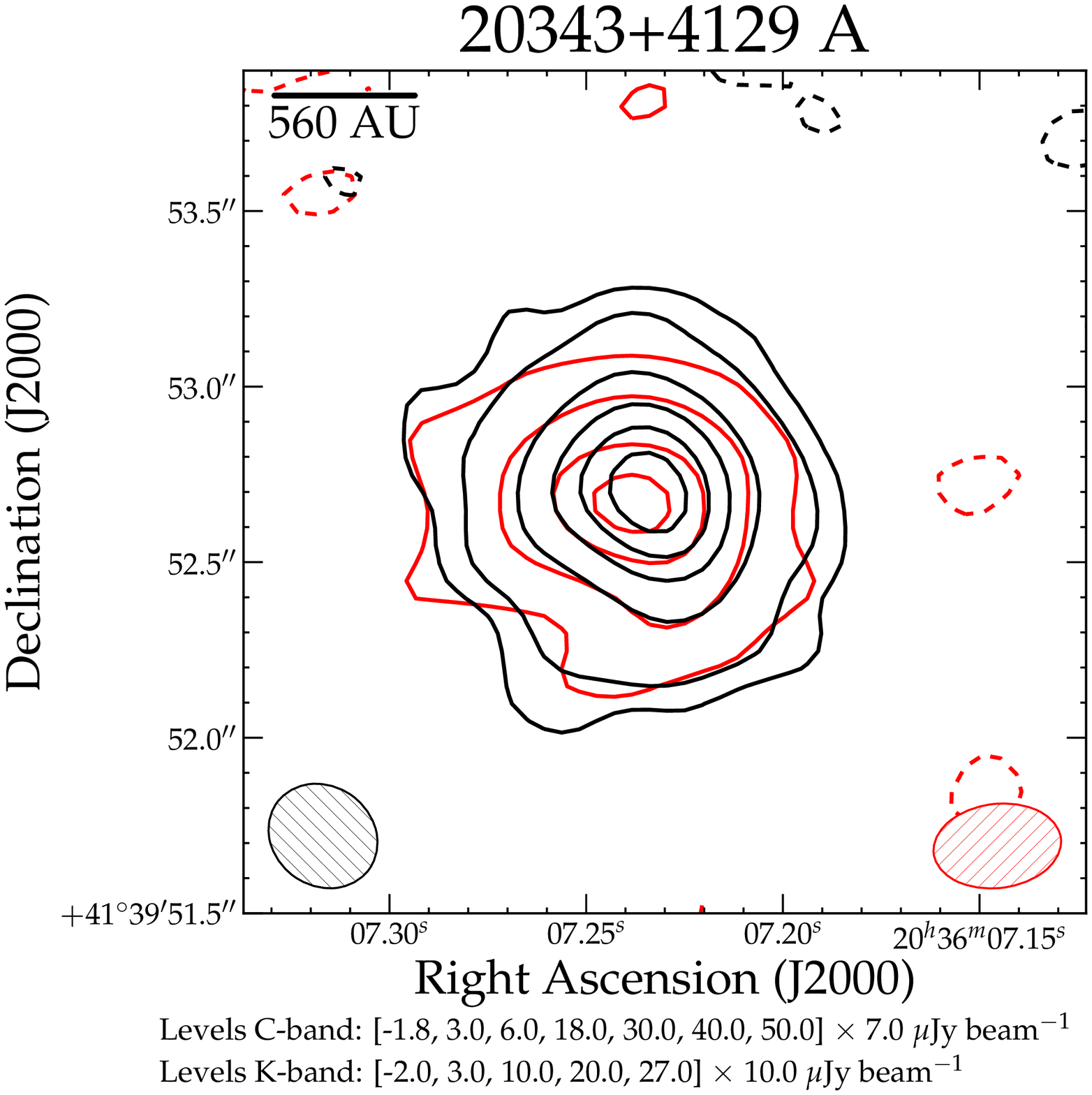} \\
  \includegraphics[width=0.44\textwidth, trim = 3 60 10 160, clip, angle = 0]{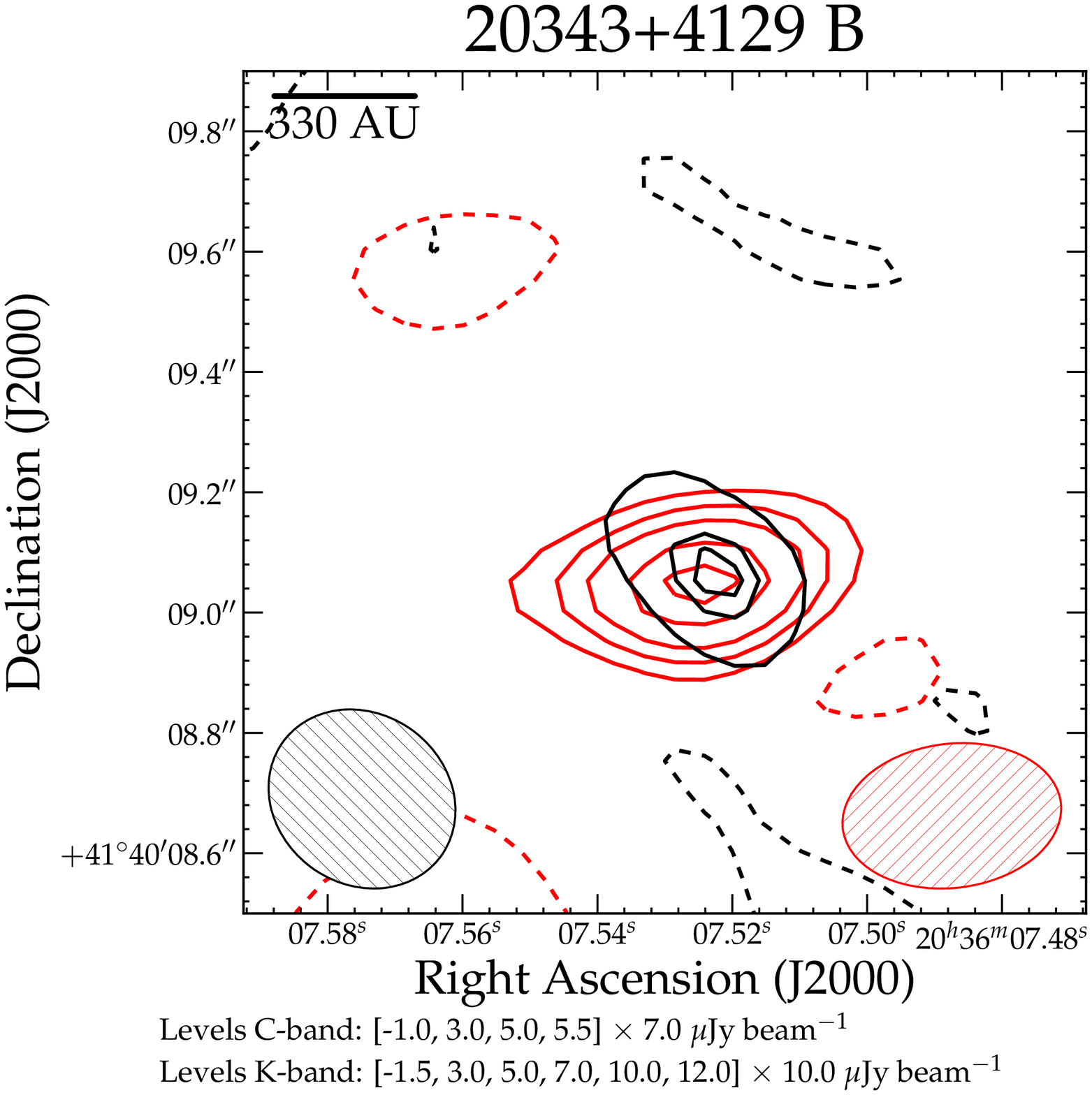} 

\end{tabular}

 \hspace*{\fill}%
\caption{\small{Continued.}}
 \label{fig:maps}
\end{figure}


\clearpage

\begin{figure}[!h]%
    \centering
    \includegraphics[width=1.0\linewidth,  trim = 3 140 10 160, clip]{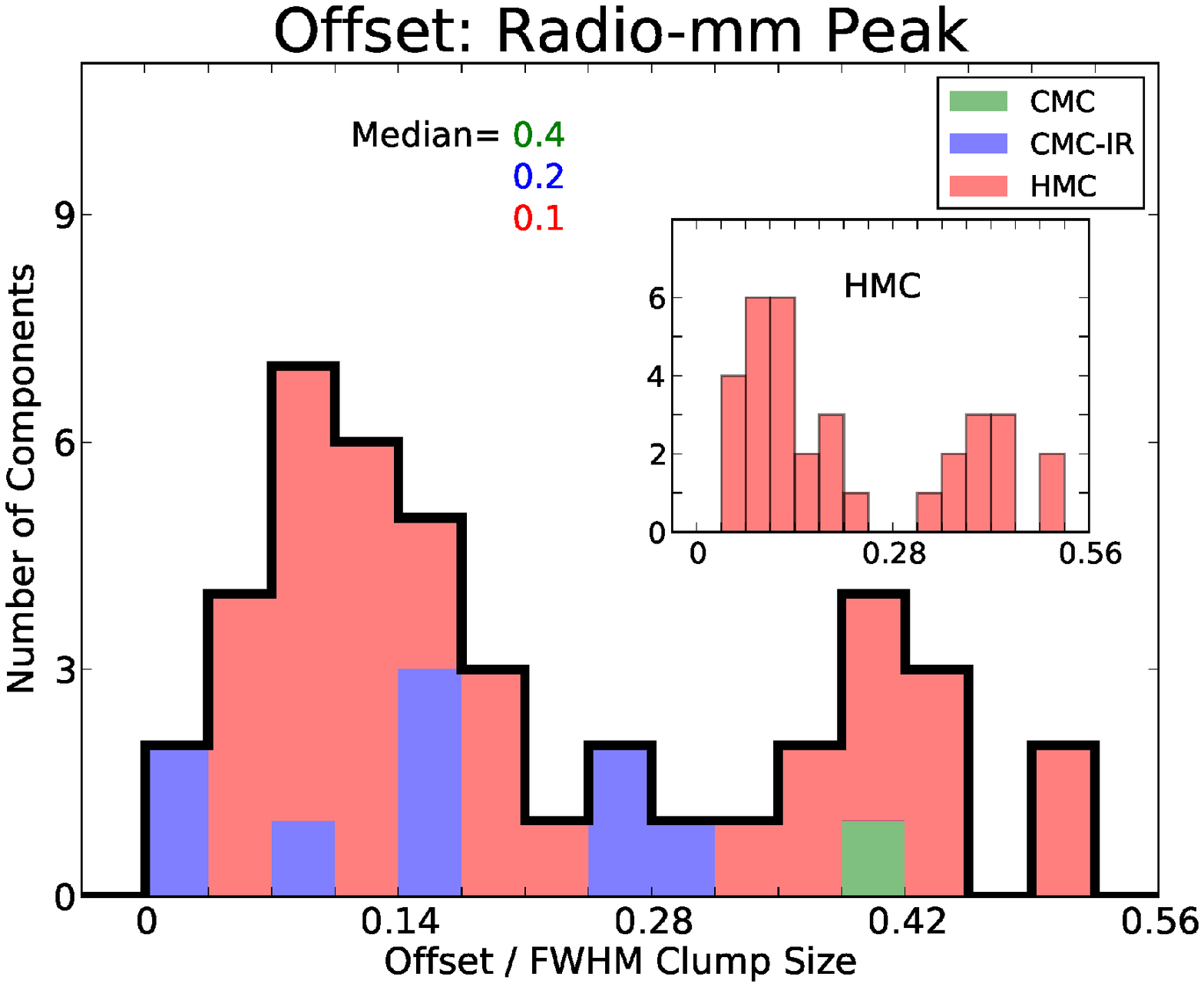}%
     \caption{\small{Position offsets between radio sources and the $1.2\,$mm peaks divided by the  FWHM size of each  clump. The inset shows the  distribution only for the HMC detections. If no $1.2\,$mm data was available, we used the median FWHM size for the specific type of clump. The median FWHM sizes for each type of 
     clump in this sample are 31$^{\prime \prime}$, 25$^{\prime \prime}$ and 18$^{\prime \prime}$ for CMCs, CMC--IRs and HMCs, respectively. The average cm positional accuracy is better than 0.1$^{\prime \prime}$, and the typical positional accuracy for 
     the $1.2\,$mm data is a few arcseconds. Clumps with multiple components (within a radius of 3$^{\prime \prime}$) have been treated as a single radio source in this histogram. Using  a Kolmogorov-Smirnov test to analyze the spatial offset distribution of the HMCs 
     and CMC--IRs we estimated a p-value$=$ 0.34. Thus, there is no strong evidence that these two distributions are different.}}%
    \label{offset_mm_radio}%
\end{figure}


\begin{figure}[htbp]
\centering
\begin{tabular}{cc}
\hspace*{\fill}%
 \includegraphics[width=0.5\textwidth, clip, angle = 0]{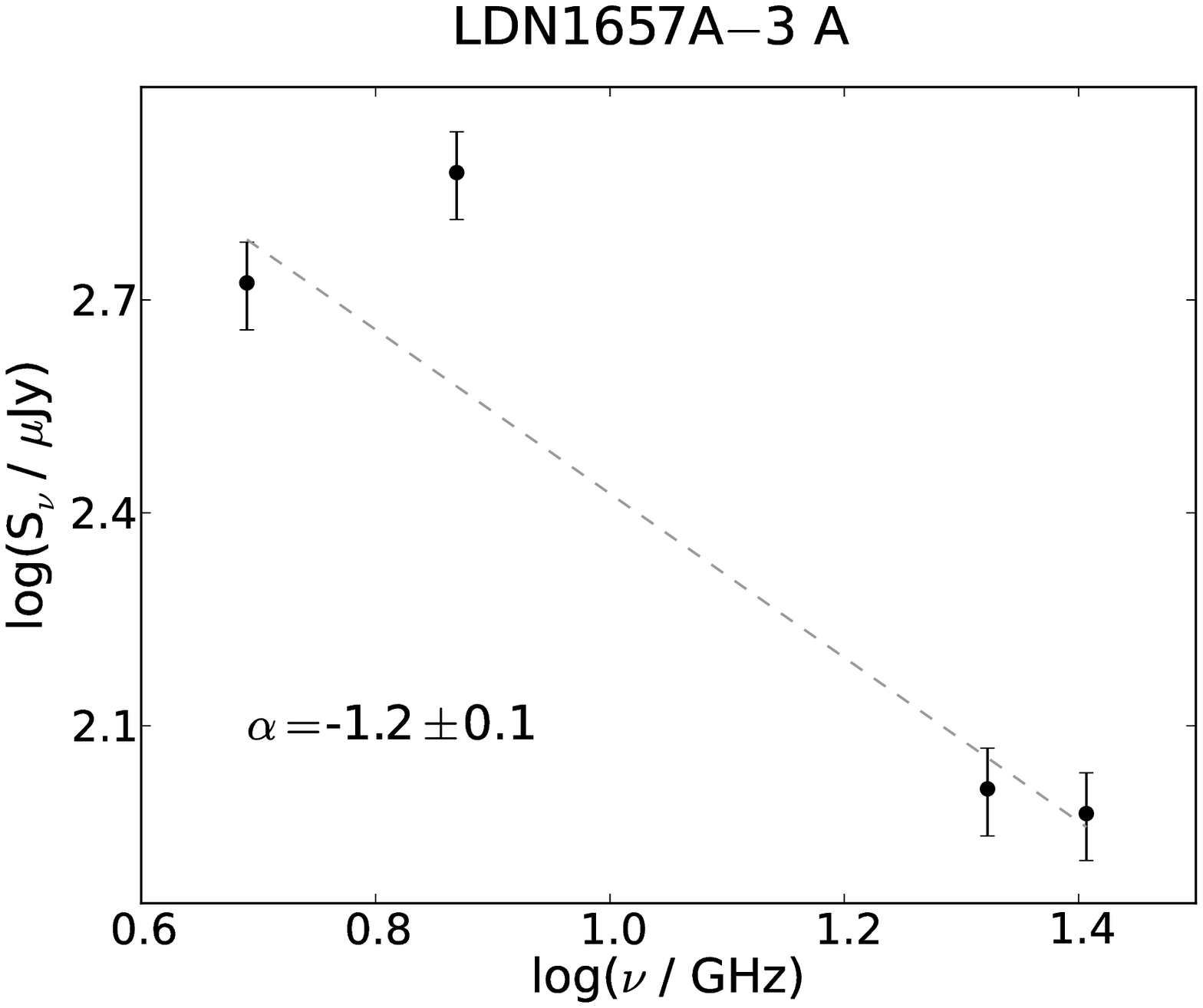} \vspace{0.5cm}&
 \includegraphics[width=0.5\textwidth, clip, angle = 0]{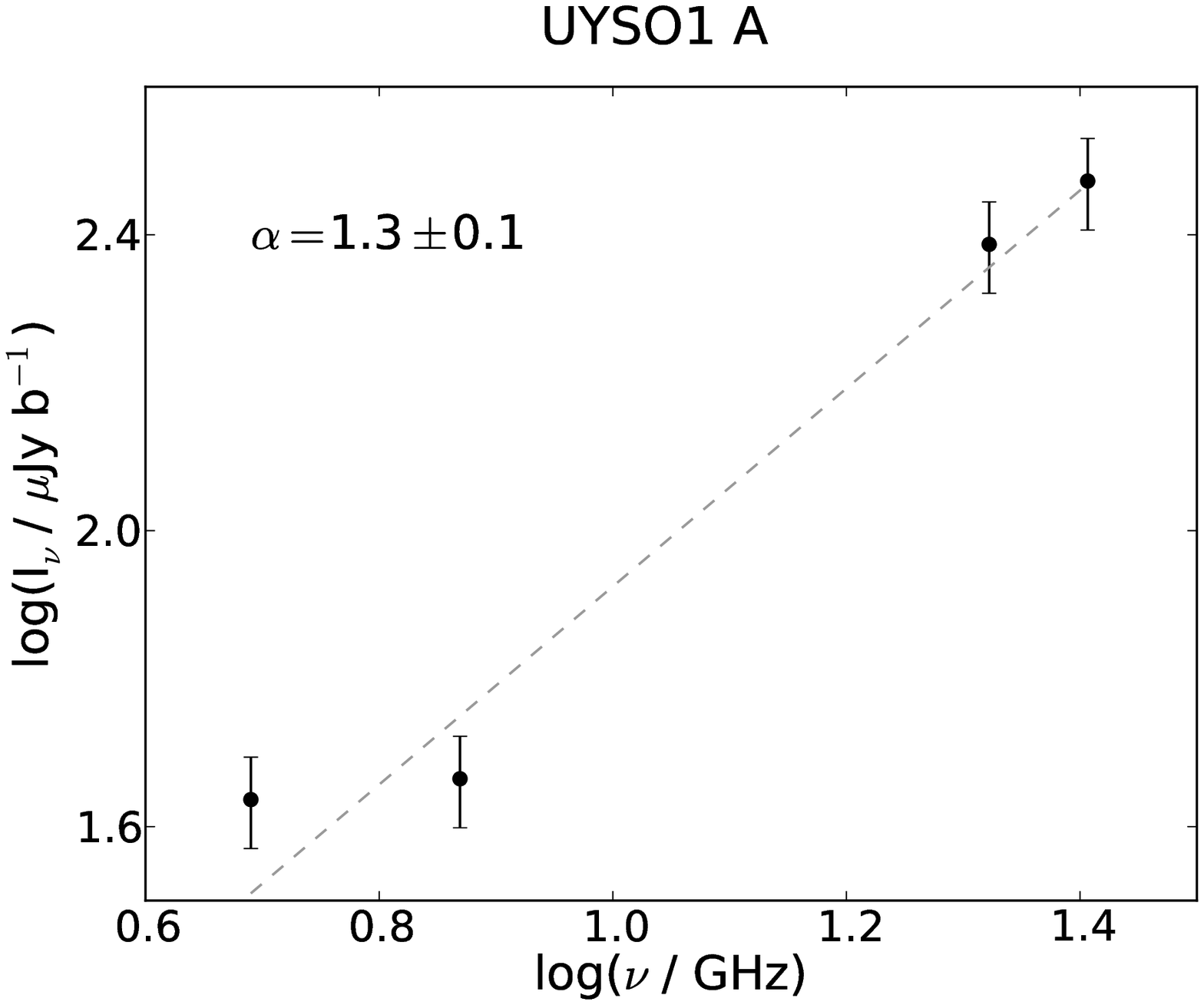} \\
  \includegraphics[width=0.5\textwidth, clip, angle = 0]{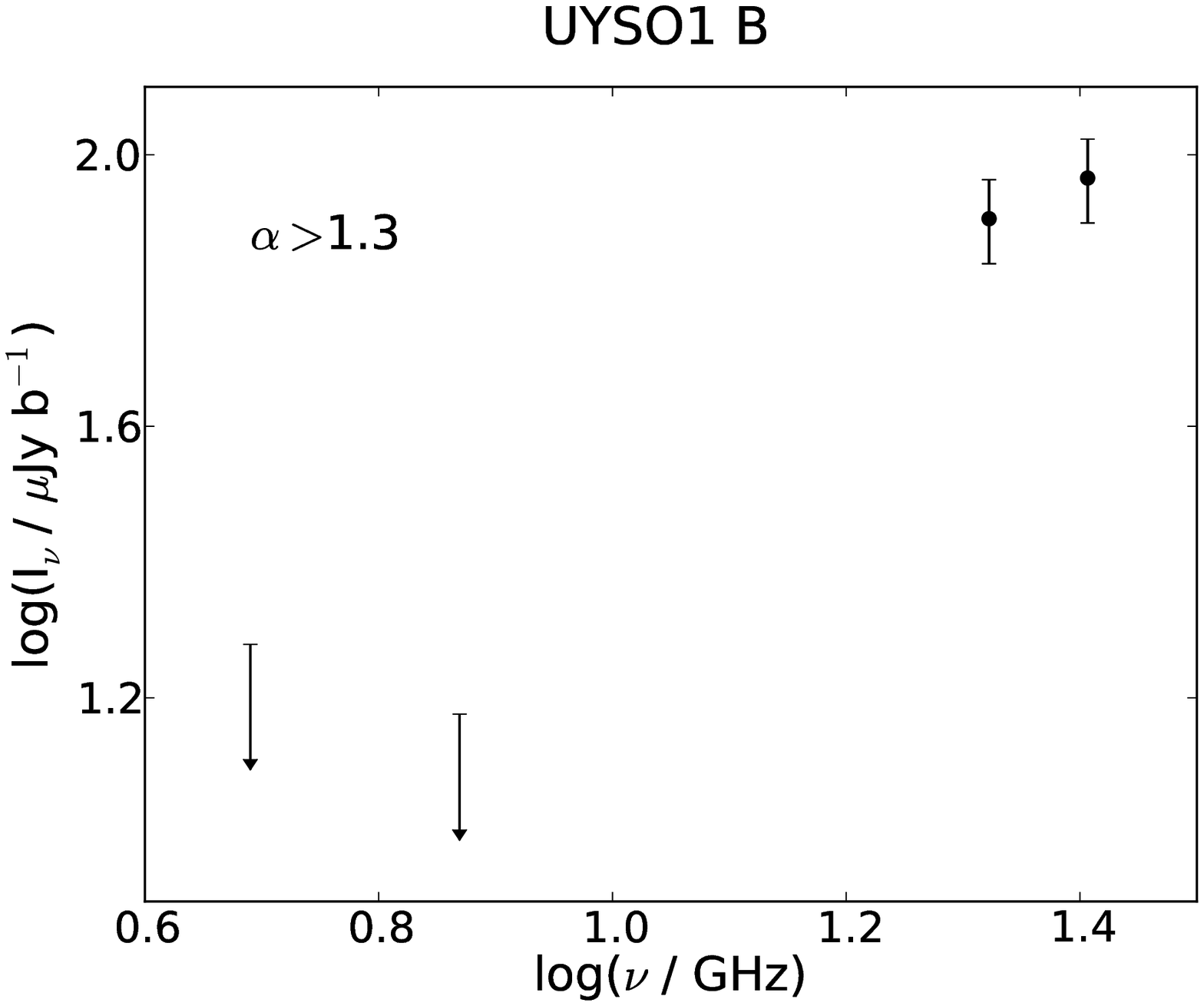} \vspace{0.5cm}&
 \includegraphics[width=0.5\textwidth, clip, angle = 0]{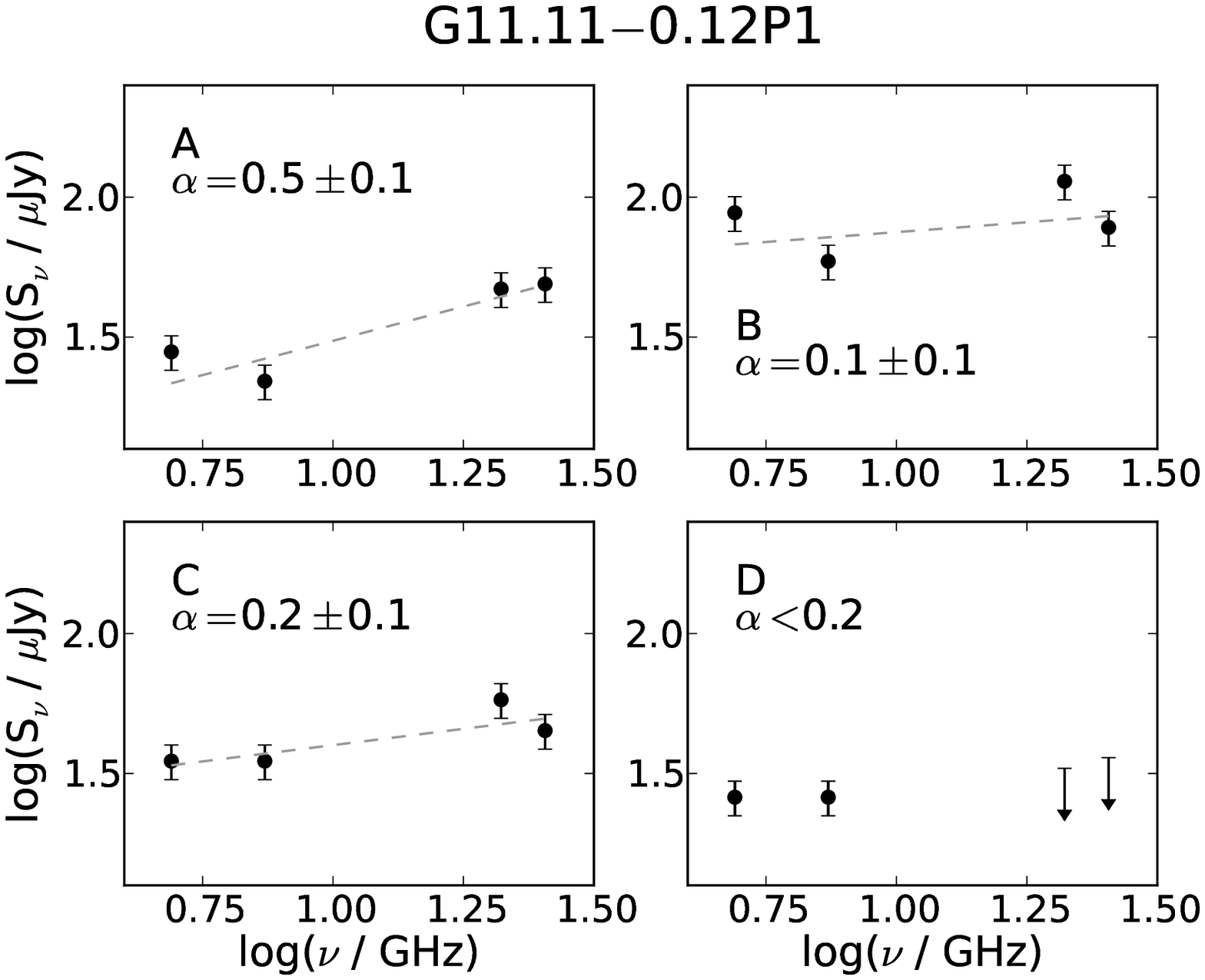} \\
  \includegraphics[width=0.5\textwidth, clip, angle = 0]{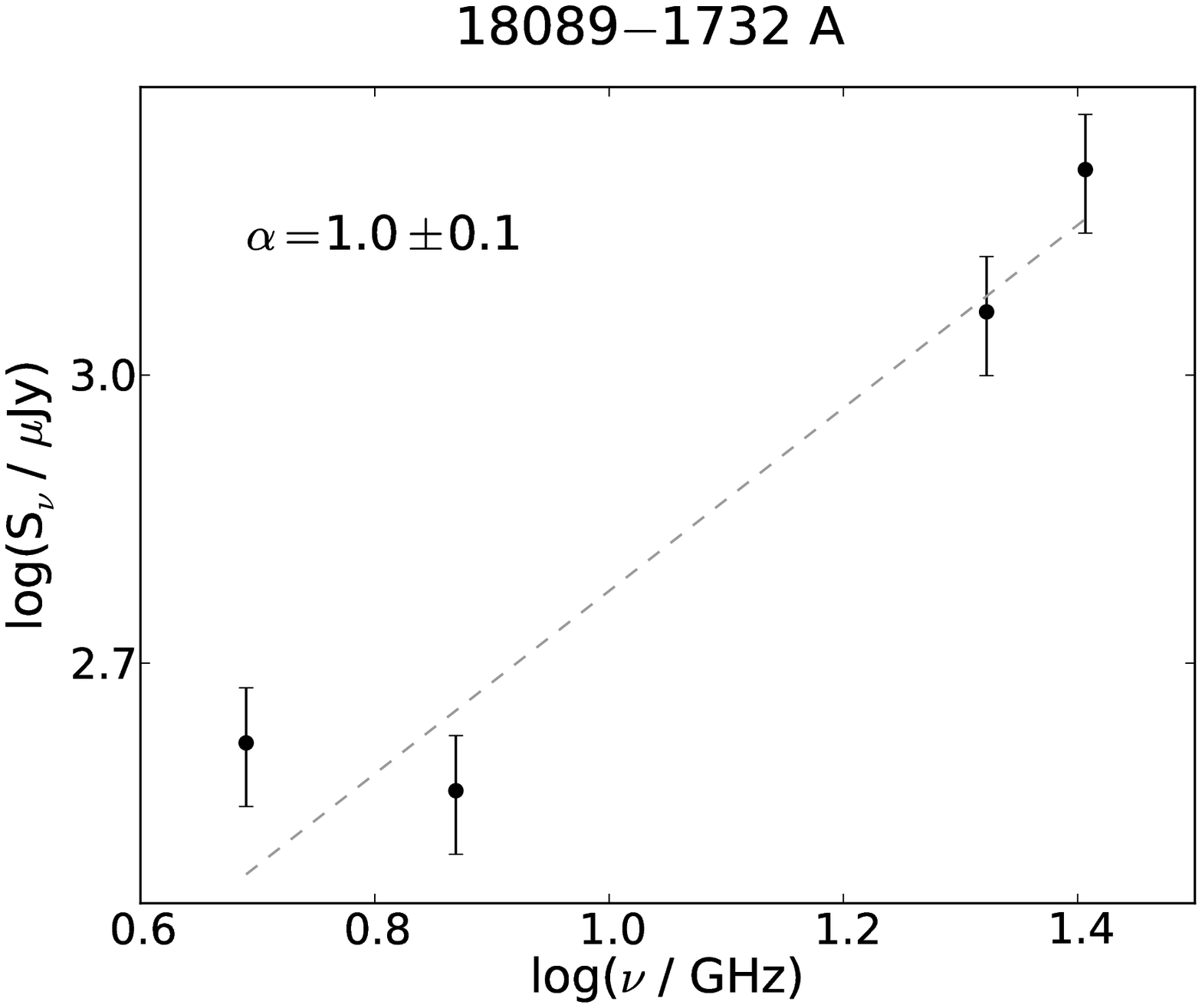} &
 \includegraphics[width=0.5\textwidth, clip, angle = 0]{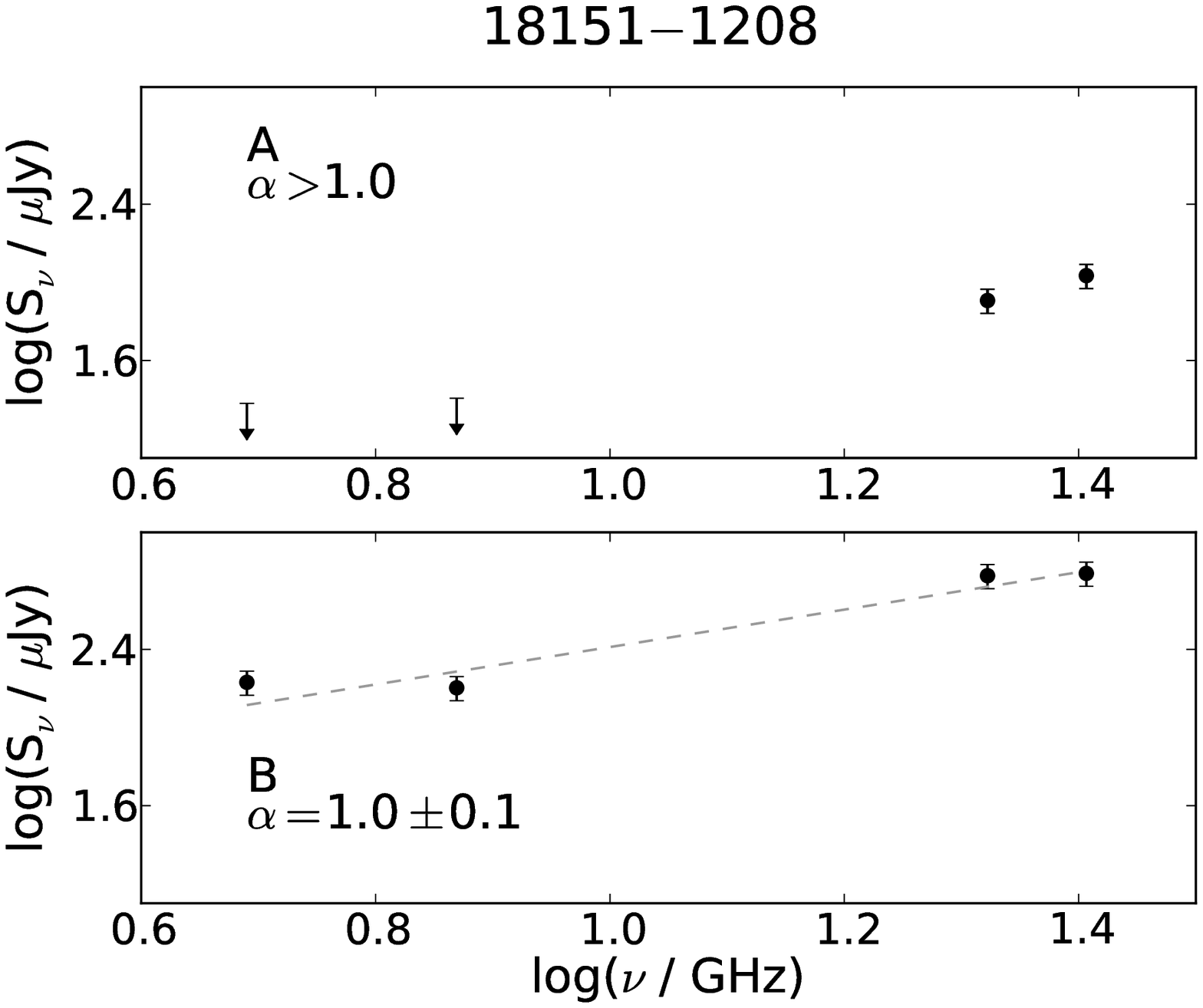} 
\end{tabular}

 \hspace*{\fill}%
     \caption{\small{Flux density as a function of frequency for each detected component. Error bars are an assumed uncertainty of 10$\%$ from the flux densities added in quadrature with an assumed 10$\%$ error in calibration. The dashed lines are the best fit to the data from a power law of the form $S_{\nu}$ $\propto$ $\nu^{\alpha}$.}}
 \label{fig:spix}
\end{figure}

\begin{figure}[htbp]
\centering
\begin{tabular}{cc}
  \ContinuedFloat
 \includegraphics[width=0.5\textwidth, clip, angle = 0]{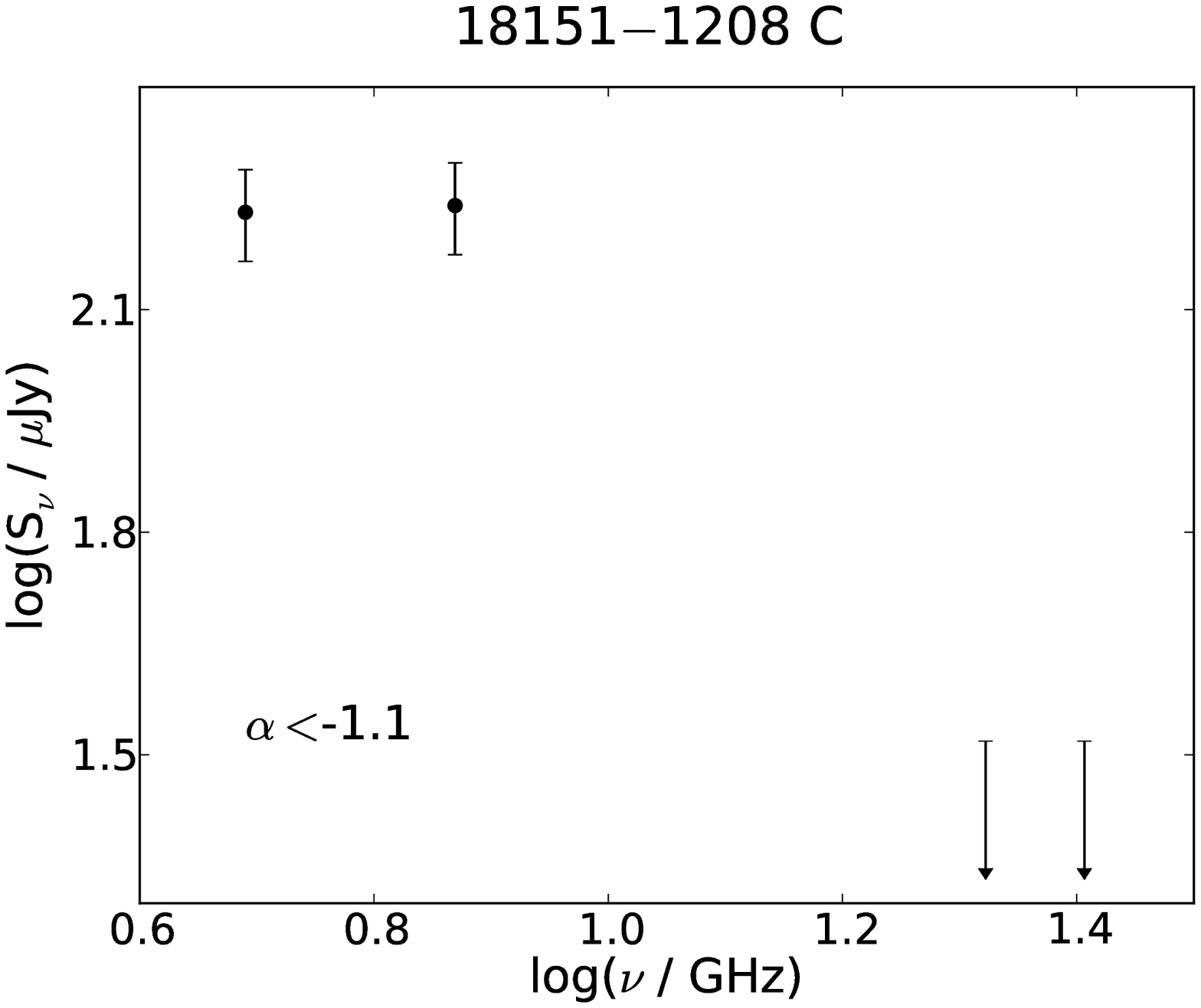} \\
 \includegraphics[width=0.7\textwidth, clip, angle = 0]{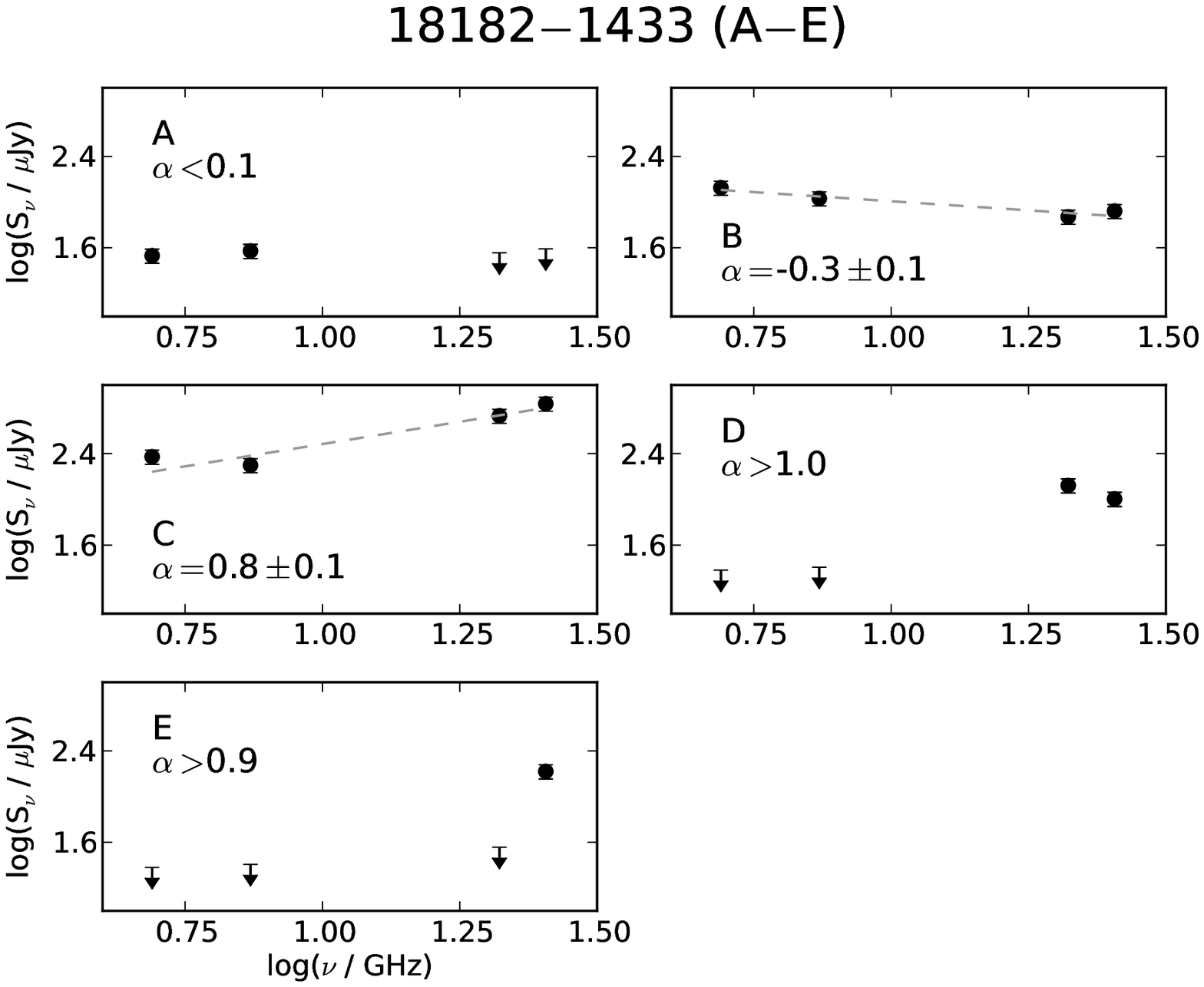} \\
  \includegraphics[width=0.5\textwidth, clip, angle = 0]{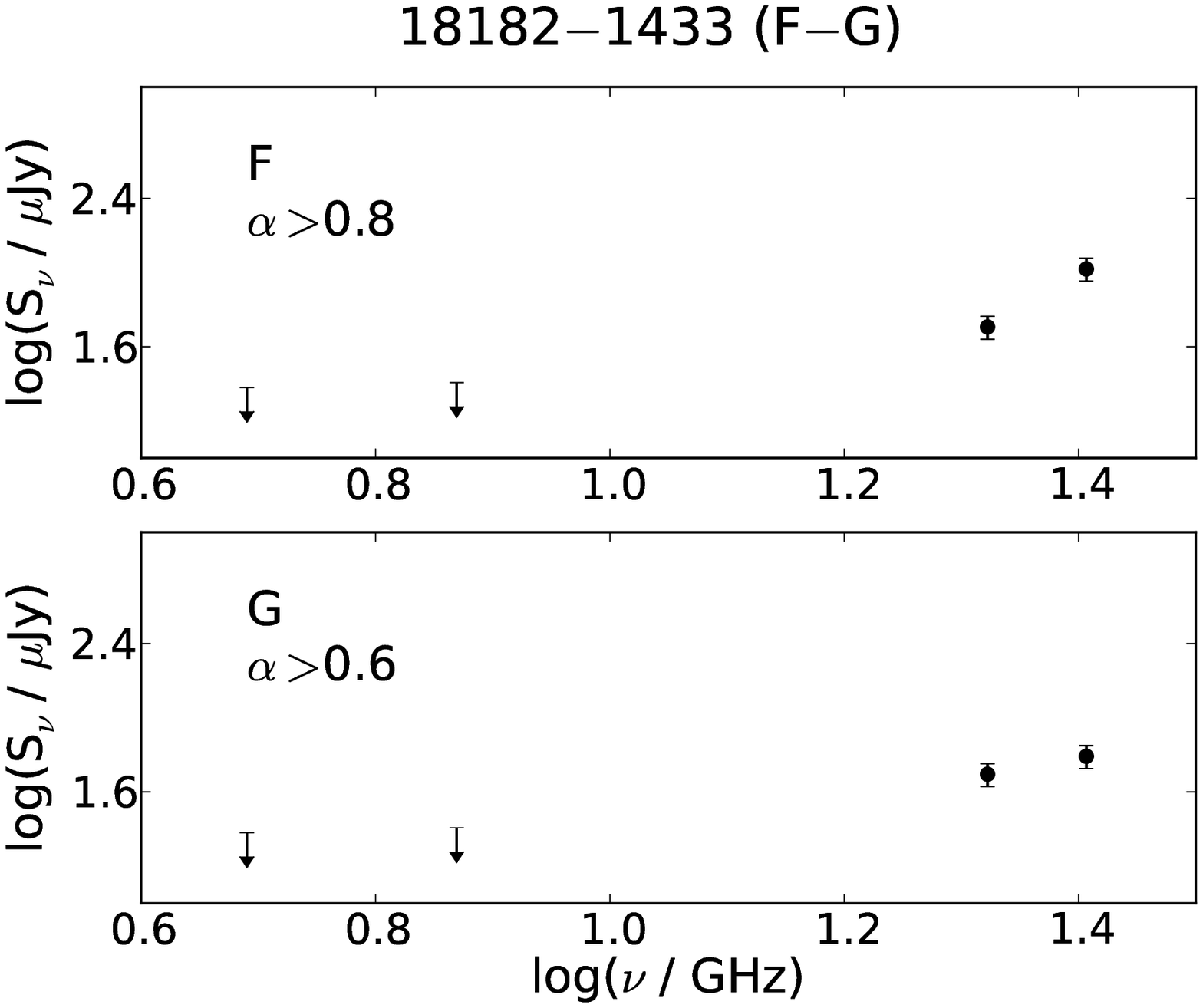}\\

\end{tabular}

 \hspace*{\fill}%
\caption{\small{Continued.}}
 \label{fig:spix}
\end{figure}

\begin{figure}[htbp]
\centering
\begin{tabular}{cc}
  \ContinuedFloat
 \includegraphics[width=0.5\textwidth, clip, angle = 0]{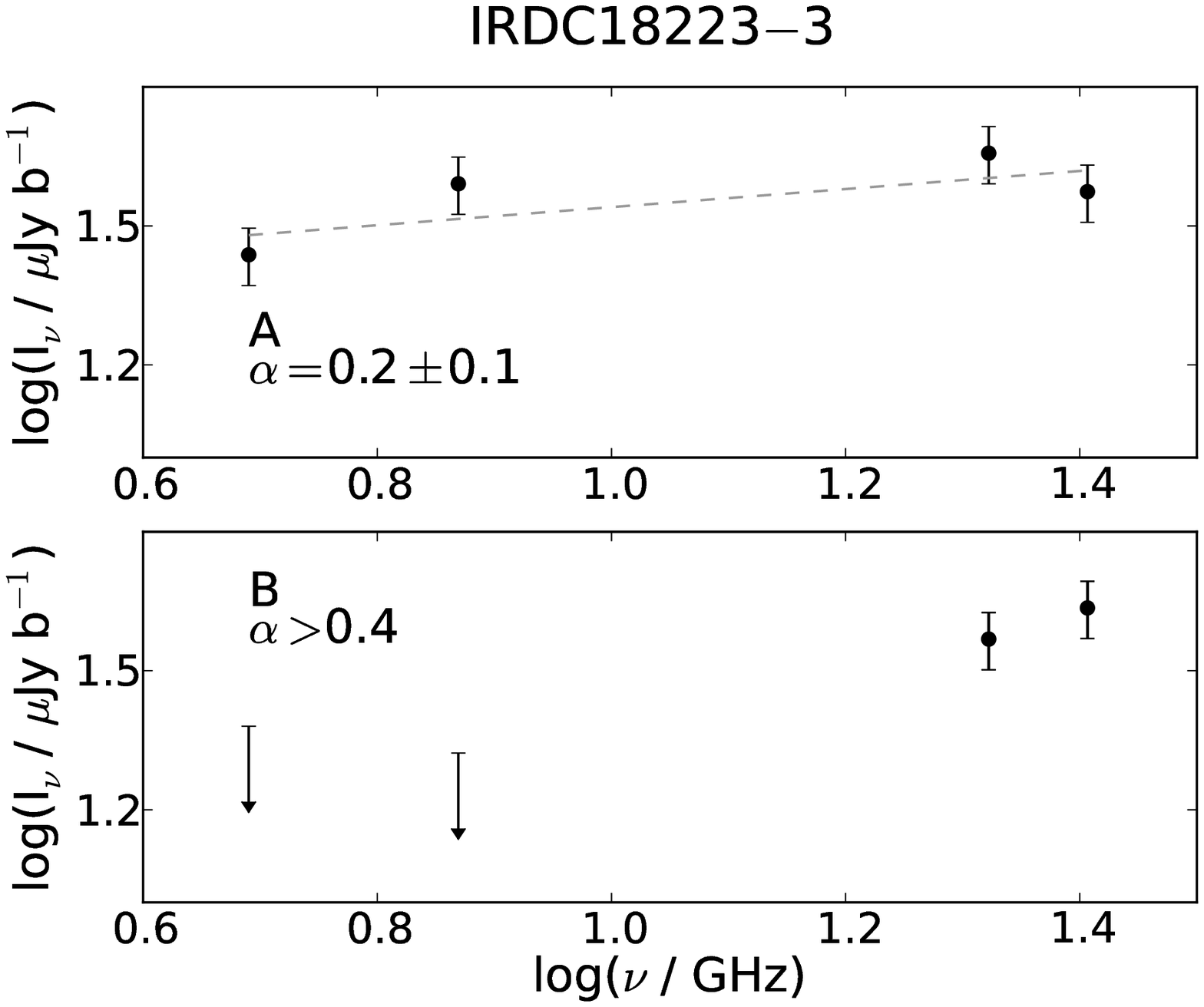} \\
 \includegraphics[width=0.7\textwidth, clip, angle = 0]{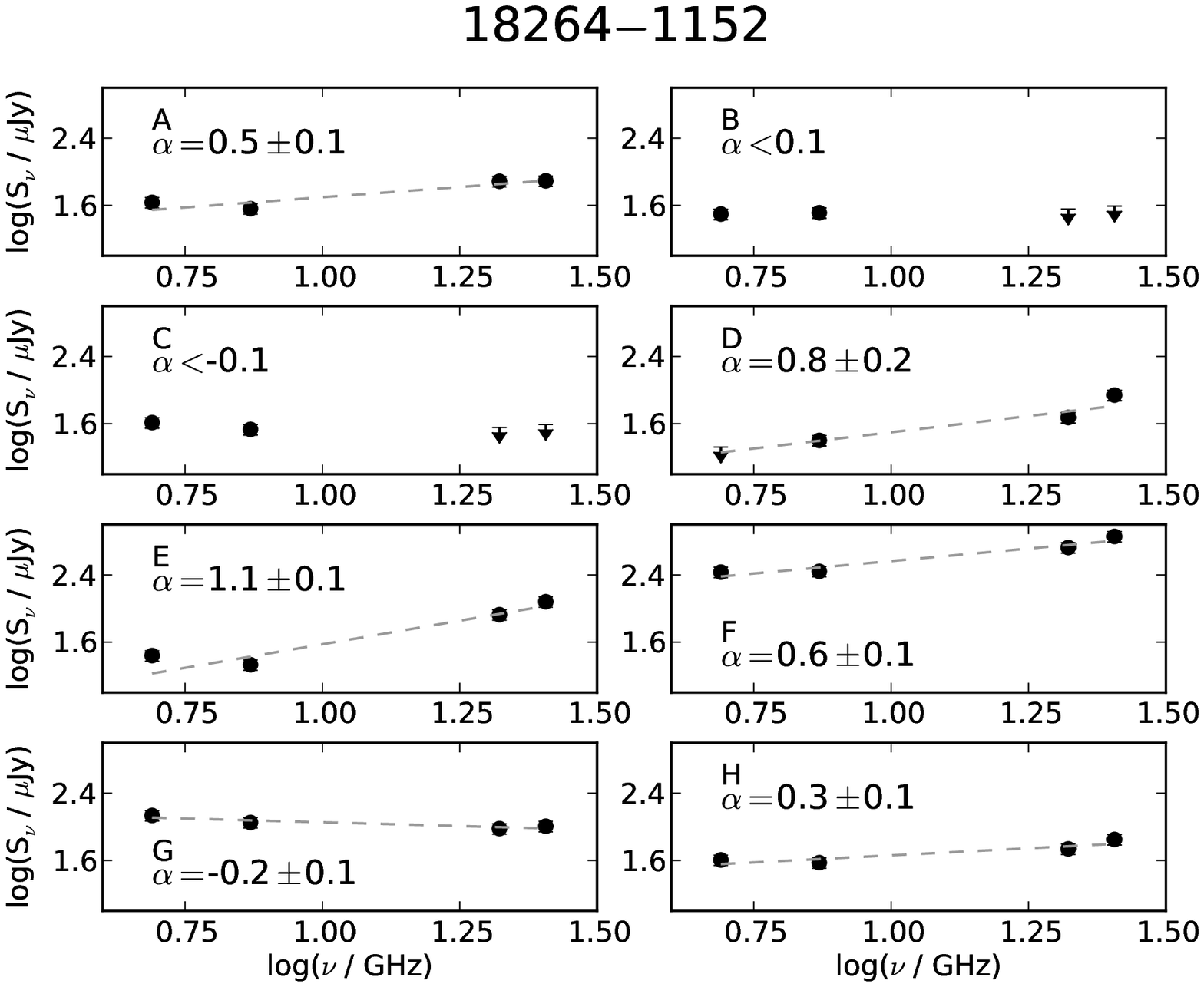} \\
  \includegraphics[width=0.5\textwidth, clip, angle = 0]{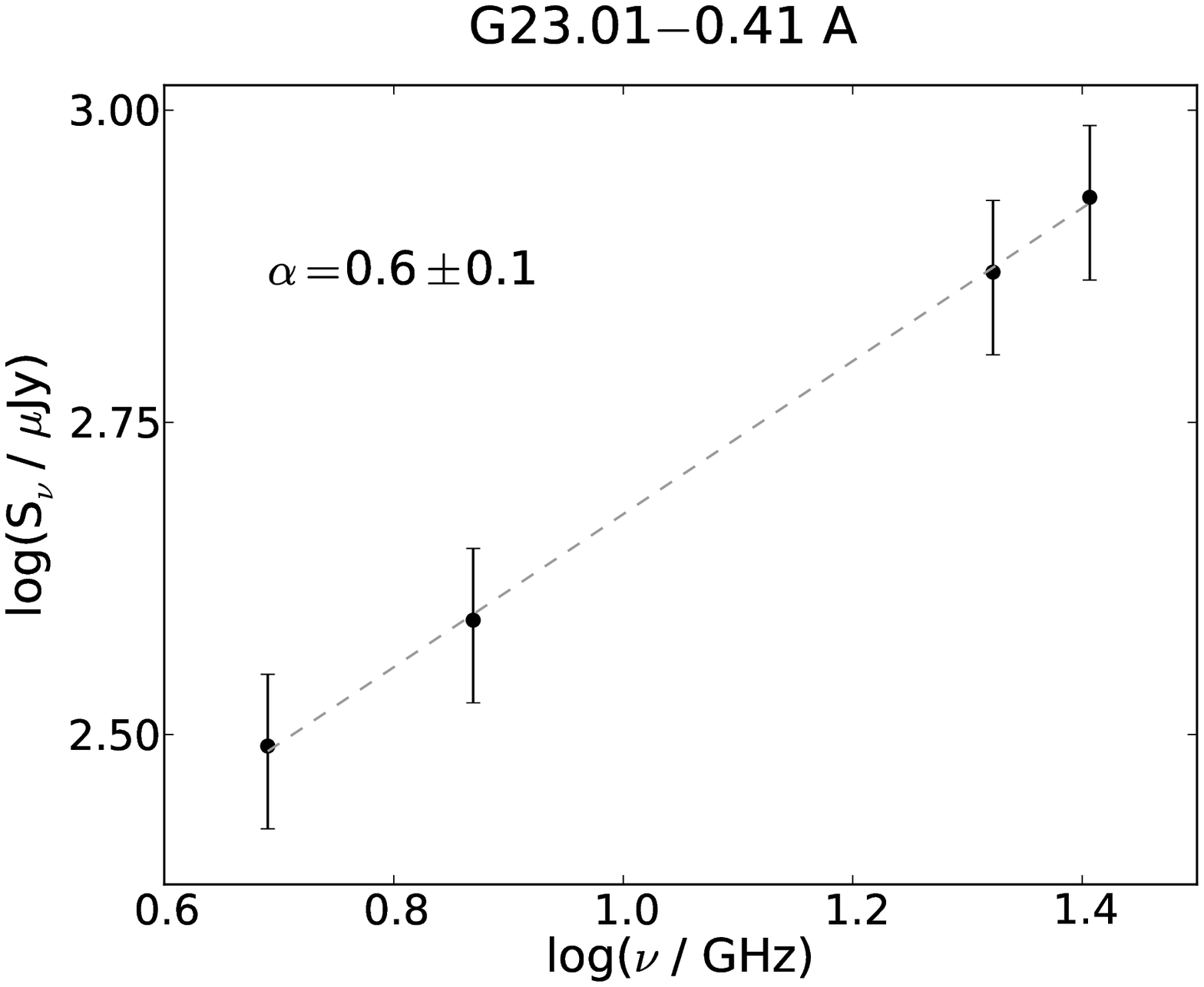} \\

\end{tabular}

 \hspace*{\fill}%
\caption{\small{Continued.}}
 \label{fig:spix}
\end{figure}

\begin{figure}[htbp]
\centering
\begin{tabular}{cc}
  \ContinuedFloat
\hspace*{\fill}%
 \includegraphics[width=0.5\textwidth, clip, angle = 0]{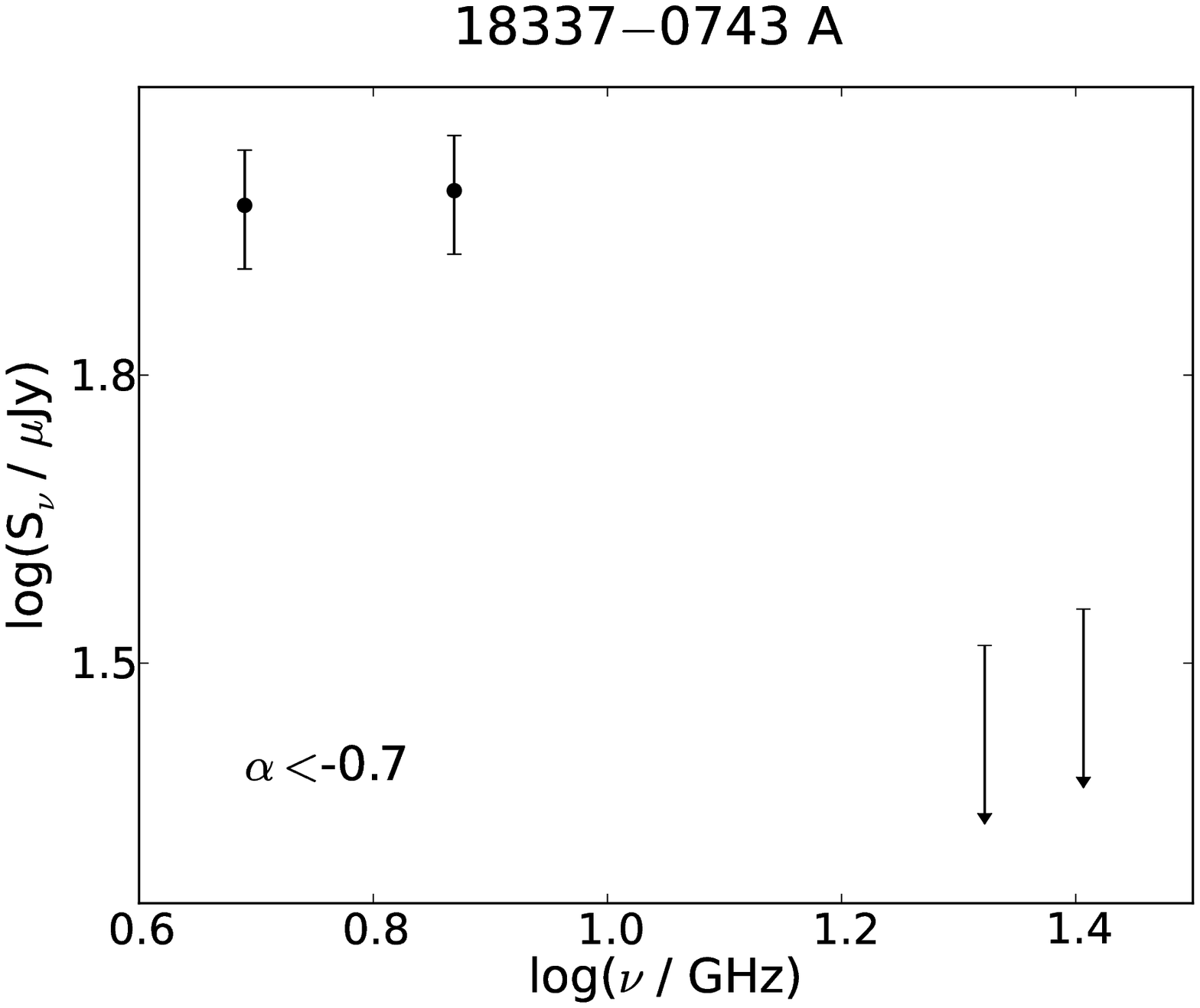} \vspace{0.5cm}&
 \includegraphics[width=0.5\textwidth, clip, angle = 0]{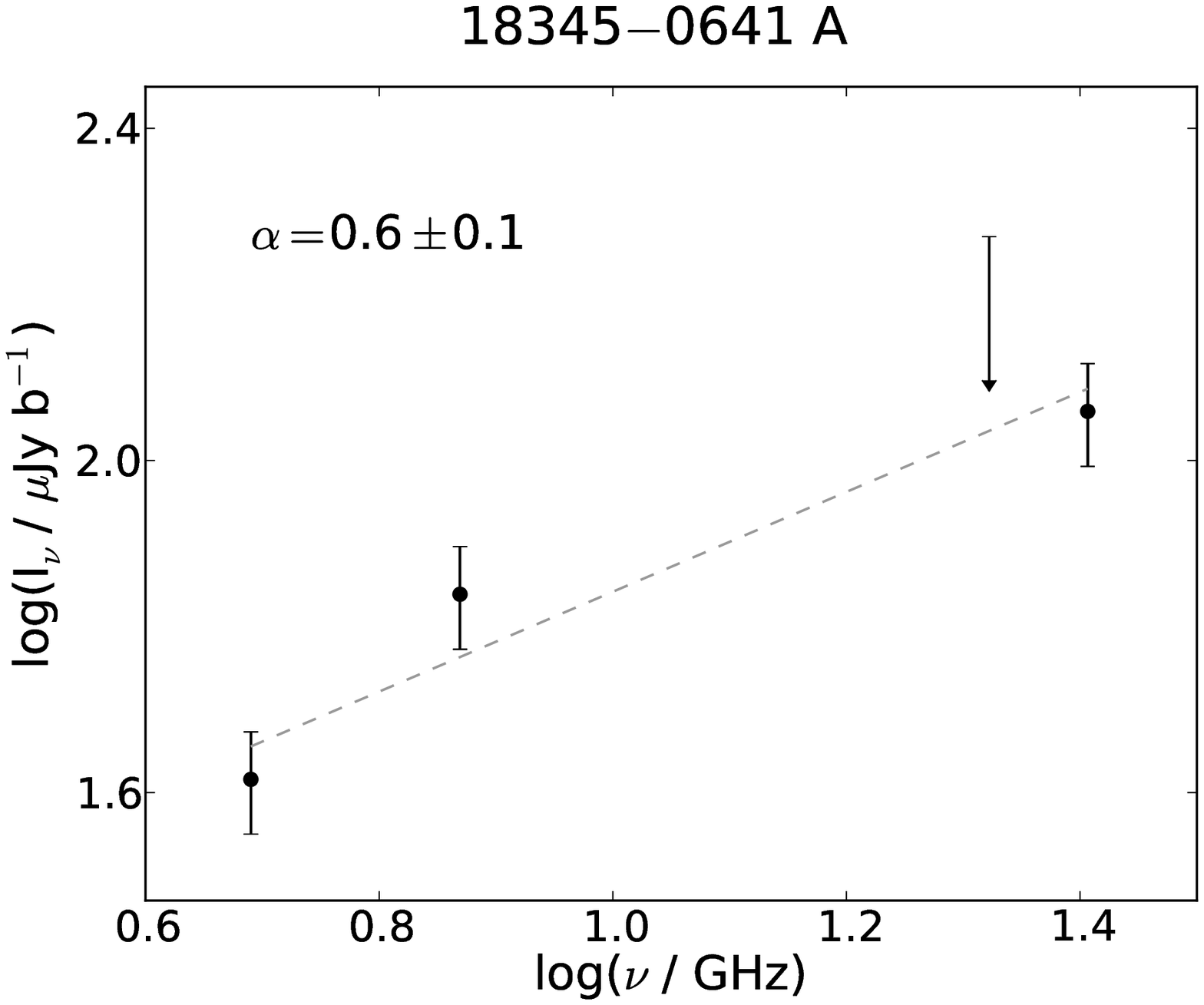} \\
  \includegraphics[width=0.5\textwidth, clip, angle = 0]{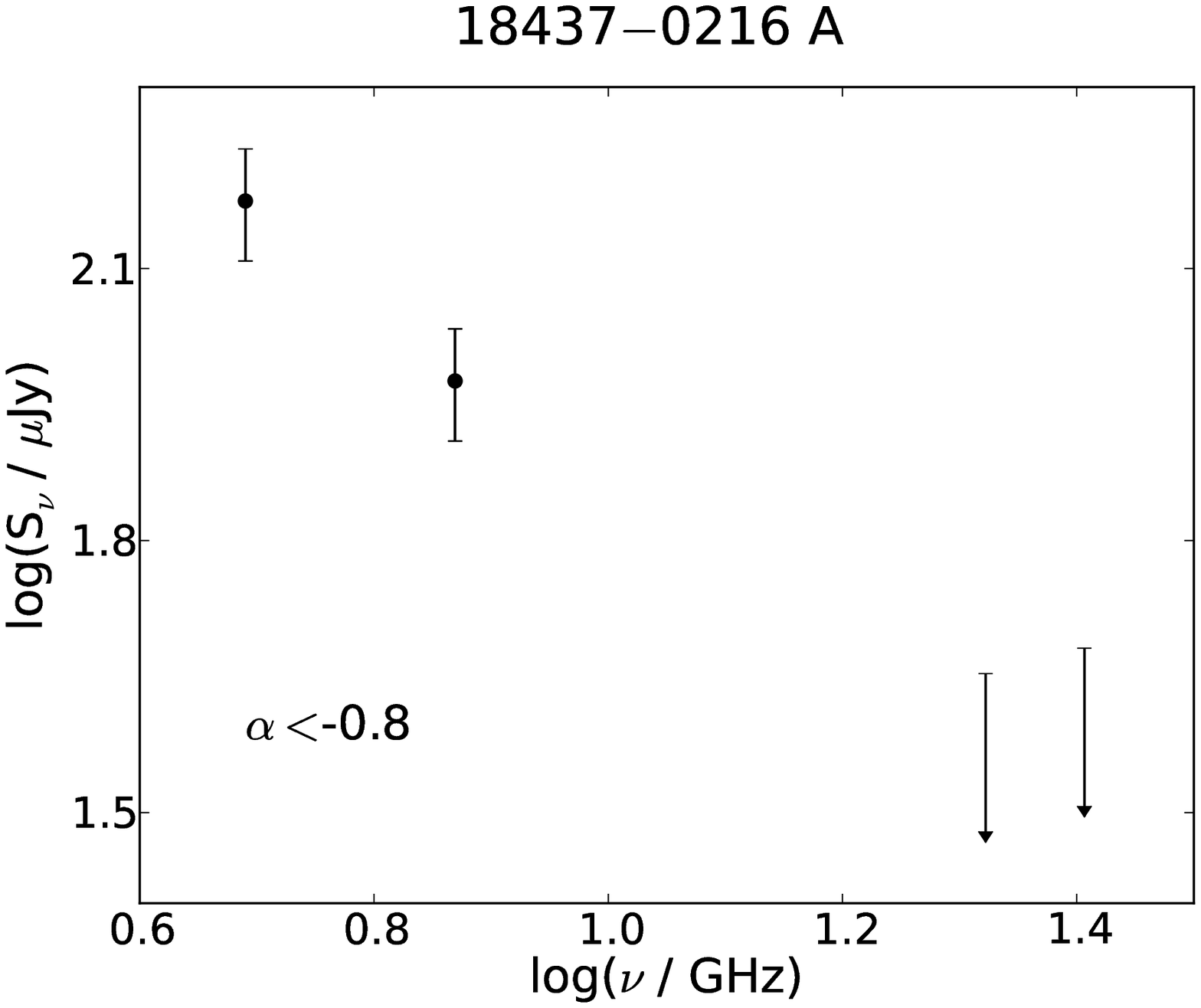} \vspace{0.5cm}&
 \includegraphics[width=0.5\textwidth, clip, angle = 0]{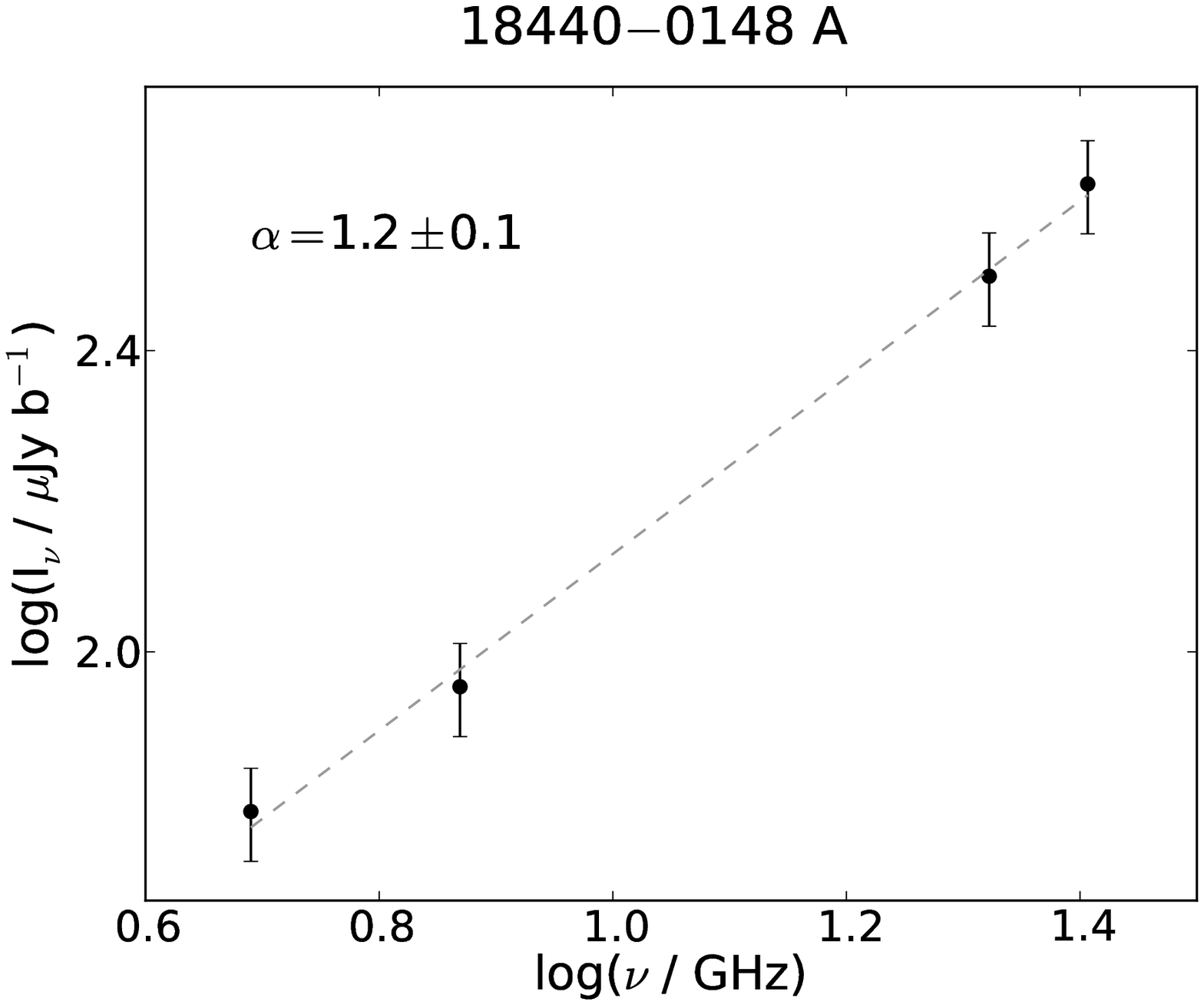} \\
  \includegraphics[width=0.5\textwidth, clip, angle = 0]{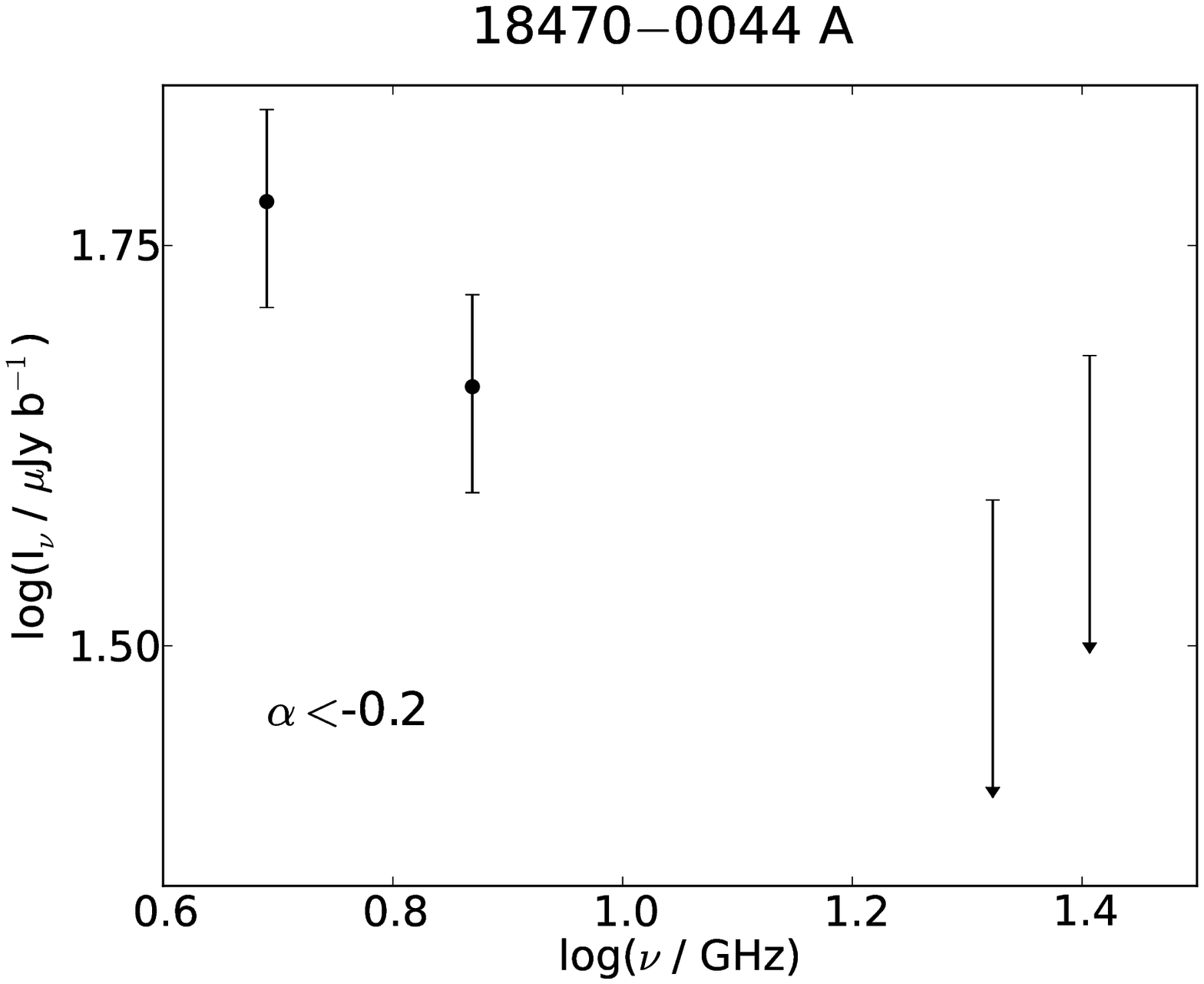} &
 \includegraphics[width=0.5\textwidth, clip, angle = 0]{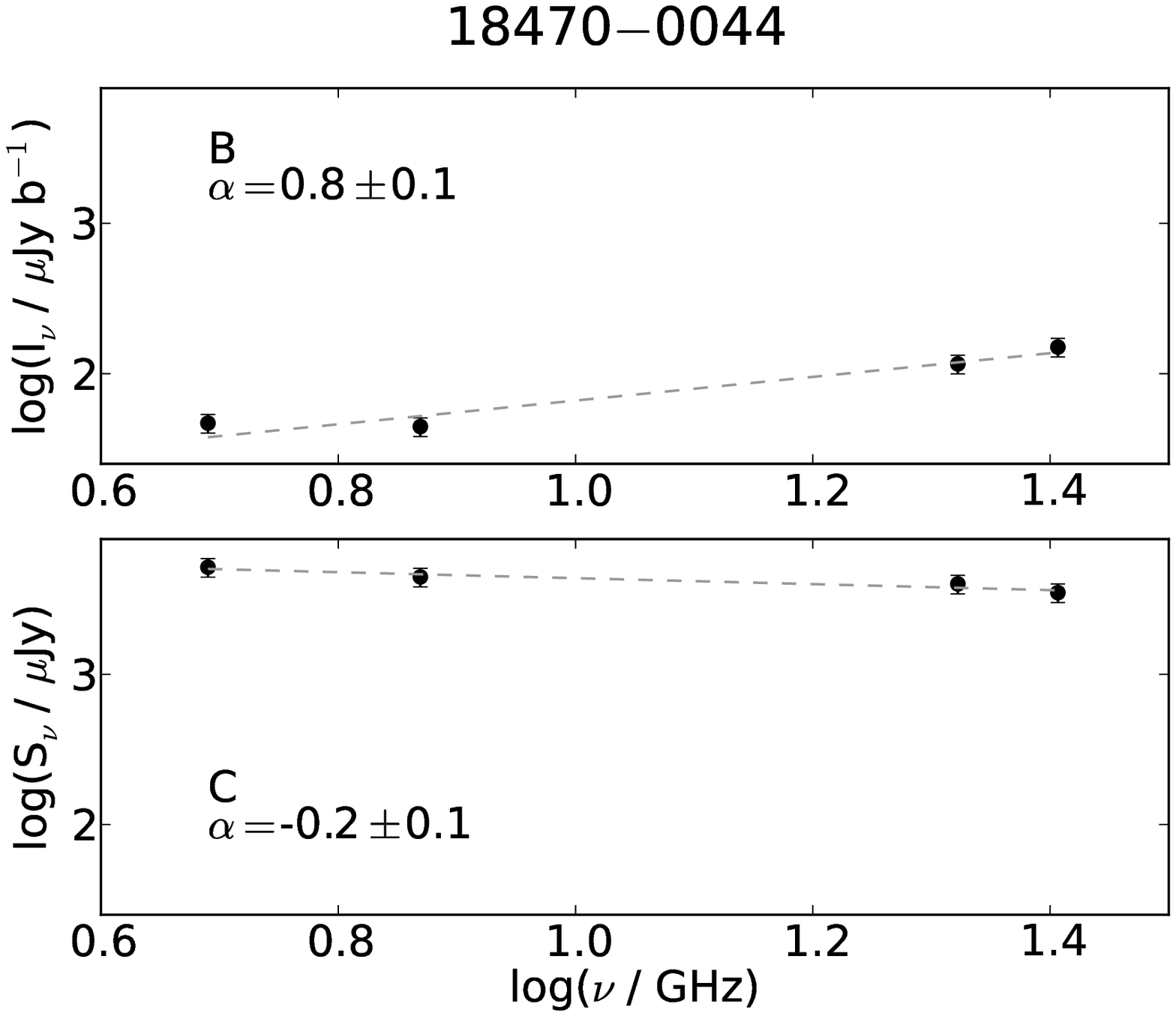} 
\end{tabular}

 \hspace*{\fill}%
\caption{\small{Continued.}}
 \label{fig:spix}
\end{figure}

\begin{figure}[htbp]
\centering
\begin{tabular}{cc}
  \ContinuedFloat
\hspace*{\fill}%
 \includegraphics[width=0.5\textwidth, clip, angle = 0]{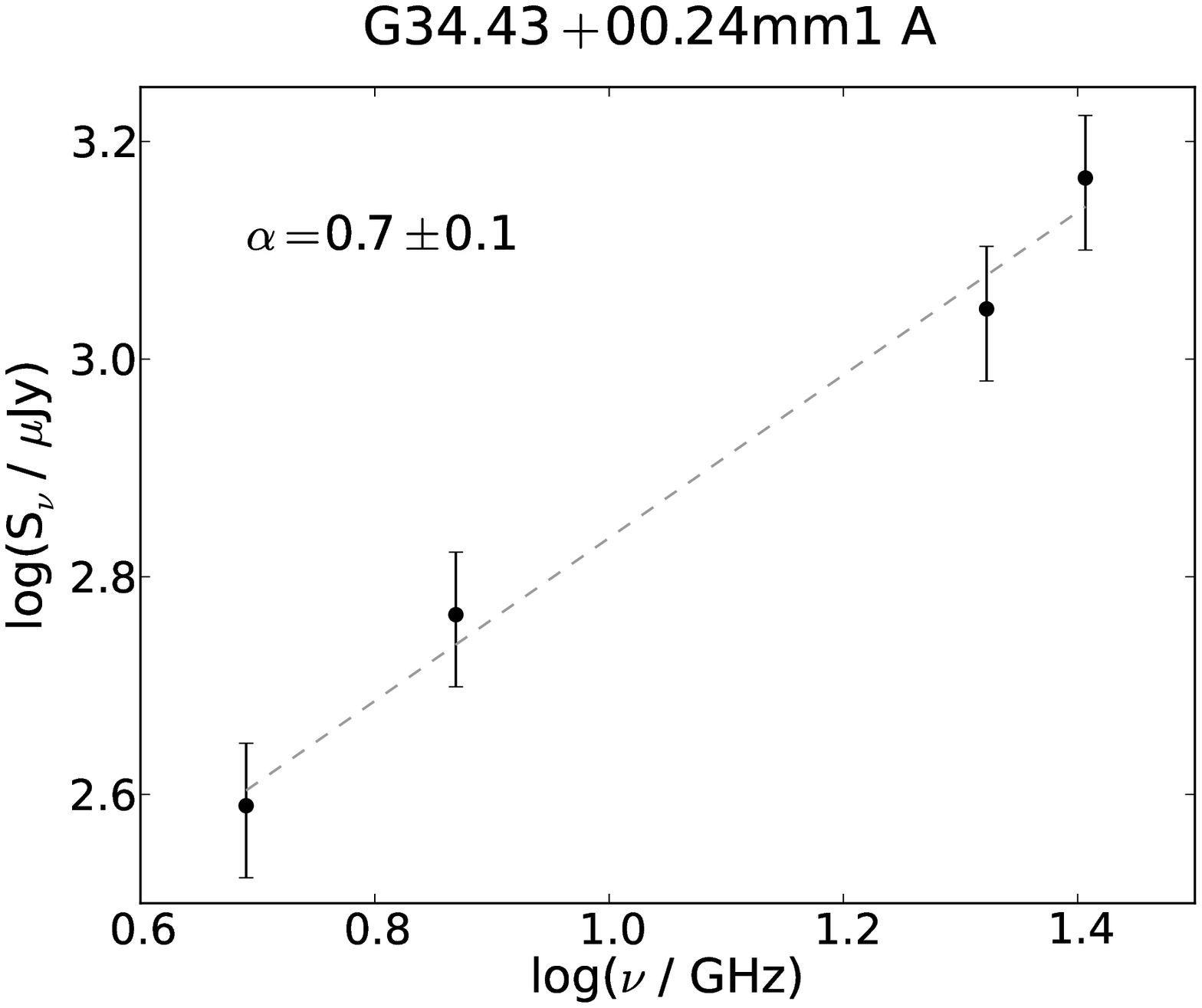} \vspace{0.5cm}&
 \includegraphics[width=0.5\textwidth, clip, angle = 0]{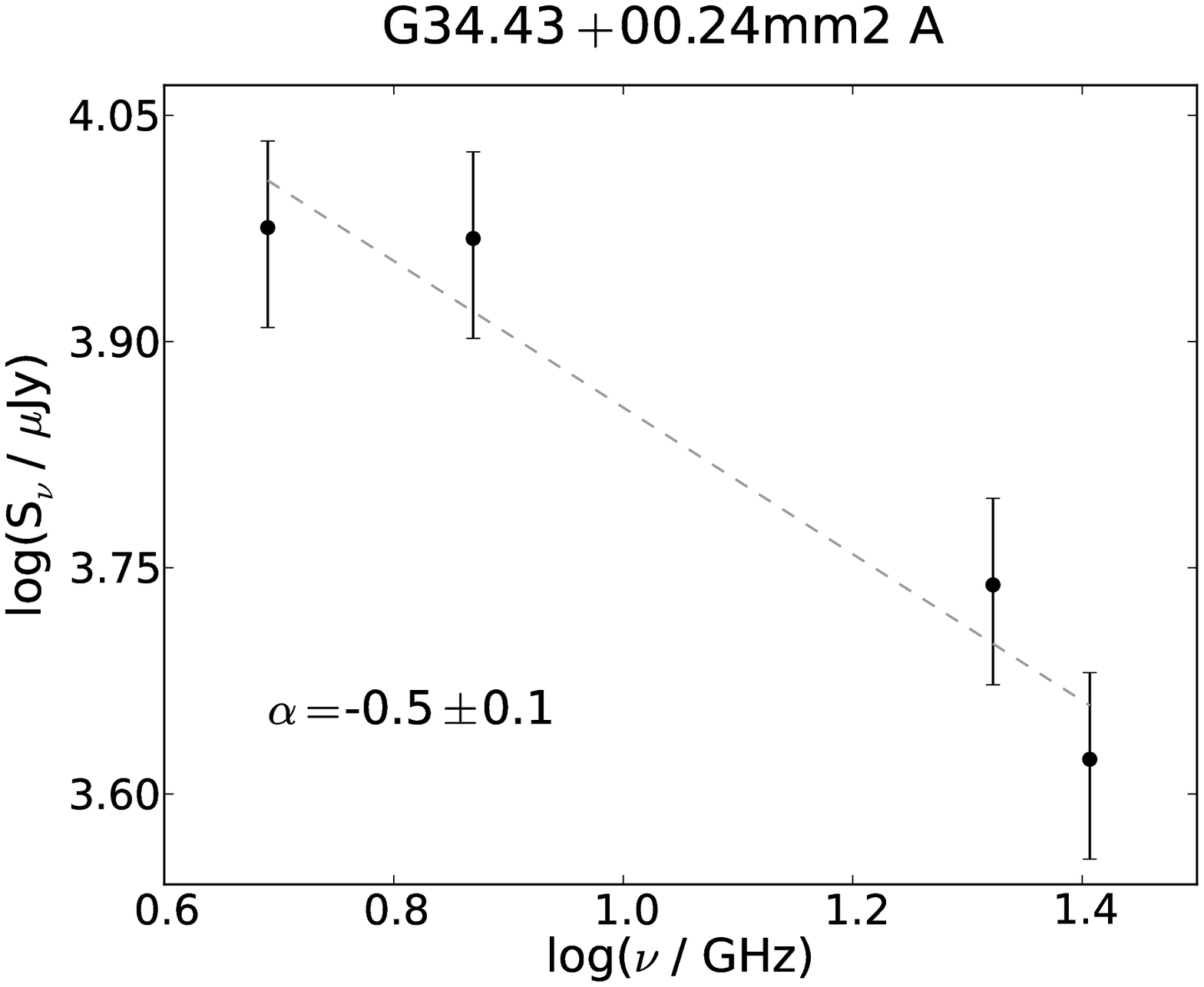} \\
  \includegraphics[width=0.5\textwidth, clip, angle = 0]{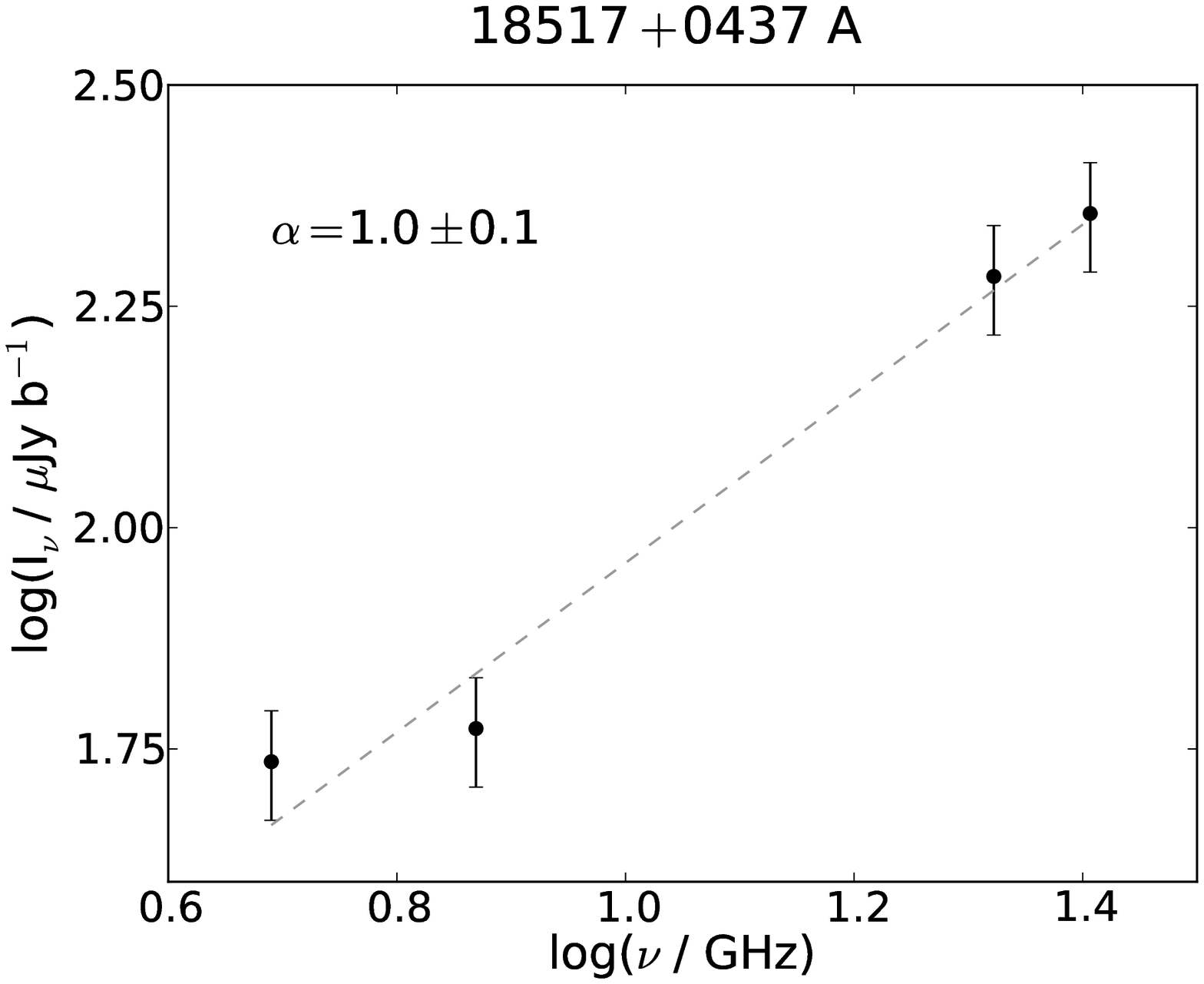} \vspace{0.5cm}&
 \includegraphics[width=0.5\textwidth, clip, angle = 0]{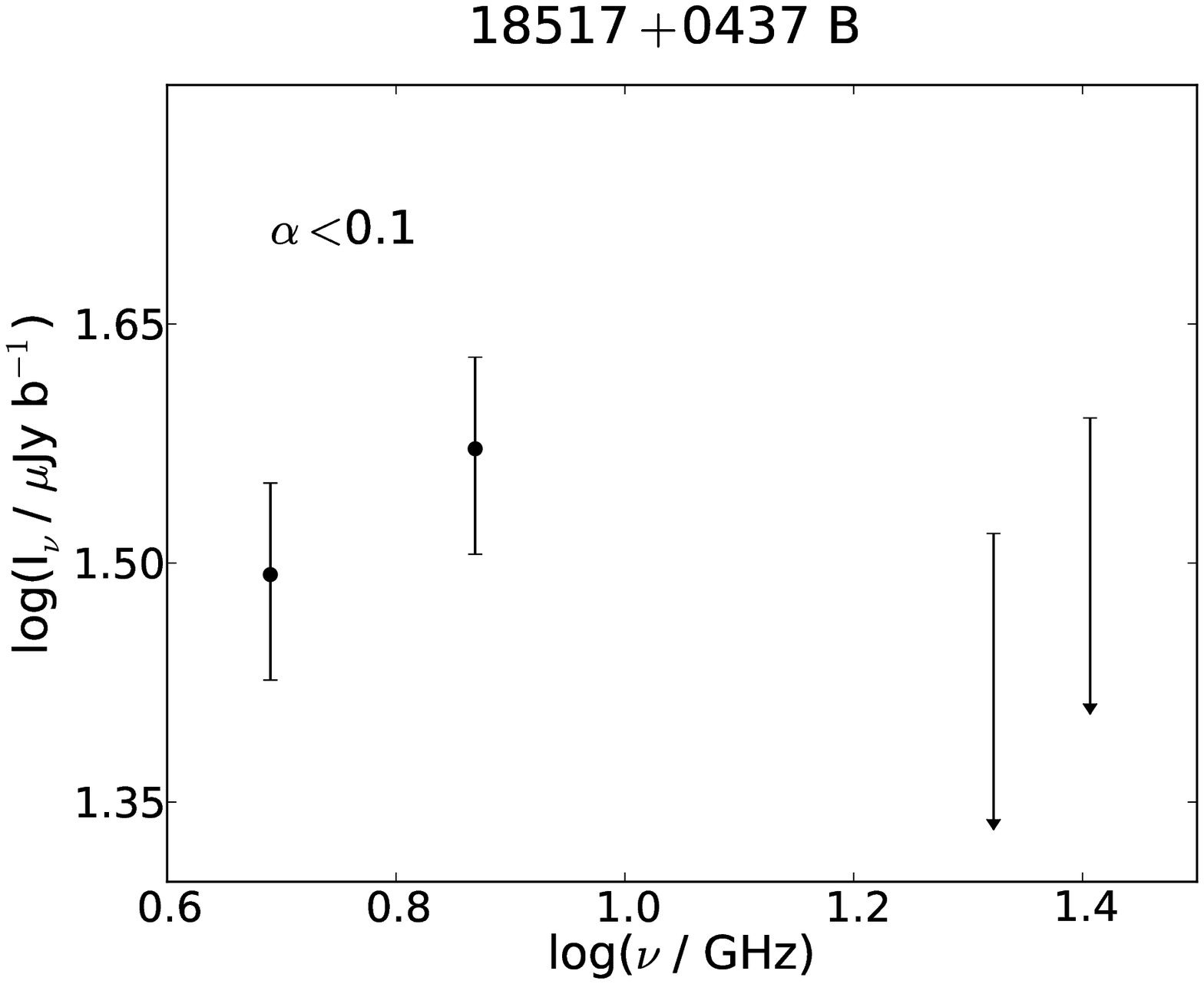} \\
  \includegraphics[width=0.5\textwidth, clip, angle = 0]{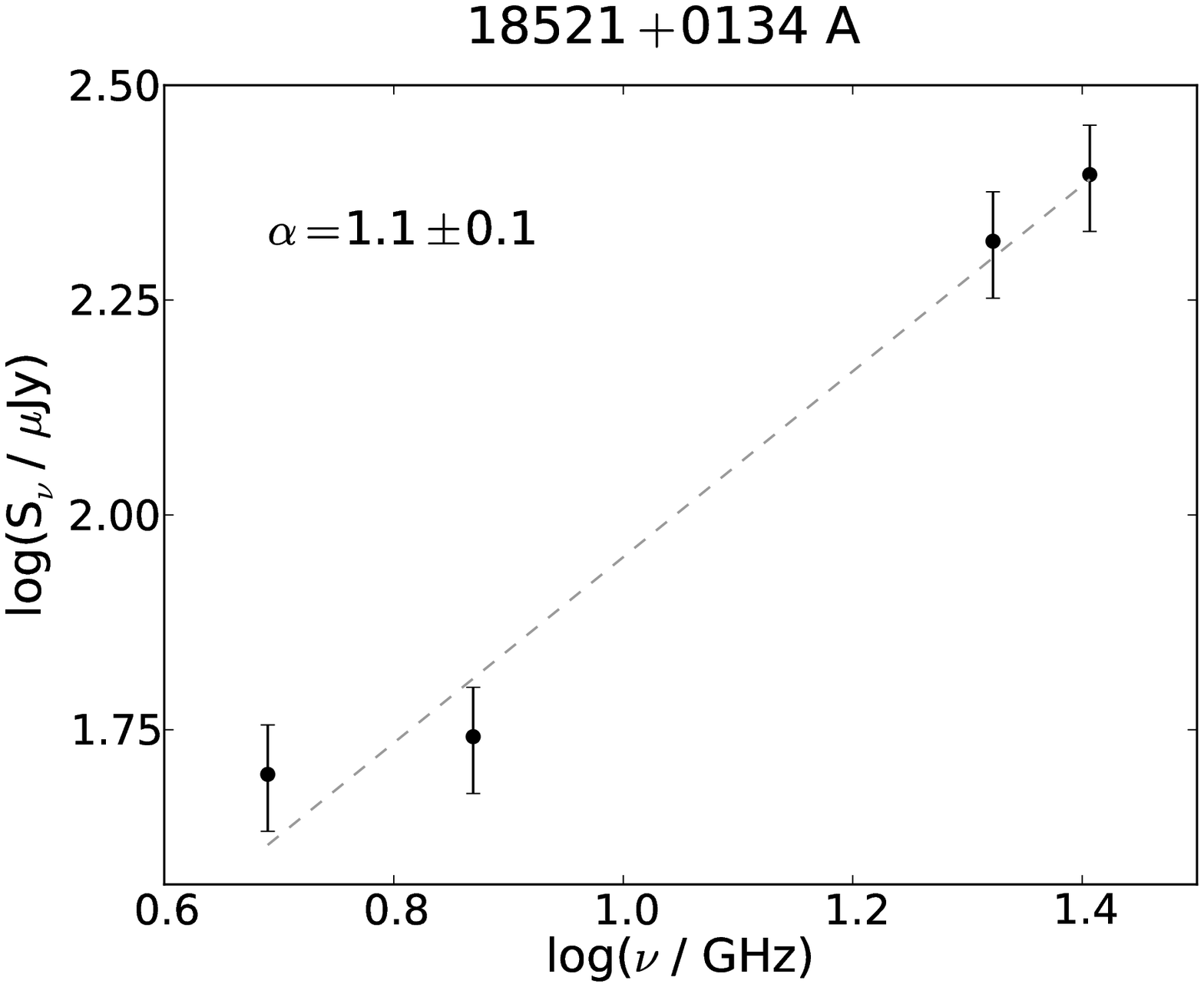} &
 \includegraphics[width=0.5\textwidth, clip, angle = 0]{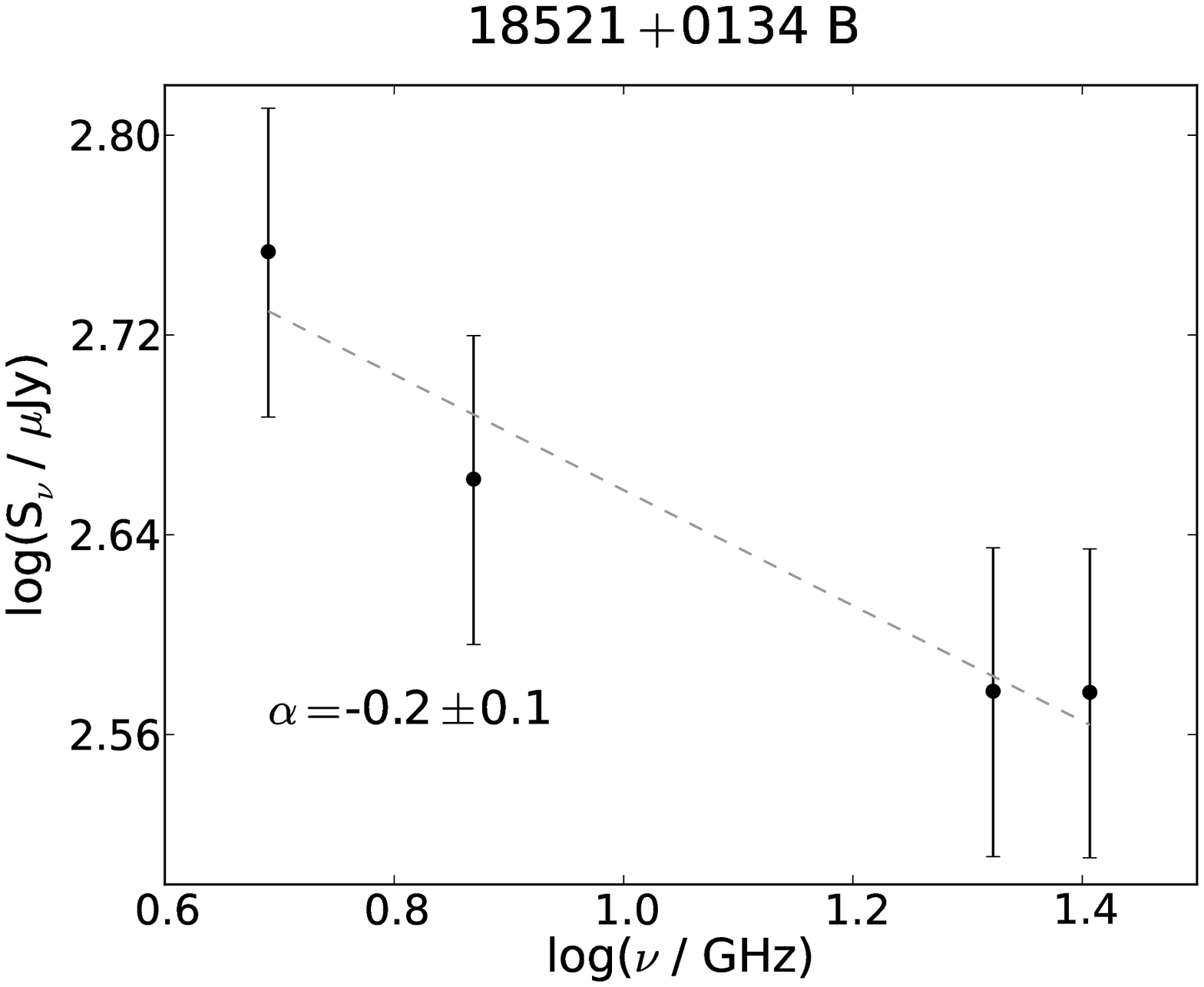} 
\end{tabular}

 \hspace*{\fill}%
\caption{\small{Continued.}}
 \label{fig:spix}
\end{figure}

\begin{figure}[htbp]
\centering
\begin{tabular}{cc}
  \ContinuedFloat
\hspace*{\fill}%
 \includegraphics[width=0.5\textwidth, clip, angle = 0]{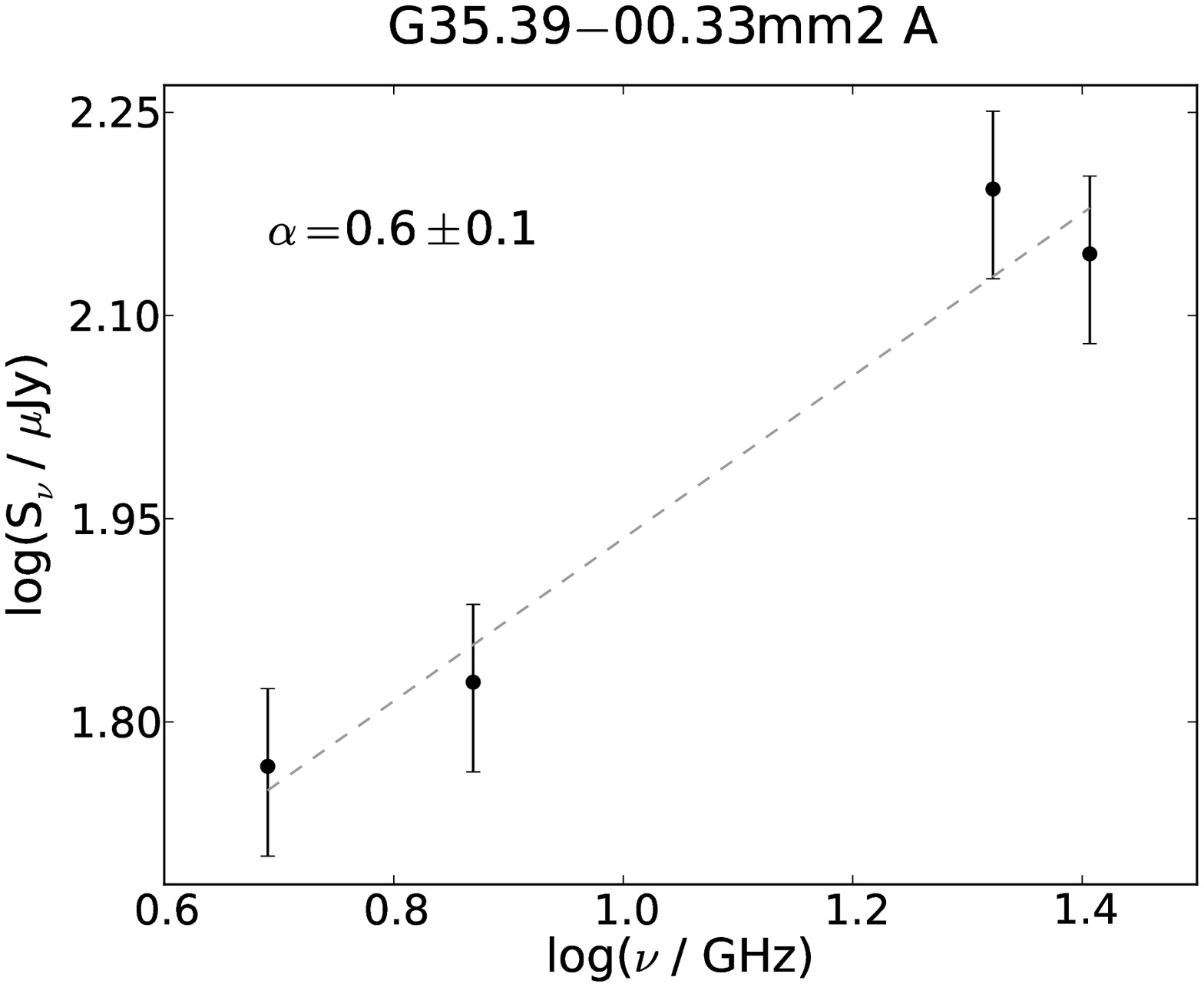} \vspace{0.5cm}&
 \includegraphics[width=0.5\textwidth, clip, angle = 0]{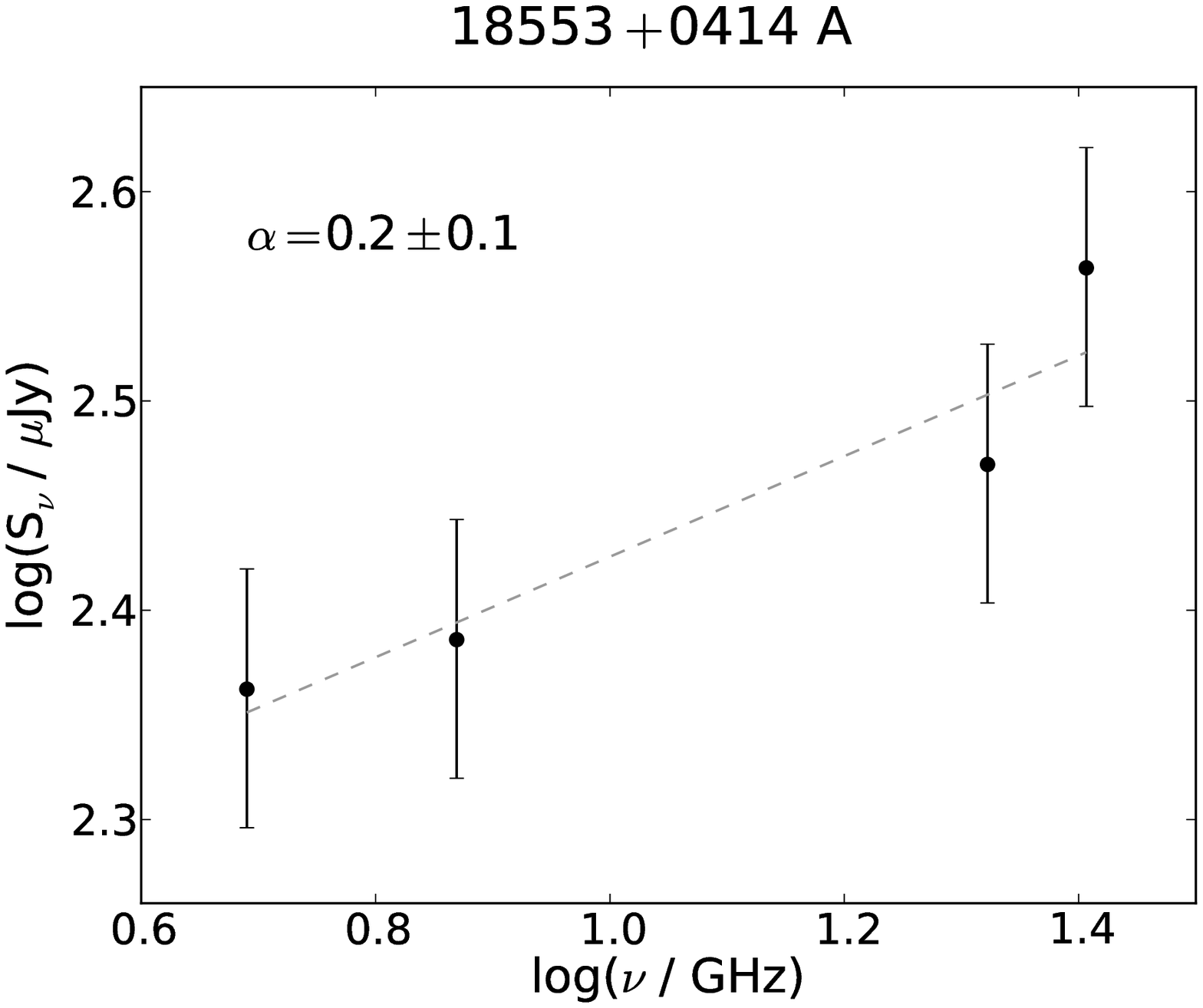} \\
  \includegraphics[width=0.5\textwidth, clip, angle = 0]{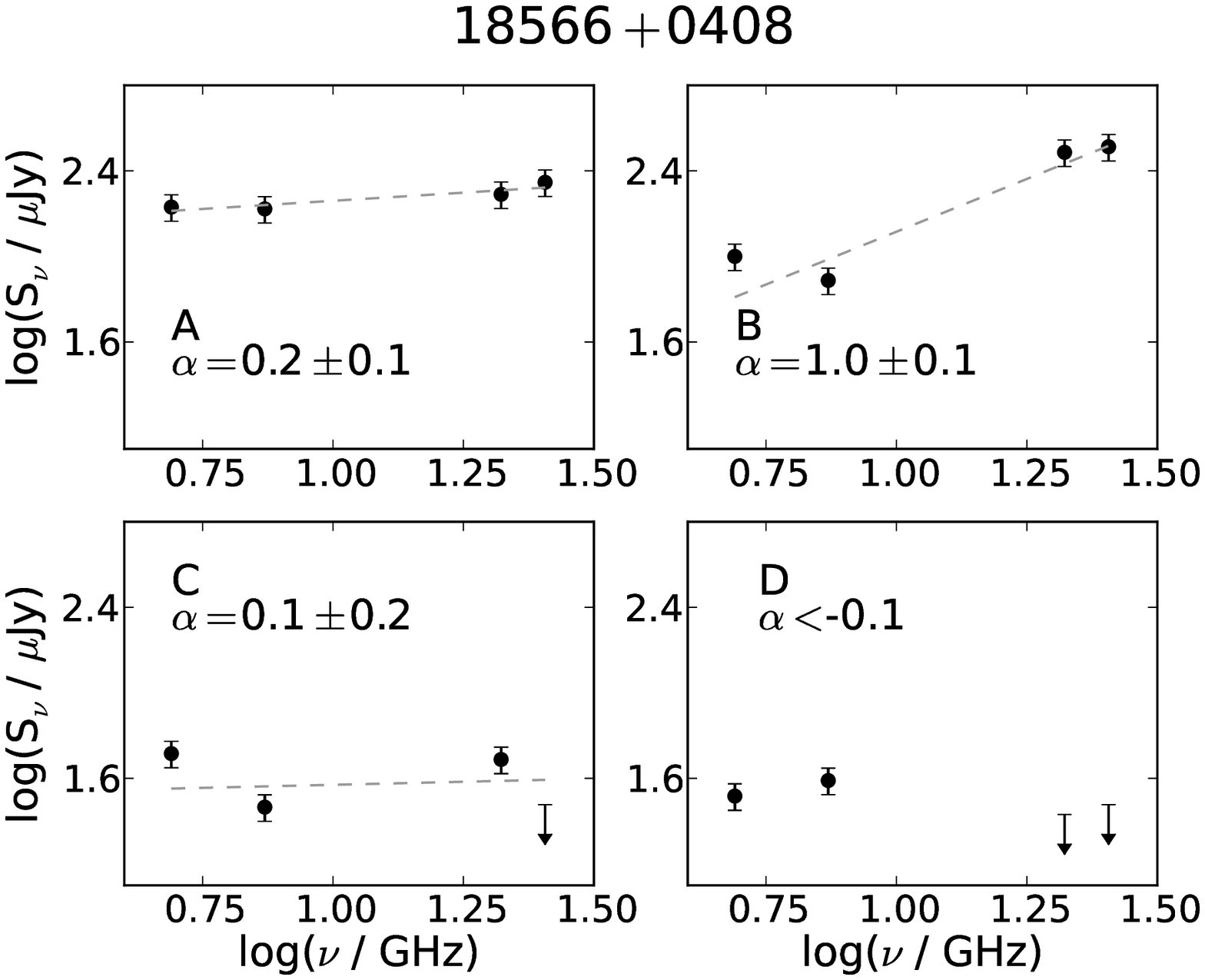} \vspace{0.5cm}&
 \includegraphics[width=0.5\textwidth, clip, angle = 0]{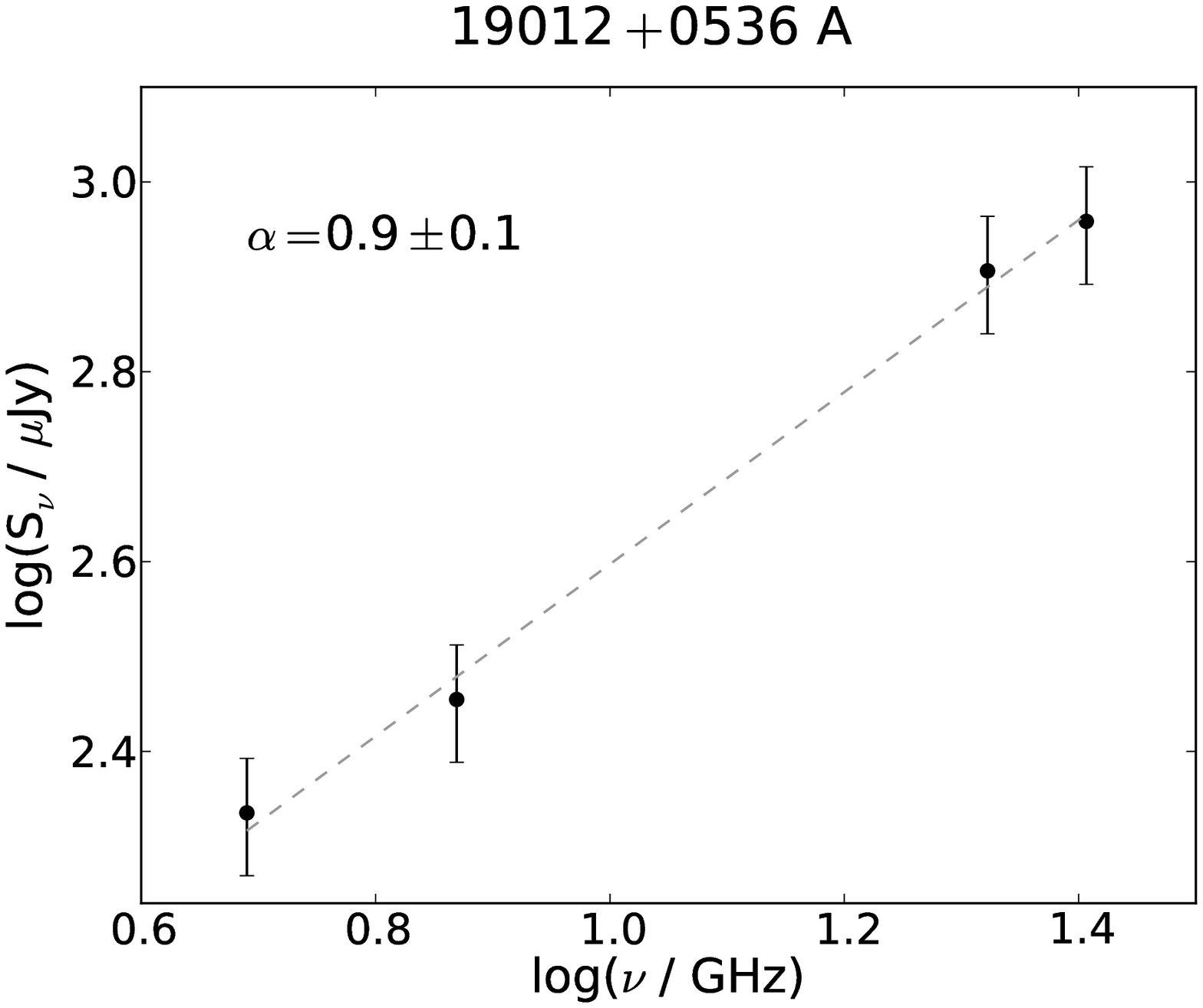} \\
  \includegraphics[width=0.5\textwidth, clip, angle = 0]{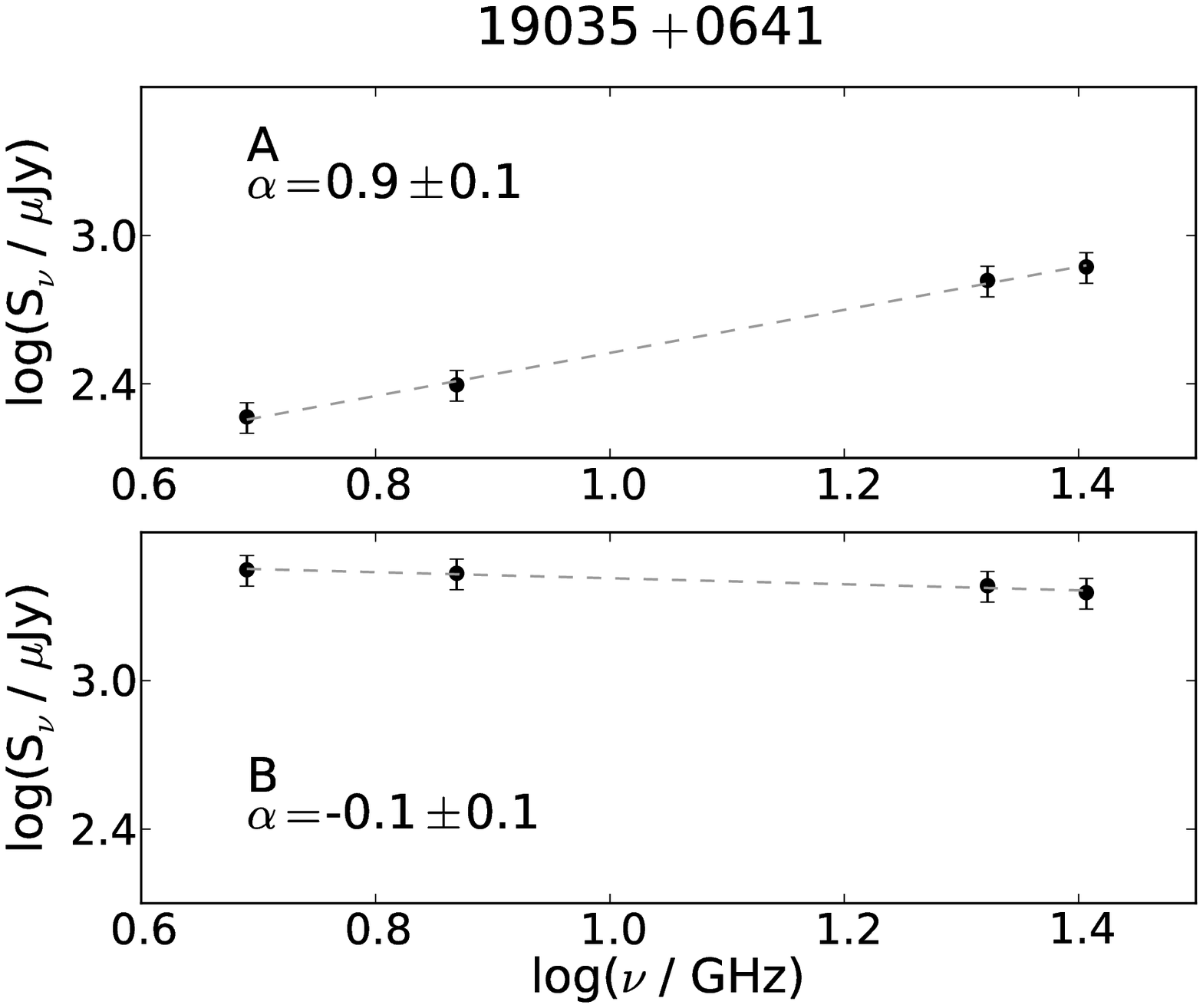} &
 \includegraphics[width=0.5\textwidth, clip, angle = 0]{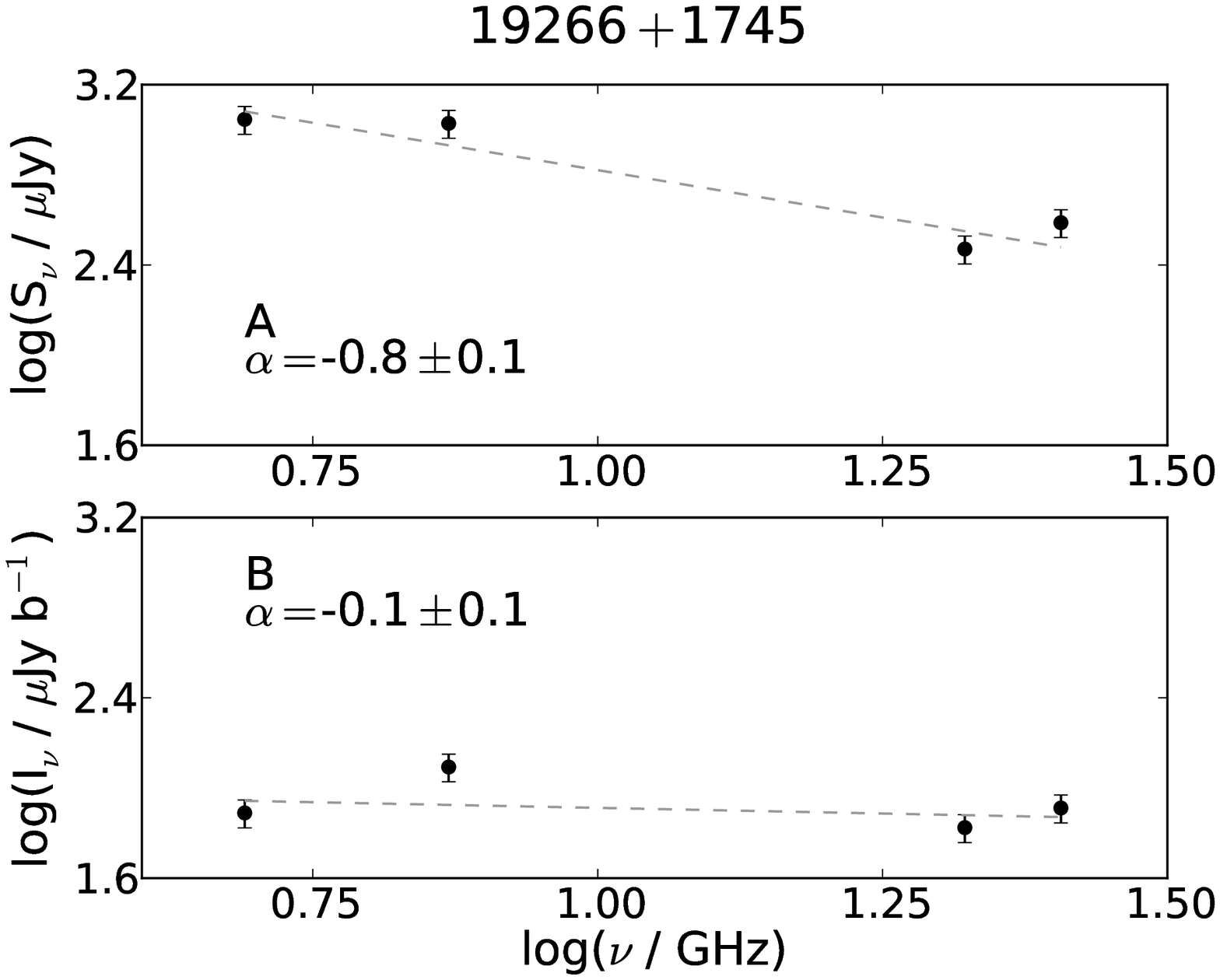} 
\end{tabular}

 \hspace*{\fill}%
\caption{\small{Continued.}}
 \label{fig:spix}
\end{figure}

\begin{figure}[htbp]
\centering
\begin{tabular}{cc}
  \ContinuedFloat
\hspace*{\fill}%
 \includegraphics[width=0.5\textwidth, clip, angle = 0]{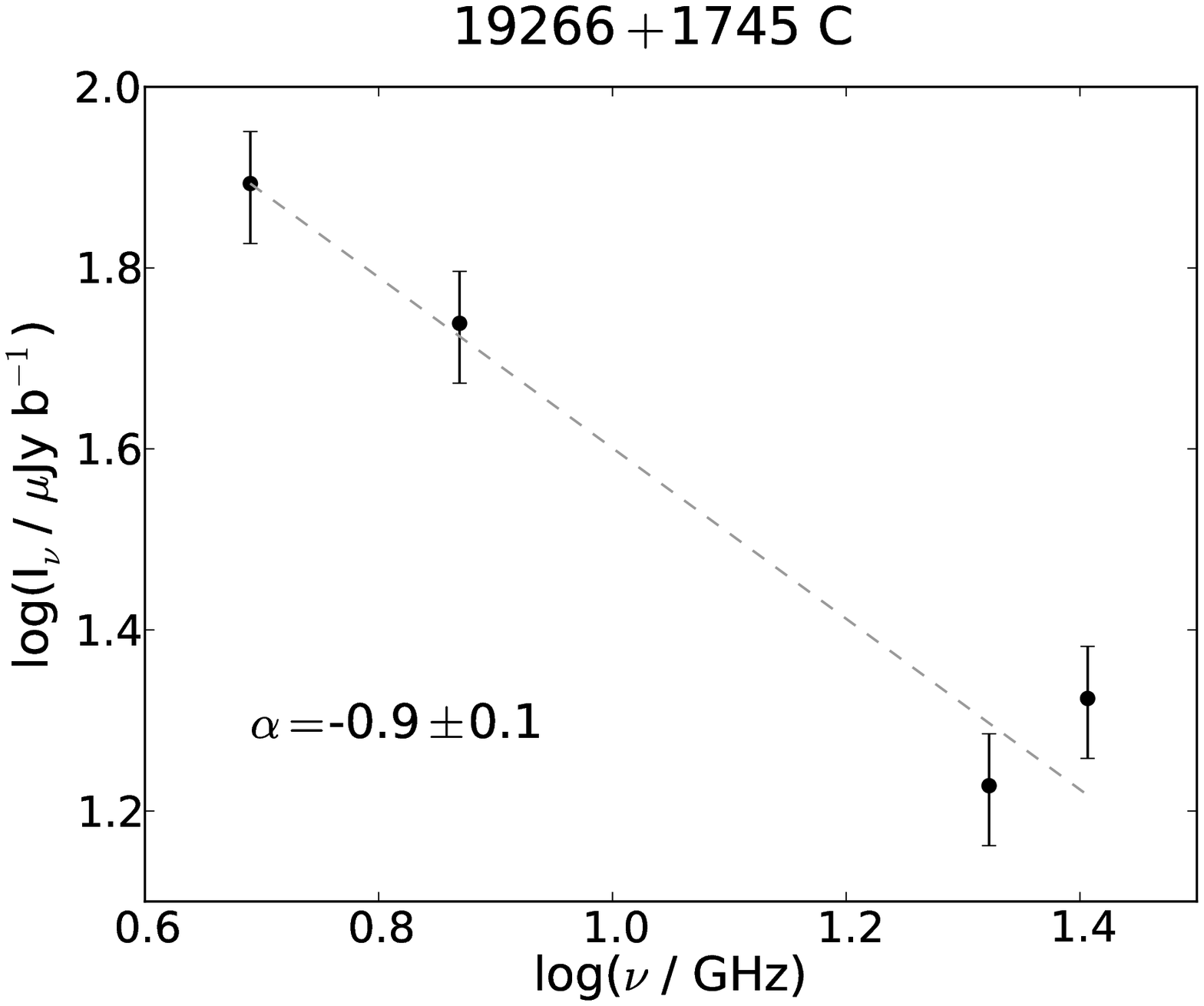} \vspace{0.5cm}&
 \includegraphics[width=0.5\textwidth, clip, angle = 0]{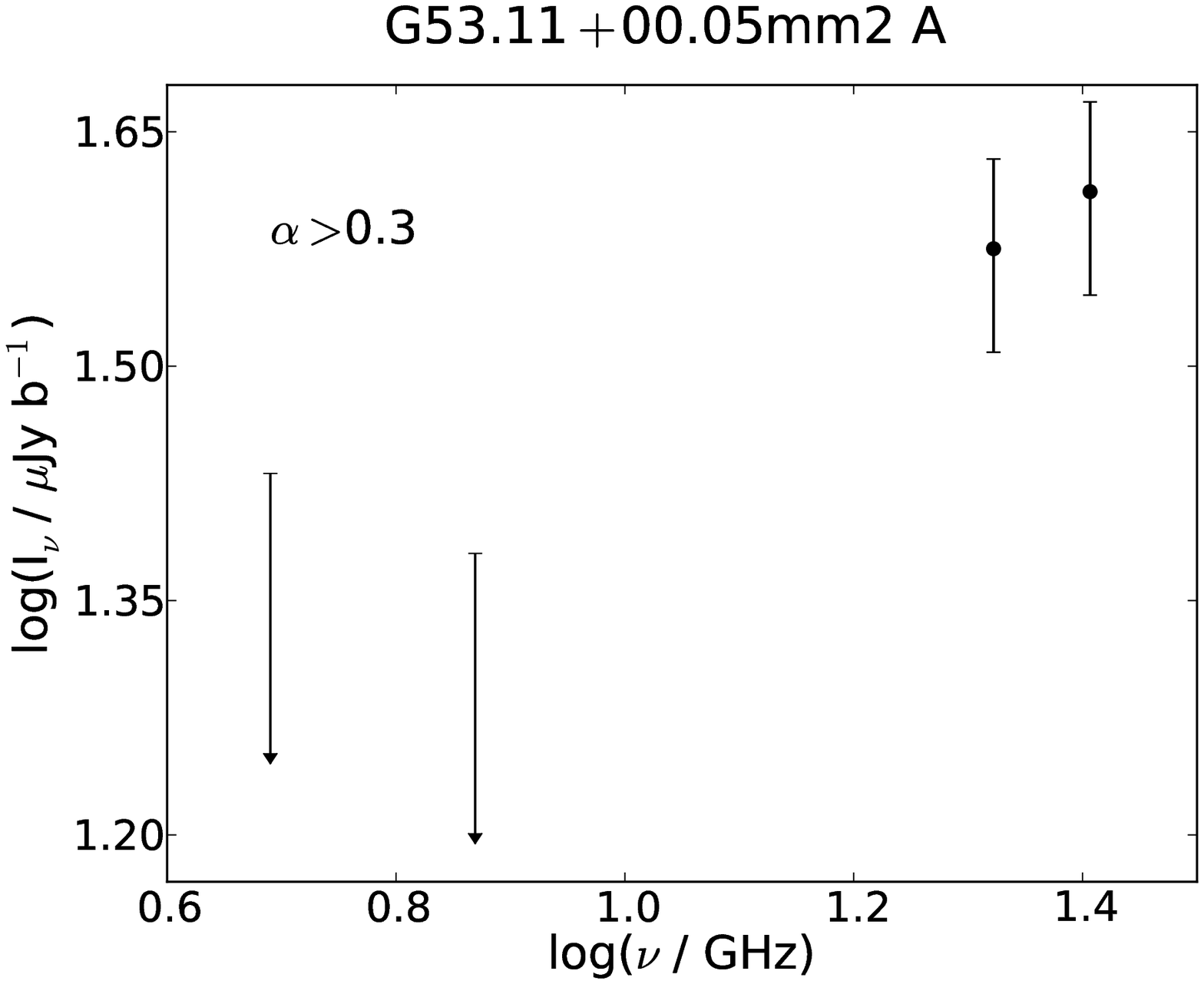} \\
  \includegraphics[width=0.5\textwidth, clip, angle = 0]{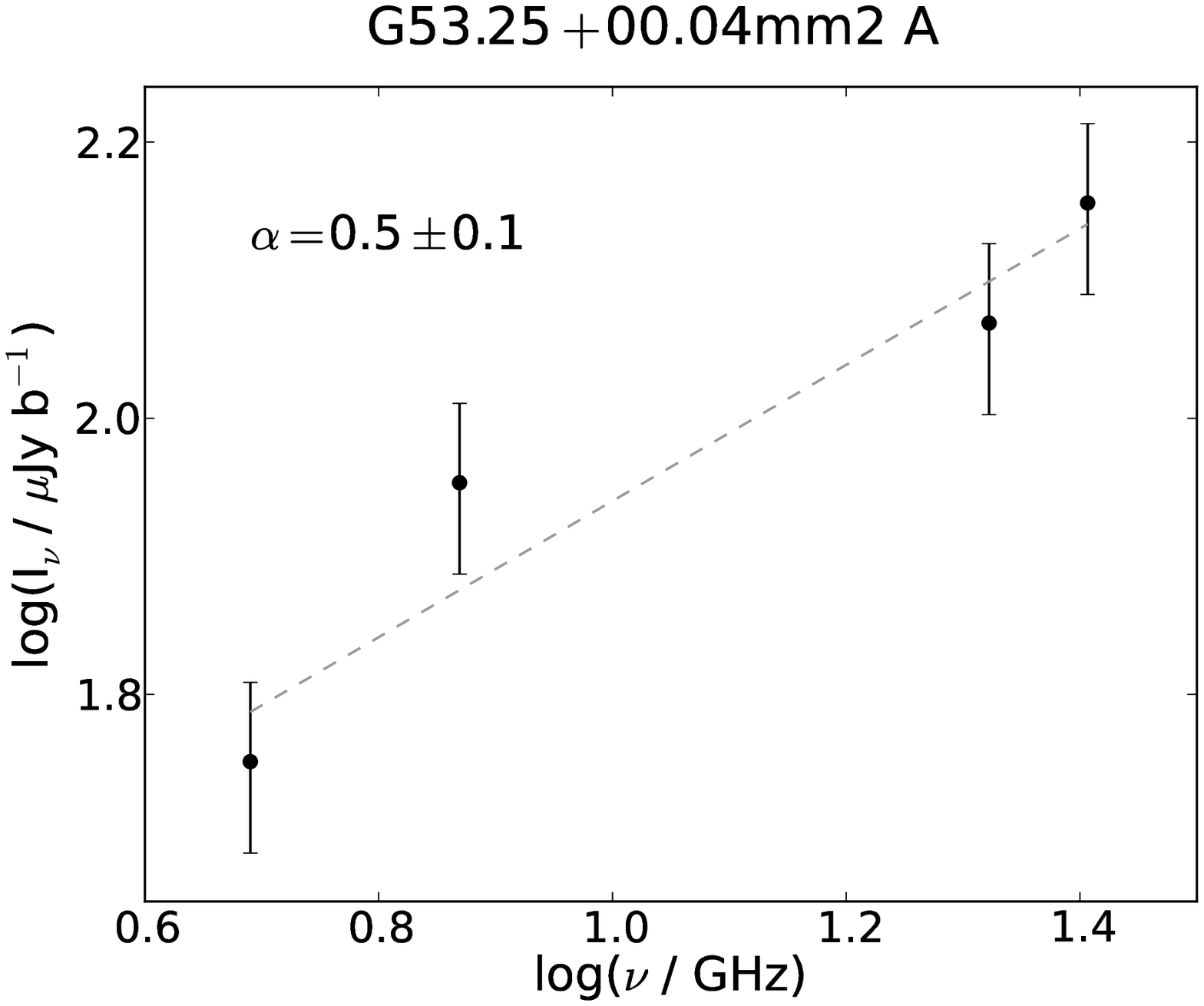} \vspace{0.5cm}&
 \includegraphics[width=0.5\textwidth, clip, angle = 0]{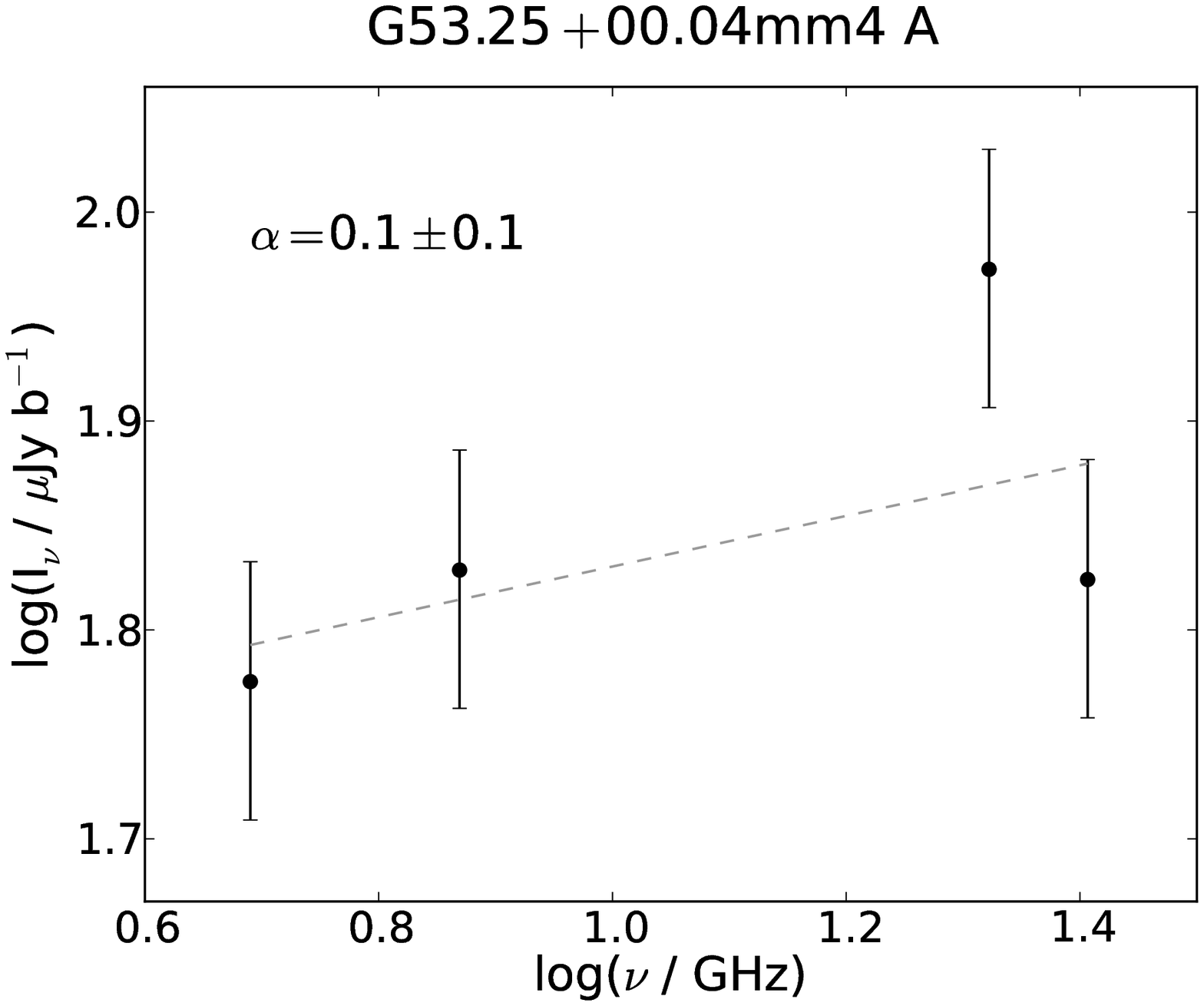} \\
  \includegraphics[width=0.5\textwidth, clip, angle = 0]{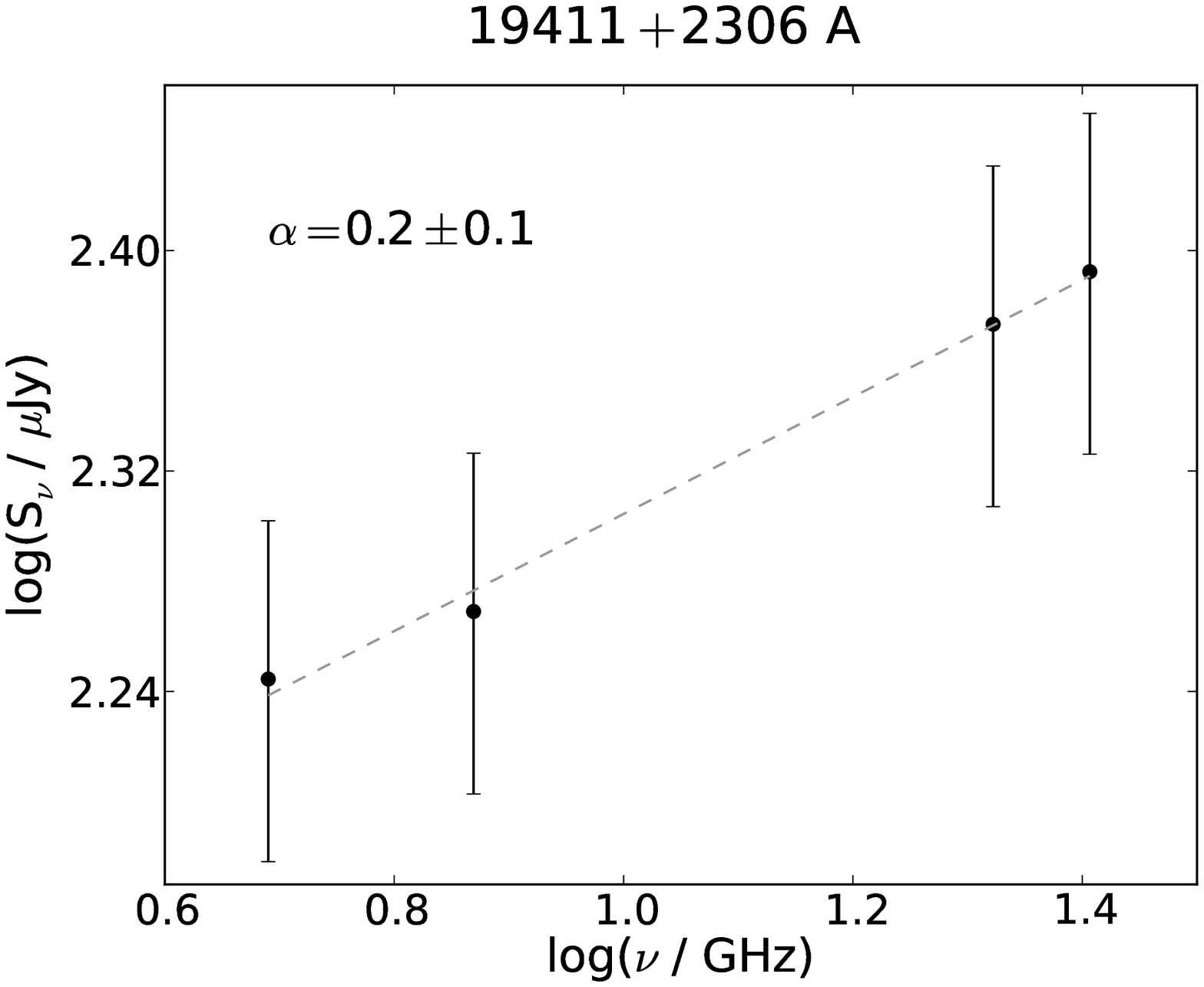} &
 \includegraphics[width=0.5\textwidth, clip, angle = 0]{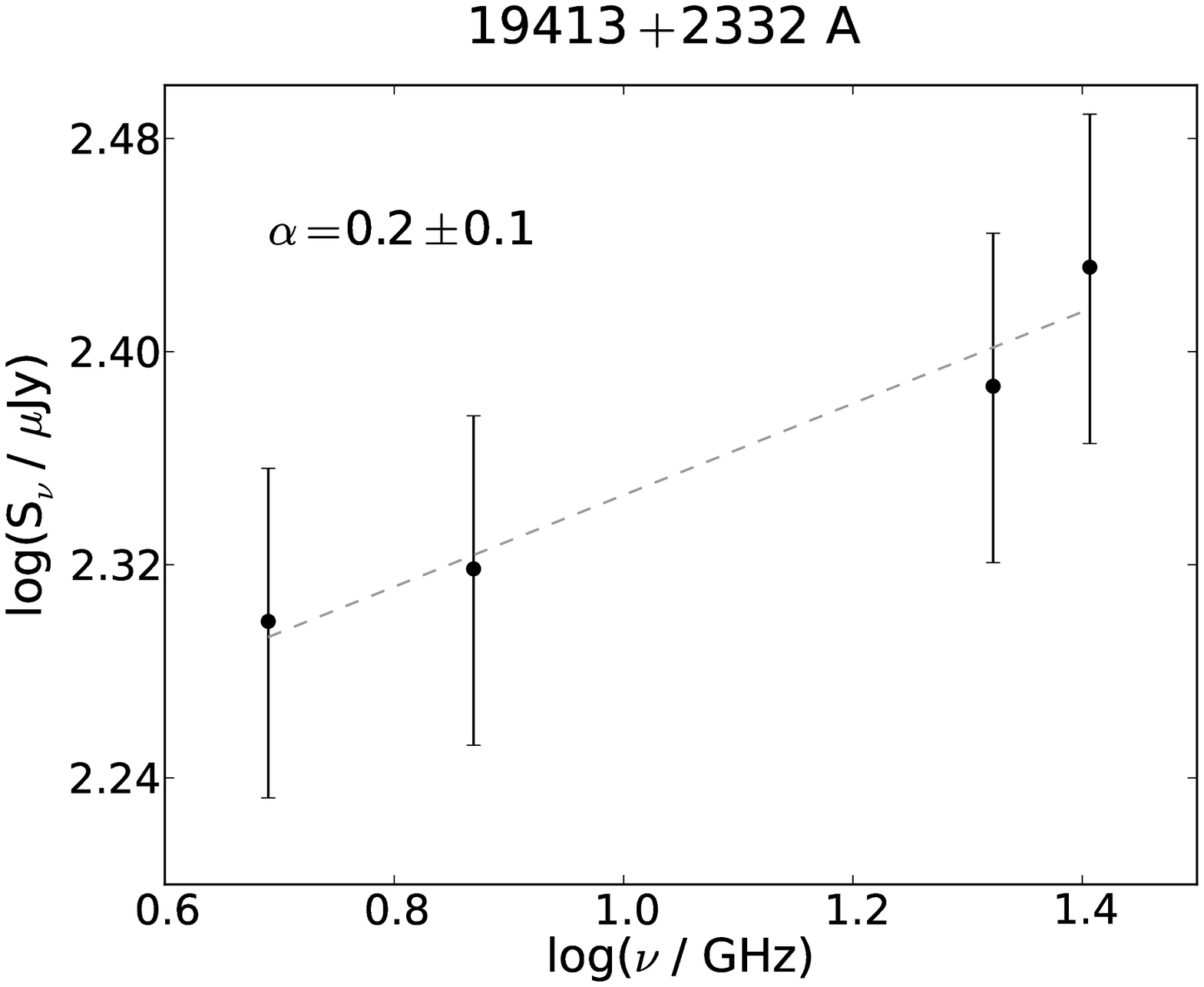} 
\end{tabular}

 \hspace*{\fill}%
\caption{\small{Continued.}}
 \label{fig:spix}
\end{figure}

\begin{figure}[htbp]
\centering
\begin{tabular}{cc}
  \ContinuedFloat
\hspace*{\fill}%
 \includegraphics[width=0.5\textwidth, clip, angle = 0]{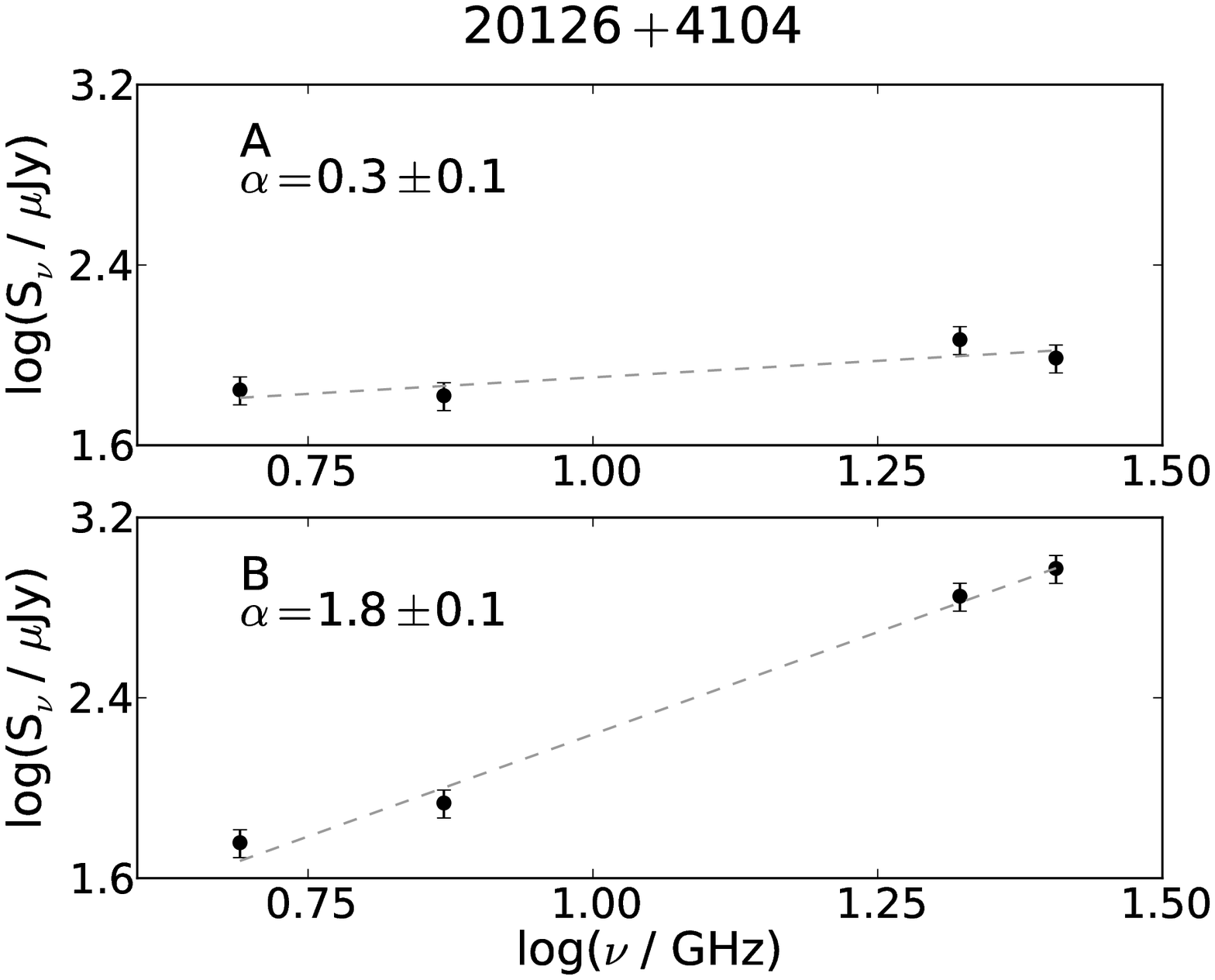} \vspace{0.5cm}&
 \includegraphics[width=0.5\textwidth, clip, angle = 0]{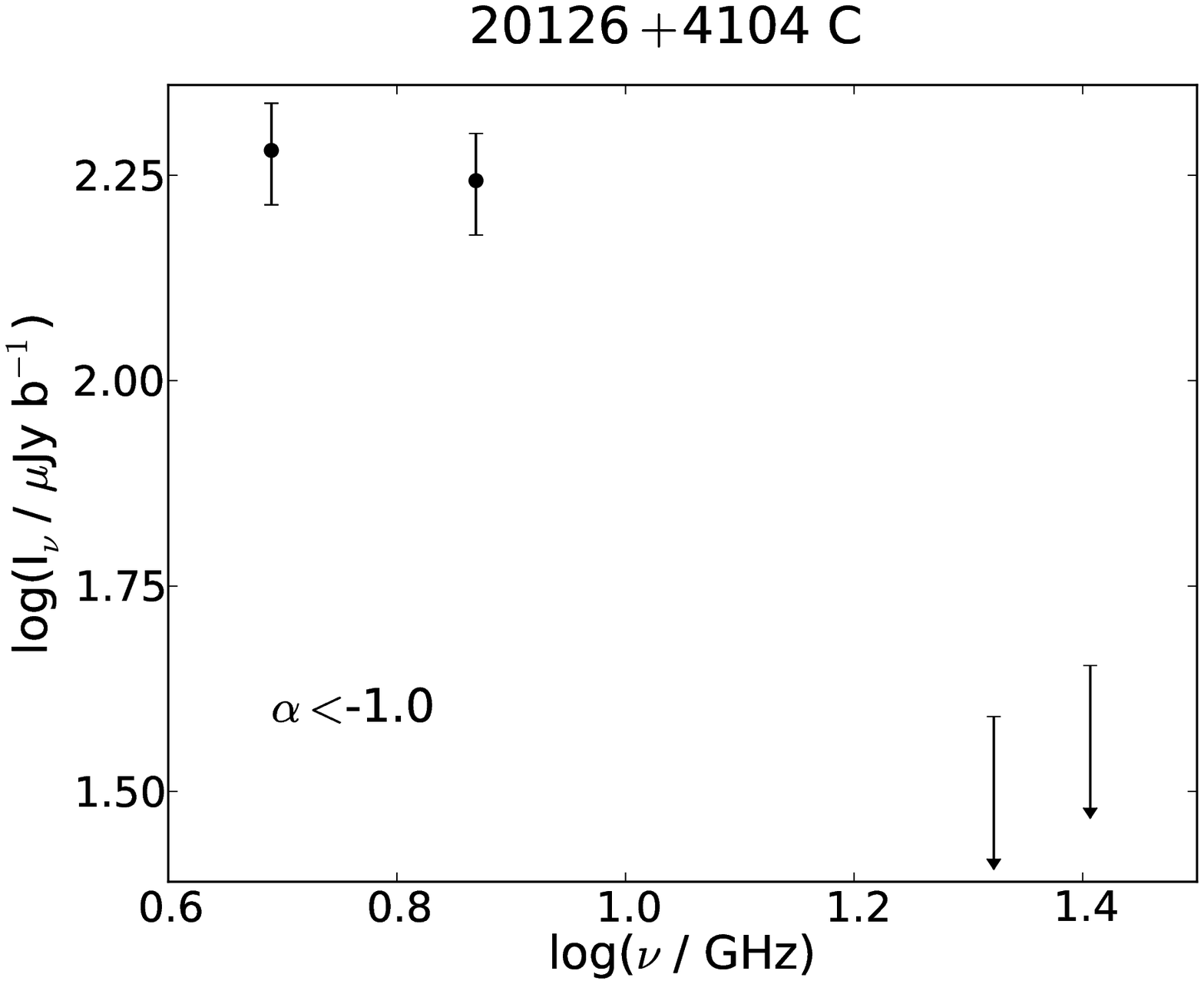} \\
  \includegraphics[width=0.5\textwidth, clip, angle = 0]{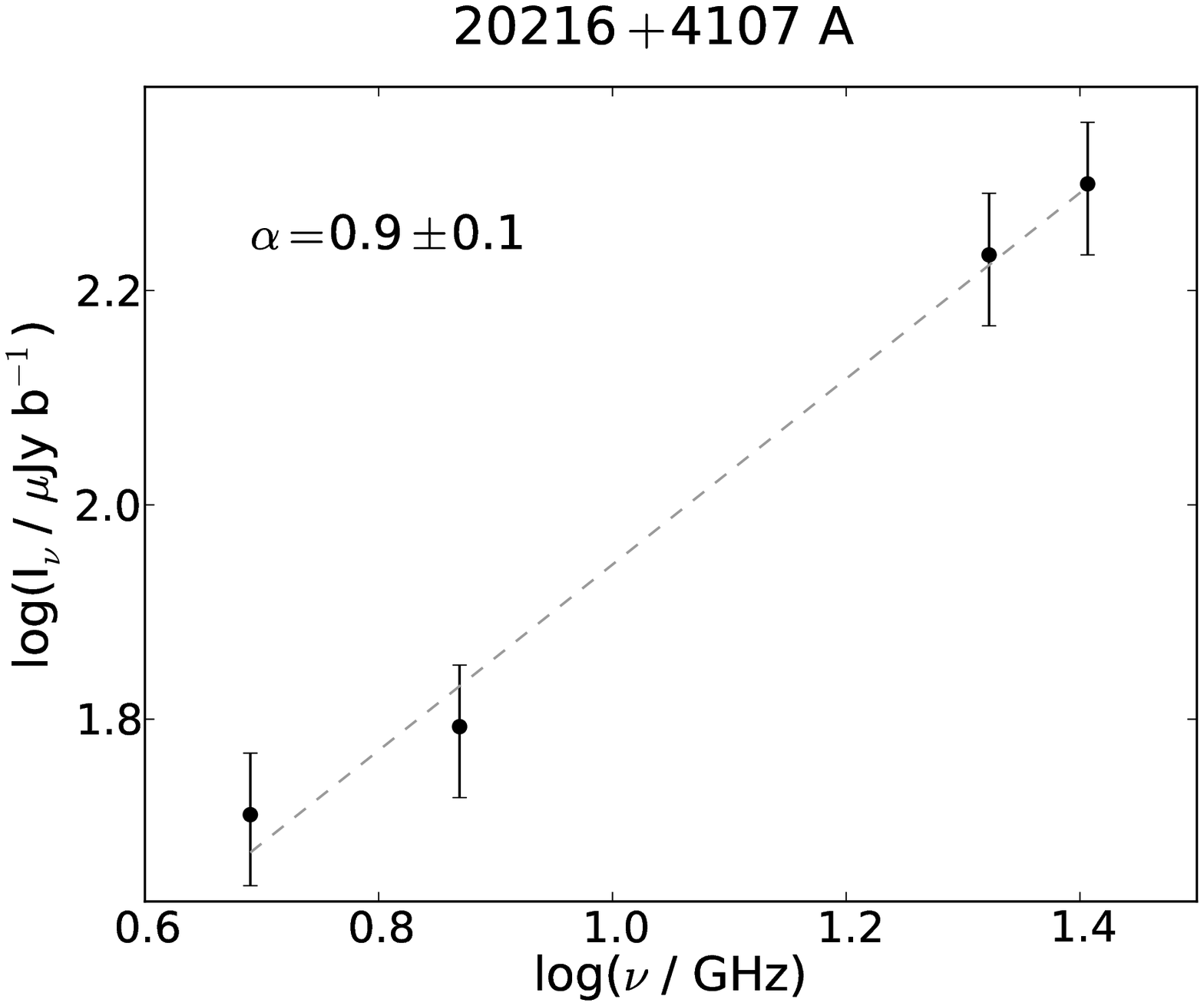} \vspace{0.5cm}&
 \includegraphics[width=0.5\textwidth, clip, angle = 0]{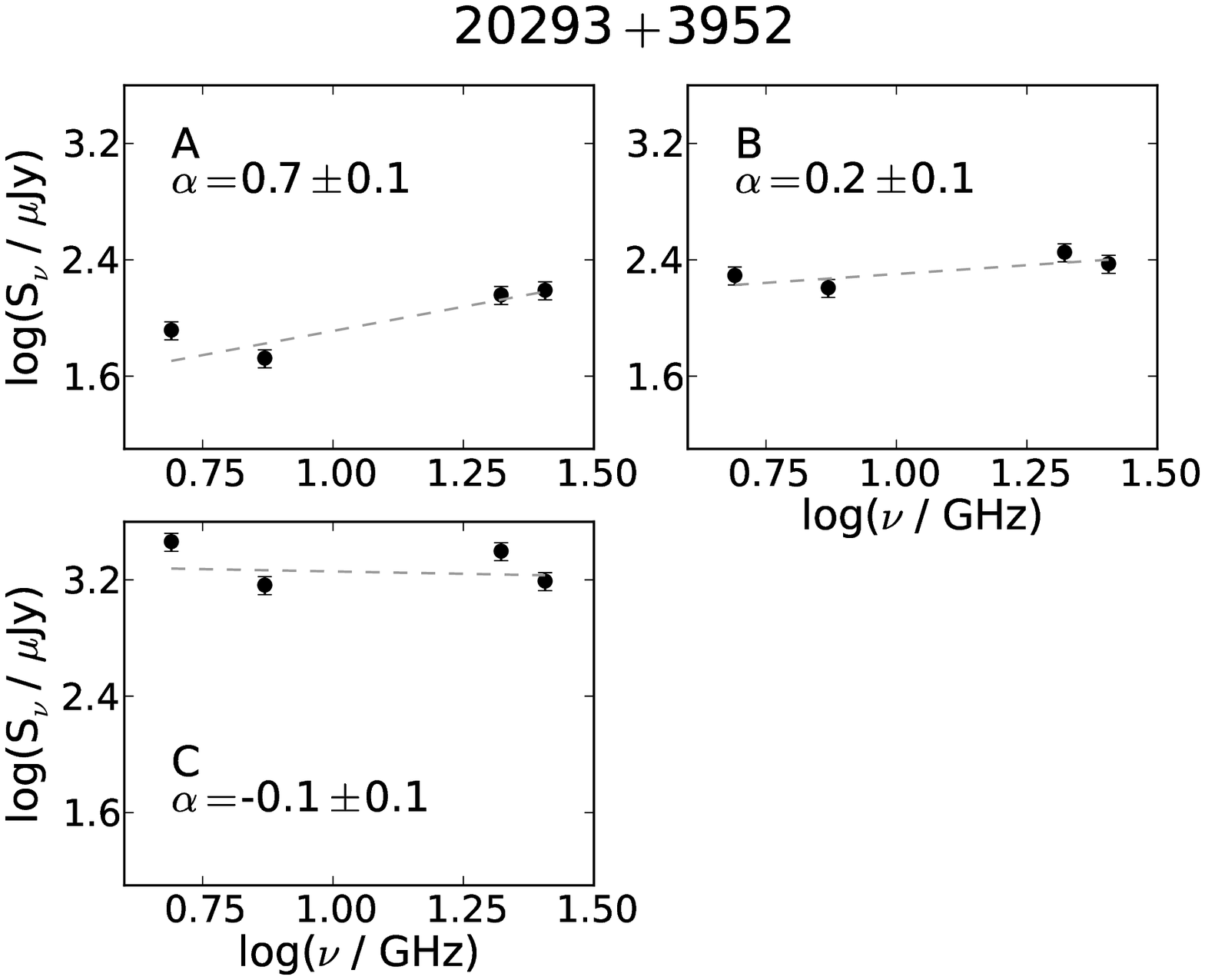} \\
  \includegraphics[width=0.5\textwidth, clip, angle = 0]{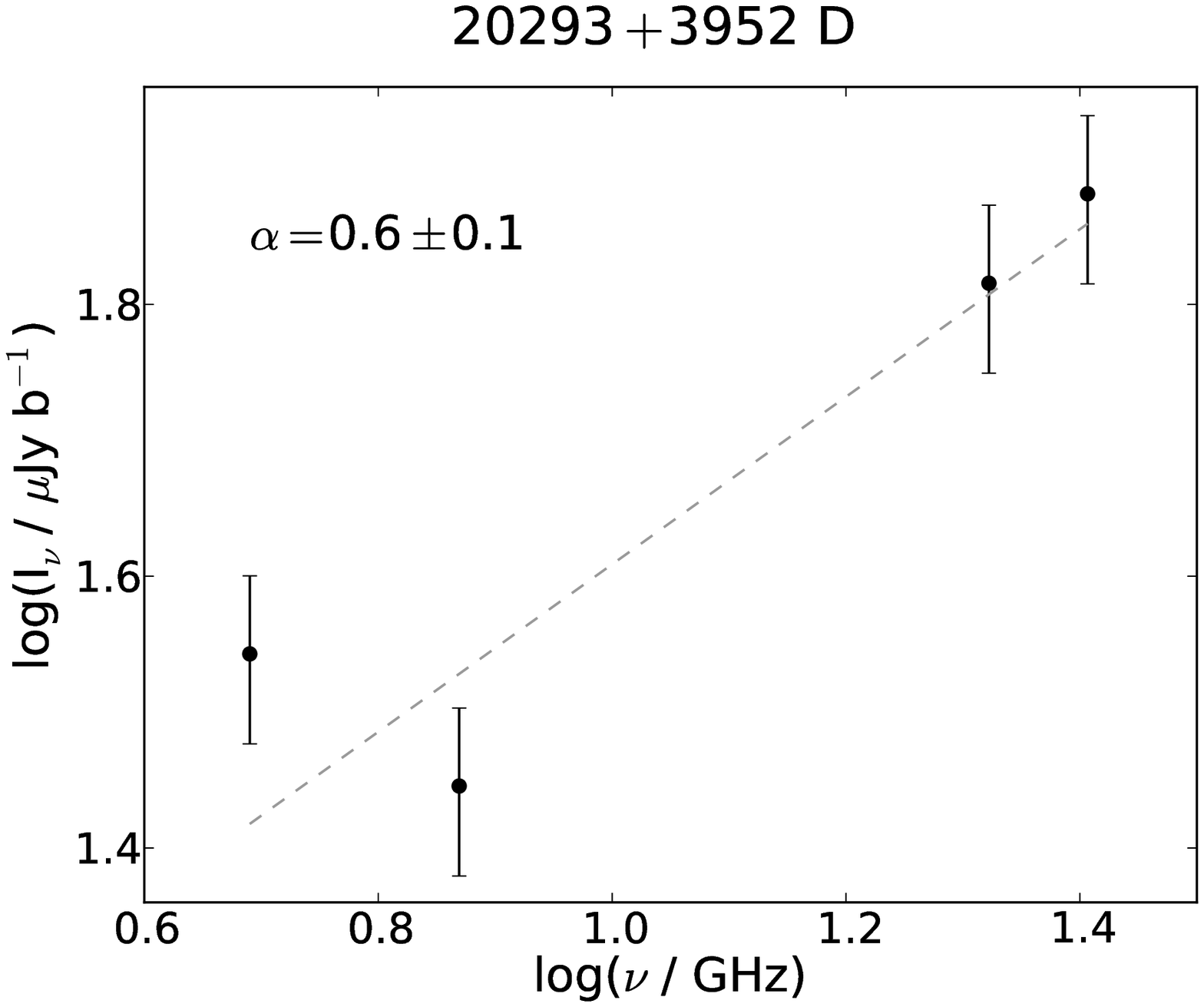} &
 \includegraphics[width=0.5\textwidth, clip, angle = 0]{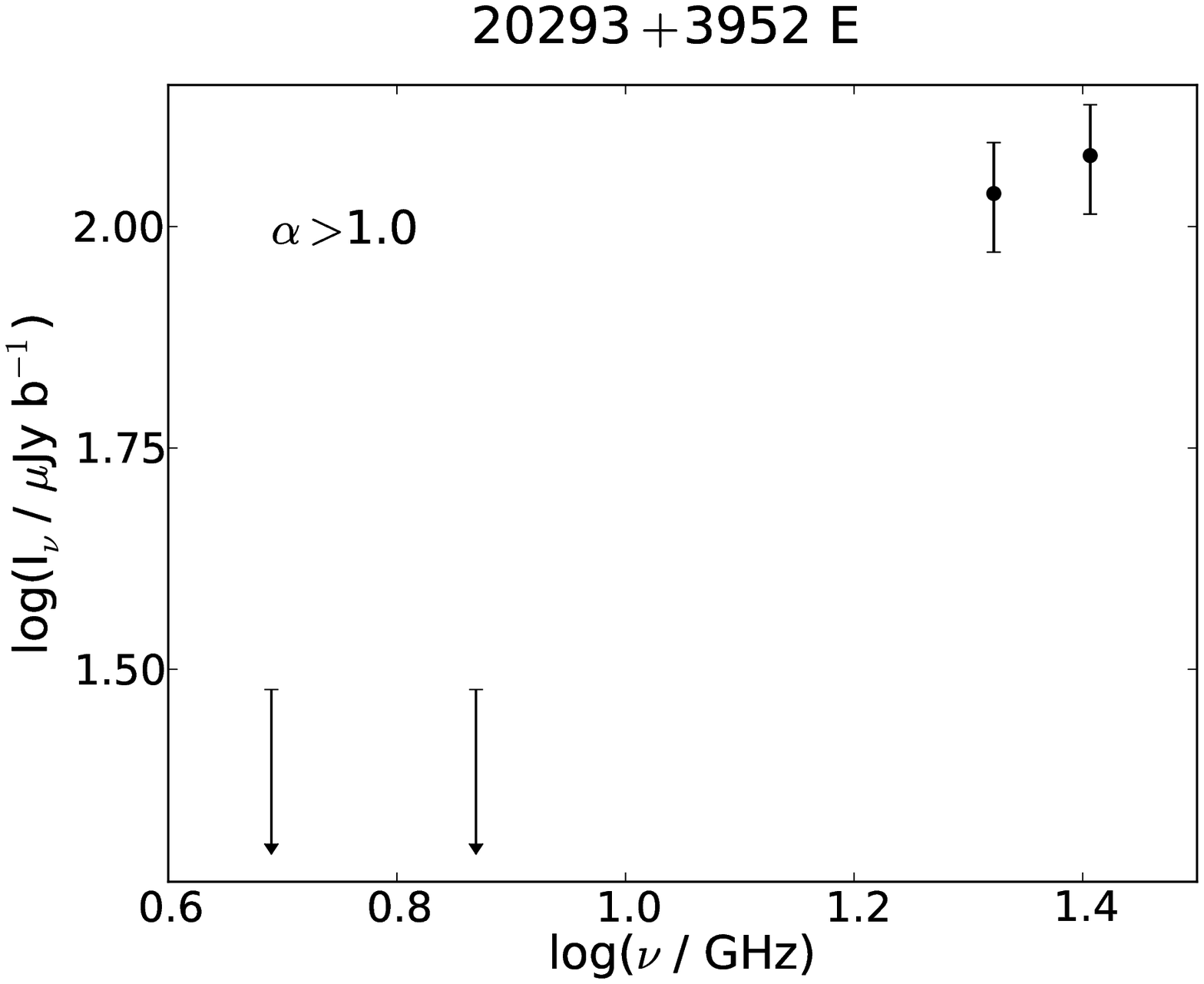} 
\end{tabular}

 \hspace*{\fill}%
\caption{\small{Continued.}}
 \label{fig:spix}
\end{figure}

\begin{figure}[htbp]
\centering
\begin{tabular}{cc}
  \ContinuedFloat
\hspace*{\fill}%
 \includegraphics[width=0.5\textwidth, clip, angle = 0]{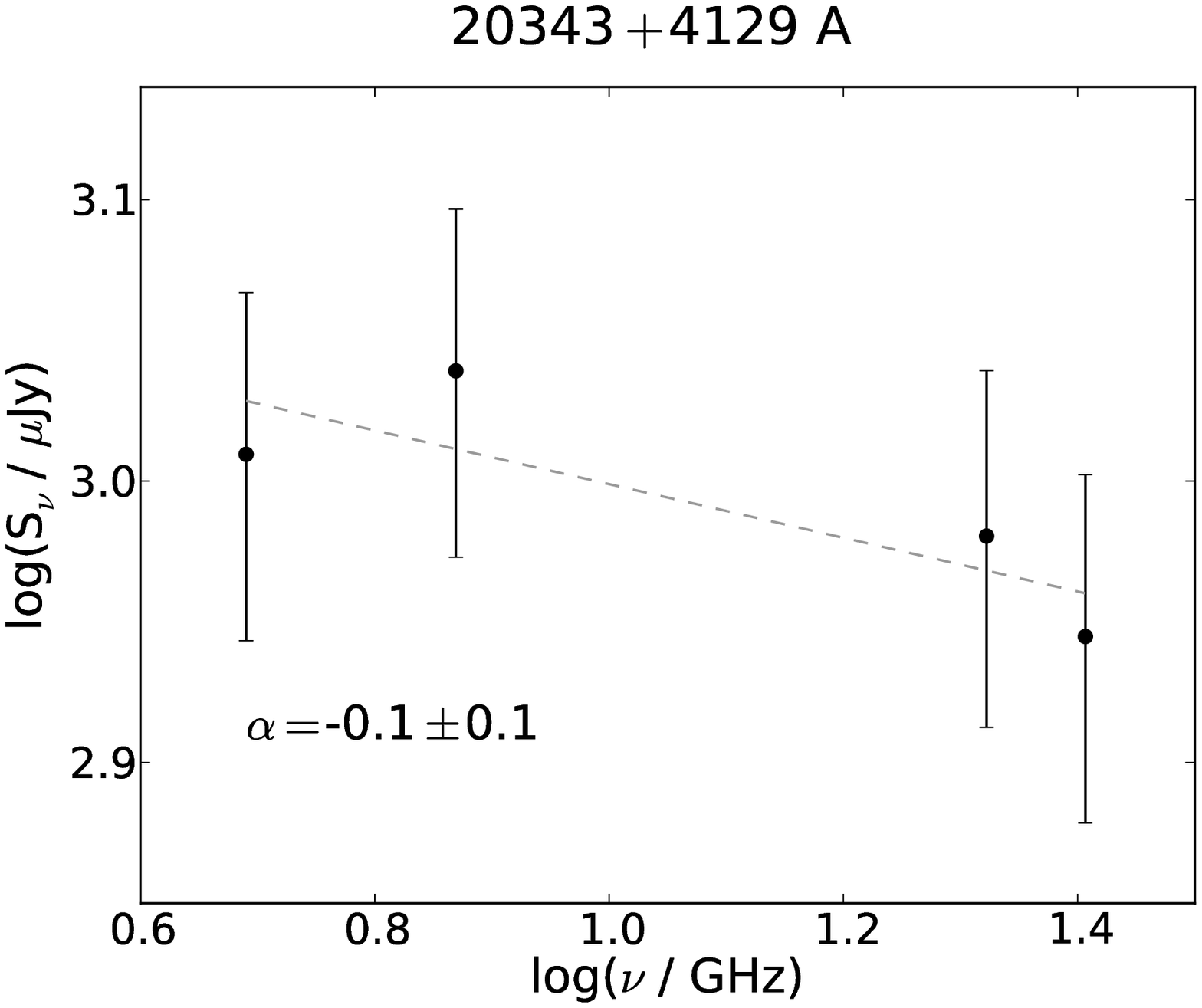} &
 \includegraphics[width=0.5\textwidth, clip, angle = 0]{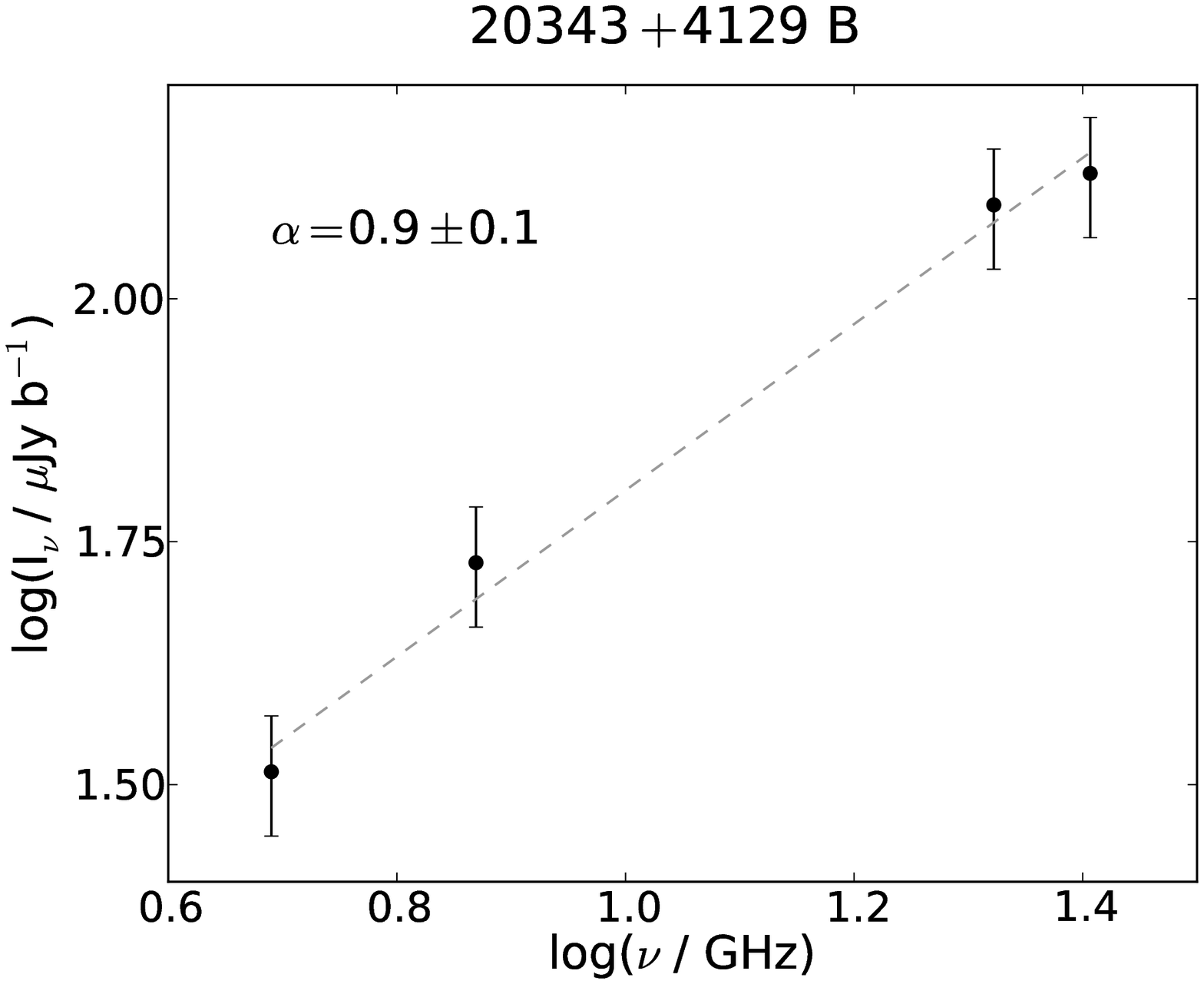} 
\end{tabular}

 \hspace*{\fill}%
\caption{\small{Continued.}}
 \label{fig:spix}
\end{figure}

\clearpage


\begin{figure}[!h]%
    \centering
    \includegraphics[width=1.0\linewidth,  trim = 3 140 10 160, clip]{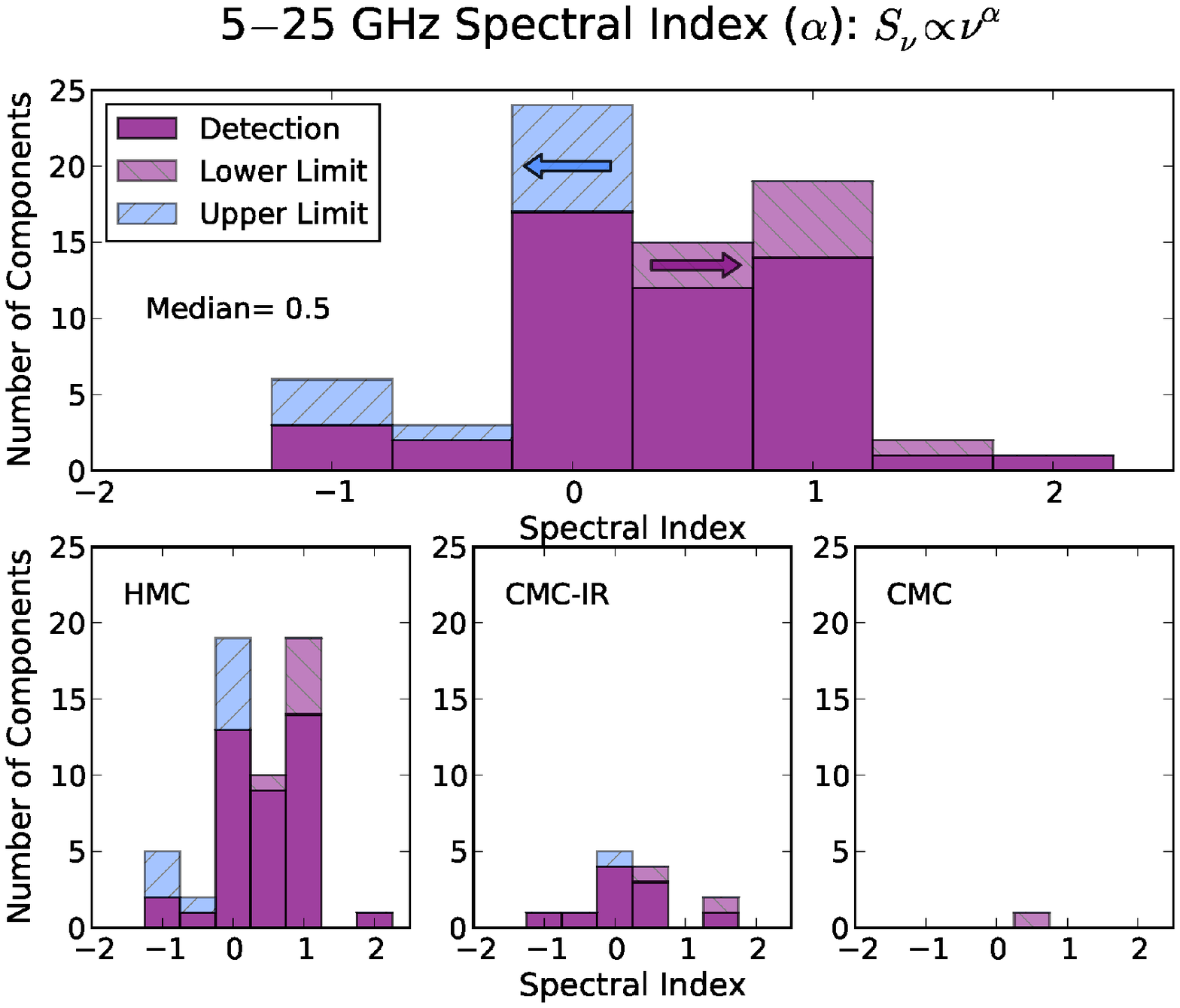}%
     \caption{\small{Spectral index ($\alpha$) distribution of  70 radio sources associated with mm--clumps, binned in increments of 0.5 in spectral index. If there is no detection at 6 cm, the estimated value of $\alpha$ is a lower limit, and
     and if there is no detection at 1.3 cm, the estimated value for $\alpha$ corresponds to an upper limit. The median value refers only to the detections. The distributions at the bottom show the results independently for each type of clump. }}%
    \label{alpha_hist}%
\end{figure}

\end{document}

%% file: latex2_v3.txt
LDN1657A$-$3       & C    & 07 05 00.7 & $-$12 16 45    &  225.399   & $-$2.608  &  0.54$\times$0.28, $-$34.4      & 4.0    & D   & CMC$-$IR     & 1.0                   &  \nodata &  \rxn \label{rxn:Forbrich2004}    \\[3pt]  
		    & K    &            &                &            &           &  0.45$\times$0.29, $-$4.7       & 12.5   &     &              &        \\[9pt] 	

UYSO1              & C    & 07 05 10.8 & $-$12 18 59    &  225.451   & $-$2.588  &  0.54$\times$0.27, $-$35.4      &  4.0   & D   & CMC$-$IR     & 1.0                   & 0.24     &  \osref{rxn:Forbrich2004} \rxn \label{rxn:Linz2010}     \\[3pt]  
		    & K    &            &                &            &           &  0.45$\times$0.29, $-$4.4       & 13.0   &                    &      &           \\[9pt]  
		                                           
G11.11$-$0.12P1    & C    & 18 10 28.4 & $-$19 22 29    &  11.109    & $-$0.114  &  0.49$\times$0.27, $-$7.9       &  5.0   & D   & CMC$-$IR     & 3.6                   & 1.3      &  \rxn \label{rxn:Pillai2006} \rxn \label{rxn:Henning2010}  \\[3pt]  
		    & K    &            &                &            &           &  0.67$\times$0.27, $-$33.4      &  8.0   &                    &      &            \\[10pt]  

18089$-$1732       & C    & 18 11 51.3 & $-$17 31 29    &  12.889    &  0.489    &  0.46$\times$0.29, $-$19.8      &  6.0   & D   & HMC          & 2.34   & 13.5\tablenotemark{a}  &  \rxn \label{rxn:Xu2011}  \rxn \label{rxn:Sridharan2002}  \\[3pt]  
		    & K    & 		 &                &            &           &  0.53$\times$0.29, $-$22.4      &  10.0  &     &              &      \\

%% file: latex1_v3.txt
LDN1657A$-$3      & A                 &      4.9    & 07 05 00.65      & $-$12 16 44.9    & 530                       & 528                       & C         & y\tablenotemark{a} &  $-$1.2(0.1)  \\[3pt]

                  &                   &      7.4    & 07 05 00.65      & $-$12 16 44.9    & 758                       & 732                       &           &        &          \\[3pt]

                  &                   &     20.9    & 07 05 00.65      & $-$12 16 45.1    & 103                       & 88                        &           &        &          \\[3pt]

                  &                   &     25.5    & 07 05 00.65      & $-$12 16 44.9    & 95                        & 90                        &           &        &          \\[9pt]

                  & G225.394+2.597    &      4.9    & 07 05 02.53      & $-$12 16 11.5    & 315                       & 302                       & C         & n\tablenotemark{a} &  $-$0.4(0.1)  \\[3pt]

                  &                   &      7.4    & 07 05 02.53      & $-$12 16 11.5    & 172                       & 172                       &           &        &          \\[3pt]

                  &                   &     20.9    & 07 05 02.53      & $-$12 16 11.5    & 146                       & 146                       &           &        &          \\[3pt]

                  &                   &     25.5    & 07 05 02.53      & $-$12 16 11.4    & 137                       & 132                       &           &        &          \\